\RequirePackage{fix-cm}
\documentclass[aps,prd,twocolumn]{revtex4}

\usepackage{soul}
\RequirePackage{graphicx,booktabs, multirow}
\RequirePackage{amsmath}
\RequirePackage{amssymb}
\RequirePackage{bm}
\RequirePackage{color}
\RequirePackage[pdfencoding=auto, psdextra]{hyperref}
\RequirePackage{verbatim}
\allowdisplaybreaks 
\usepackage{natbib} 
\usepackage{orcidlink} 
\newcommand{\ba}{\begin{eqnarray}}
\newcommand{\ea}{\end{eqnarray}}

\begin{document}
\bibliographystyle{unsrtnat} 

\title{ Radiative decays of the 1$P$, 1$D$, 2$S$, and 2$P$ $\Lambda_c$ and 1$D$, 2$S$, and 2$P$ $\Xi_c$ charmed baryons }

\author{R. Gamboa-Goni~\orcidlink{0000-0003-1026-7633}}
\affiliation{Tecnologico de Monterrey, Escuela de Ingenieria y Ciencias, General Ramon Corona 2514,
Zapopan 45138, Mexico}

\author{Ailier Rivero-Acosta~\orcidlink{0000-0001-7651-3927}}\email[Corresponding author: ]{ailierrivero@gmail.com} 
\affiliation{Departamento de F\'isica, DCI, Campus Le\'on, Universidad de Guanajuato, Loma del Bosque 103, Lomas del Campestre, 37150, Le\'on, Guanajuato, Mexico}
\affiliation{INFN, Sezione di Genova, Via Dodecaneso 33, 16146 Genova, Italy}

\author{H. Garc{\'i}a-Tecocoatzi~\orcidlink{0000-0002-8767-9786}}\email[Corresponding author: ]{hugo.garcia.t@tec.mx}
\affiliation{Tecnologico de Monterrey, Escuela de Ingenieria y Ciencias, General Ramon Corona 2514,
Zapopan 45138, Mexico}

\author{A. Gutierrez-Rodriguez~\orcidlink{0000-0003-4431-3159}}
\affiliation{Unidad Acad\'emica de Física, Universidad Autónoma de Zacatecas, Solidaridad, esquina la Bufa S/N, Zacatecas, 98060, Mexico }

\author{A. Ramirez-Morales~\orcidlink{0000-0001-8821-5708}}
\affiliation{Unidad Acad\'emica de Física, Universidad Autónoma de Zacatecas, Solidaridad, esquina la Bufa S/N, Zacatecas, 98060, Mexico }

\author{E. Santopinto~\orcidlink{0000-0003-3942-6554}}
\affiliation{INFN, Sezione di Genova, Via Dodecaneso 33, 16146 Genova, Italy }

\author{Carlos Alberto Vaquera-Araujo~\orcidlink{0000-0001-8578-9263}}
\affiliation{Secretar\'ia de Ciencia, Humanidades, Tecnolog\'ia e Innovaci\'on,  Insurgentes Sur 1582. Colonia Cr\'edito Constructor, Benito Ju\'arez, 03940, Ciudad de M\'exico, Mexico}
\affiliation{Departamento de F\'isica, DCI, Campus Le\'on, Universidad de
  Guanajuato, Loma del Bosque 103, Lomas del Campestre 37150, Le\'on, Guanajuato, Mexico}
\affiliation{Dual CP Institute of High Energy Physics, 28045, Colima, Mexico}

\begin{abstract}
We analyze the radiative decays of the 1$P$, 1$D$, 2$S$, and 2$P$ $\Lambda_c$ and 1$D$, 2$S$, and 2$P$ $\Xi_c$ charmed baryons, belonging to the flavor anti-triplet ($\bf {\bar 3}_{\rm F}$), using the constituent quark model. We compute electromagnetic transitions from ground and $P$-wave states to ground states, as well as from \mbox{second-shell} states to both ground and $P$-wave final states. Electromagnetic decay widths are especially valuable for identifying resonances when multiple states share similar mass and total decay width.  We give branching ratios of several electromagnetic decay widths that can confirm the assignment of the $\Xi_c(3055)$ reported by LHCb. Likewise, we provide branching-ratio predictions that can help guide the assignment of the $\Xi_c(3080)$ states, and discuss their possible interpretation either as $1D$ states with $\mathbf{J^P}=5/2^{+}$ or as $2S$ states with $\mathbf{J^P}=1/2^{+}$. For the first time, this work provides calculations of electromagnetic decays for $D_\rho$-wave states, $\rho-\lambda$ mixed configurations, and $\rho$-mode radially excited states in singly charmed baryons of the flavor anti-triplet. Both experimental and model-dependent uncertainties are taken into account throughout our analysis.

\end{abstract}

\maketitle

\section{Introduction }  
The recent observation of numerous excited states in the heavy-baryon sector has considerably improved our understanding of singly heavy baryons. Charmed baryons, in particular, provide a prominent example of this progress. In addition to the determination of their masses and decay widths, increasing attention has been devoted to measuring their quantum numbers, which play a crucial role in establishing a consistent connection between theoretical predictions and experimental observations. Examples include studies by the \mbox{BESIII} collaboration on the $\Lambda_c^+$ state, where its spin was determined to be \mbox{$\mathbf{J}=1/2$}, with a significance exceeding 6$\sigma$ \cite{BESIII:2020kap}; by the Belle collaboration on the $\Xi_c(2970)^+$ state, for which a spin-parity of \mbox{$\mathbf{J^{P}}=1/2^{+}$} was favored \cite{Belle:2020tom}; and more recently by the LHCb collaboration in 2025, where the spin-parity for the $\Xi(3055)^{+,0}$ baryon was determined to be \mbox{$\mathbf{J^{P}}=3/2^{+}$}, with a significance greater than 6.5$\sigma$ \cite{LHCb:2024eyx}.

Similarly, the LHCb collaboration excluded the \mbox{$\mathbf{J}=1/2$} assignment for the $\Omega_c(3050)^0$ and $\Omega_c(3065)^0$ states with significances of 2.2$\sigma$ and 3.6$\sigma$, respectively \cite{LHCb:2021ptx}. Moreover, the BaBar collaboration~\cite{BaBar:2008get} ruled out the $\mathbf{J}=3/2$ hypothesis for the $\Sigma_c(2455)^0$ baryon with a significance surpassing 4$\sigma$, favoring instead a $\mathbf{J}=1/2$ assignment, in agreement with quark model expectations.

A brief overview of the most relevant experimental and theoretical developments is presented in Ref.~\cite{Davila-Rivera:2025exk}, where some of the authors of the present work studied the electromagnetic decay (EMD) widths of the $\Sigma_c$, $\Xi'_c$, and $\Omega_c$ charmed baryons belonging to the flavor $\mathbf{6}_{\rm F}$ multiplet. Further discussions on this topic can be found in the review articles~\cite{klempt2010baryon,cheng2022charmed,chen2017review}. In the present work, we focus on reviewing the main results for singly charmed baryons belonging to the flavor antitriplet.



The first  singly charmed baryon was identified as $\Lambda_{c}^{+}$ in 1975 by the Brookhaven National Laboratory (BNL) group  \cite{Cazzoli:1975et}, with a mass of $2426$ MeV. In the following years, new states were discovered including $\Sigma_c^{++}$~\cite{Knapp:1976qw} and $\Sigma_c^{+}$~\cite{BEBCTSTNeutrino:1980ktj}.  The WA62 Collaboration observed the ground state of the $\Xi_c$ in 1983, with a measured mass of $2460~\mathrm{MeV}$~\cite{Biagi:1983en}, and in 1985 reported evidence for the $\Omega_c^0$ baryon~\cite{Biagi:1984mu}. In 1993, Ref.~\cite{ARGUS:1993vtm} reported the observation of a charmed baryon with a mass of $2626.6$~MeV  ($90\%$ C.L.) in DORIS II at DESY. Theoretical estimates assign this mass to $P$-wave $\Lambda_c(2630)$ or $\Lambda_c(2640)$ states with $\mathbf{J^P} = 1/2^-$ or $3/2^-$. The $\Lambda_c(2595)^+$ and $\Lambda_c(2625)^+$ baryons, with respective masses of 2593 MeV and \mbox{2625 MeV}, were observed in 1995 by the CLEO Collaboration~\cite{CLEO:1994oxm}.  In 1997, Ref.~\cite{ARGUS:1997snv} studied and confirmed the existence of the excited charmed baryon $\Lambda_c(2595)^{+}$. By the end of the XX century, CLEO  introduced the states  $\Xi_{c}(2815)^{+}$ and  $\Xi_{c}(2815)^{0}$, with $\mathbf{J^P} = 3/2^-$, which are the charmed-strange analogs of $\Lambda_{c}(2625)^{+}$~\cite{CLEO:1999msf}, as well as two excited $\Lambda_c^+$ states with masses of about $3105~\mathrm{MeV}$ and $3221~\mathrm{MeV}$ (90\% CL)~\cite{CLEO:2000mbh}. In 2001, CLEO also provided evidence for 
the $\mathbf{J^P} = 1/2^-$ $\Xi_{c}(2790)^0$ and $\Xi_{c}(2790)^+$ states, analogs of the $\Lambda_c(2595)^+$~\cite{CLEO:2000ibb}. In the first decade of the XXI century, the Belle Collaboration observed two new $\Xi_c$ baryons ~\cite{Belle:2006edu}, and confirmed the existence of the $\Xi_c(2815)$ baryon ~\cite{Belle:2008yxs}. 

In 2011 CDF reported the properties of several $\Lambda_c$ and $\Sigma_c$ baryons~\cite{CDF:2011zbc}, and LHCb discovered in 2017 five $\Omega_c^0$ baryons~\cite{LHCb:2017uwr}. Later that year, Belle observed five new resonant states \cite{Belle:2017ext} and confirmed four of the states observed previously by LHCb.

More recently, the LHCb Collaboration reported the observation of three excited $\Xi_c^0$ baryons~\cite{LHCb:2020iby}, confirmed four $\Omega_c^0$ states ~\cite{LHCb:2021ptx}, and reported two more $\Omega_c$ excited states~\cite{LHCb:2023sxp}. In 2025, the same collaboration~\cite{LHCb:2024eyx} reported the first determination of the spin-parity of the $\Xi_c(3055)^{+,-}$ baryons, supporting the hypothesis that such baryons correspond to the first $D$-wave $\lambda$-mode excitation of the $\Xi_c$ flavor antitriplet. Finally, in 2025, LHCb ~\cite{LHCb:2025mge} observed four $\Xi_c^{'+}$ states with high significance.

The spectroscopy of singly charmed baryons has been studied within different theoretical frameworks. In particular, QCD sum rules was employed in Ref.~\cite{Groote:1996em}, bootstrap quark models in Ref.~\cite{Gerasyuta:1999pc}, and heavy-quark symmetry corrections in Ref.~\cite{Wang:2003zp}. Lattice QCD, chiral perturbation theory, and relativized quark models have also been applied~\cite{Brown:2014ena,Padmanath:2013bla,Bahtiyar:2020uuj,Yang:2021lce,Pan:2023hwt,Yu:2022ymb,Wang:2023wii}. Predictions for the $\Xi_c$ and $\Xi_b$ masses were presented in Ref.~\cite{Duraes:2007te}.

Strong decays of charmed baryons were also analyzed using several techniques, including Heavy Hadron Chiral Perturbation Theory~\cite{Pirjol:1997nh}, constituent and light-front quark models~\cite{Chiladze:1997ev,Tawfiq:1998nk,Ivanov:1998qe}, nonrelativistic quark models with heavy-quark symmetry wavefunctions~\cite{Albertus:2005zy}, combined heavy quark and chiral symmetries~\cite{Cheng:2006dk}, Bethe–Salpeter approaches ~\cite{Guo:2007qu}, and the $^3P_0$ model~\cite{Ye:2017yvl,Zhao:2017fov}. Recent works studied the structure and decays of $2P$-wave charmed baryons like $\Lambda_c(2940)^+$ and $\Xi_c(3123)^+$~\cite{Yang:2023fsc,Yu:2023bxn}, with extensions to bottom baryons in the $\bf {\bar 3}_{\rm F}$ representation~\cite{Wang:2024rai}.

Different proposals for the nature of excited charmed baryons can be found in the literature, including molecular interpretations for $\Lambda_c(2593)^+$~\cite{Blechman:2003mq} and $\Lambda_c(2940)^+$~\cite{Ortega:2012cx}, positive-parity assignments for $\Xi_c(2980)$ and $\Xi_c(3080)$~\cite{Cheng:2015naa}, $\lambda$-mode excitations for $\Xi_c(2930)$~\cite{Aliev:2018ube}, $\Xi_c(2923)^0$, $\Xi_c(2939)^0$, $\Xi_c(2965)^0$ ~\cite{Lu:2020ivo},  systematic patterns among $\bf {\bar 3}_{\rm F}$ and $\bf 6_{\rm F}$ states~\cite{Chen:2021eyk}, and unquenched quark models applied to $\Lambda_c(2910)^+$ and $\Lambda_c(2940)^+$~\cite{Zhang:2022pxc}.

Research on electromagnetic decays (EMDs) has progressed significantly over the years, enhancing our understanding of hadronic structure. The first experimental observation of electromagnetic decays from excited charmed baryons, comprising $\Xi_c^{\prime +} \rightarrow \Xi_c^{+} \gamma$ and \mbox{$\Xi_c^{\prime 0} \rightarrow \Xi_c^{0} \gamma$,} was conducted in 1999  by CLEO~\cite{CLEO:1998wvk}. In 2006, BABAR measured $\Omega_c^{*0} \rightarrow \Omega_c^{0} \gamma$ ~\cite{BaBar:2006pve}, and two years later, Belle verified this result ~\cite{Solovieva:2008fw}. More recently, Belle announced the first detection of EMDs originating from $P$-wave $\Xi_c$ states: $\Xi_c(2790)^{0} \rightarrow \Xi^{0}_c \gamma$ and $\Xi_c(2815)^{0} \rightarrow \Xi^{0}_c \gamma$, with statistical significances of $3.8 \sigma$ and $8.6 \sigma$, respectively ~\cite{Belle:2020ozq} .

Theoretical advances in EMDs of singly charmed baryons can be found in Refs.~\cite{Cheng:1992xi,Banuls:1999br,Jiang:2015xqa,Wang:2018cre,Zhu:1998ih,Wang:2010xfj,Wang:2009cd,Aliev:2009jt,Aliev:2011bm,Aliev:2014bma,Aliev:2016xvq,Bernotas:2013eia,Shah:2016nxi,Gandhi:2019xfw,Gandhi:2019bju,Yang:2019tst,Kim:2021xpp,Chow:1995nw,Gamermann:2010ga,Zhu:2000py,Luo:2025jpn,Luo:2025pzb,Bijker:2020tns,Ortiz-Pacheco:2023kjn,Garcia-Tecocoatzi:2025fxp,Wang:2017hej,Wang:2017kfr,Yao:2018jmc,Peng:2024pyl,Ivanov:1998wj,Ivanov:1999bk,Tawfiq:1999cf}. Most of these studies focus on ground states  \cite{Cheng:1992xi,Banuls:1999br,Jiang:2015xqa,Wang:2018cre,Zhu:1998ih,Wang:2010xfj,Wang:2009cd,Aliev:2009jt,Aliev:2011bm,Aliev:2014bma,Aliev:2016xvq,Bernotas:2013eia,Shah:2016nxi,Gandhi:2019xfw,Gandhi:2019bju,Yang:2019tst,Kim:2021xpp},  $P$-wave states \mbox{ \cite{Chow:1995nw,Gamermann:2010ga,Zhu:2000py,Luo:2025jpn,Luo:2025pzb,Bijker:2020tns,Ortiz-Pacheco:2023kjn,Garcia-Tecocoatzi:2025fxp},} and both ground and $P$-wave states \cite{Wang:2017hej,Wang:2017kfr,Yao:2018jmc,Peng:2024pyl,Ivanov:1998wj,Ivanov:1999bk,Tawfiq:1999cf} using several theoretical frameworks.  Specifically, chiral perturbation theory ($\chi$PT) was implemented in \mbox{ Refs.~\cite{Cheng:1992xi,Banuls:1999br,Jiang:2015xqa,Wang:2018cre},} while light-cone QCD (LC-QCD) was applied in \mbox{ Refs.~\cite{Zhu:1998ih,Wang:2010xfj,Wang:2009cd,Aliev:2009jt,Aliev:2011bm,Aliev:2014bma,Aliev:2016xvq}.} A modified bag model was adopted in Ref.~\cite{Bernotas:2013eia} to compute the EMDs of heavy baryons. The hypercentral constituent quark model (hCQM),  which was introduced in \cite{Ferraris:1995ui,Santopinto:1997jz,Santopinto:1998ma,Giannini:2015zia}, was used in Refs.~\cite{Shah:2016nxi,Gandhi:2019xfw,Gandhi:2019bju}, whereas the chiral quark-soliton model ($\chi$QSM) was applied in Refs.~\cite{Yang:2019tst,Kim:2021xpp}.
For $P$-wave excited $\Lambda_Q$ baryons, the chiral soliton model was considered in Ref.~\cite{Chow:1995nw}. Moreover, the Coupled-Channel Approach (CCA) was used in Ref.~\cite{Gamermann:2010ga}, and LC-QCD was employed in Refs.~\cite{Zhu:2000py,Luo:2025jpn,Luo:2025pzb} to examine the EMDs of singly charmed baryons. Another commonly used approach is the Constituent Quark Model (CQM), as seen in Refs.~\cite{Bijker:2020tns,Ortiz-Pacheco:2023kjn,Garcia-Tecocoatzi:2025fxp,Wang:2017hej,Wang:2017kfr,Yao:2018jmc,Peng:2024pyl}. Electromagnetic transitions were also studied using a relativistic three-quark model (RQM) in Refs.~\cite{Ivanov:1998wj,Ivanov:1999bk}. Additionally, heavy quark symmetry was invoked in Ref.~\cite{Tawfiq:1999cf}.

\textcolor{black}{In Ref.~\cite{Garcia-Tecocoatzi:2025fxp}, the electromagnetic decay widths for the $\Xi_c(2790)^{+,0}$ and $\Xi_c(2815)^{+,0}$ baryons were obtained, employing a nonrelativistic constituent quark model. These theoretical results are in agreement with the recent experimental data reported by the Belle Collaboration~\cite{Belle:2020ozq}.}

To date, only two studies have examined a subset of second-shell charmed baryon states within the CQM~\cite{Yao:2018jmc,Peng:2024pyl}. In both works, the Close–Copley replacement~\cite{Close:1970kt} was adopted to evaluate the transition amplitudes.
In the present study, we calculate the EMD widths of the $\Lambda_c$ and $\Xi_c$ charmed baryons, with isospin 0 and $1/2$, respectively, belonging to the flavor $\bf {\bar 3}_{\rm F}$-plet, within the constituent quark model formalism~\cite{Garcia-Tecocoatzi:2023btk,Garcia-Tecocoatzi:2025fxp,Rivero-Acosta:2025drn}, following the notation and methods of Ref.~\cite{Davila-Rivera:2025exk}. We evaluate the transition amplitudes analytically without relying on the Close–Copley replacement. We consider transitions from ground and $P$-wave states to ground states for the $\Lambda_c$ baryons, since transitions involving the $\Xi_c$ baryons at this level have been addressed in our previous work~\cite{Garcia-Tecocoatzi:2025fxp}. We also analyze transitions from all second-shell $\Lambda_c$ and $\Xi_c$ states to both ground and $P$-wave final states. This work presents the first calculation of the EMDs of $D_\rho$-wave states, $\rho$--$\lambda$ mixed states, and radially excited $\rho$-mode states for singly charmed baryons belonging to the flavor antitriplet. The paper is organized as follows: Section~\ref{EMdecaywidths} briefly presents the quark model used to determine the mass spectra
and outline the formalism used to calculate the EMD widths. Section~\ref{Results} addresses our results along with a discussion. Finally, conclusions are summarized in Section~\ref{Conclusions}.

\section{Electromagnetic decay widths} 
\label{EMdecaywidths}


\begin{figure}[h]
    \centering
    \includegraphics[width=0.2\textwidth]{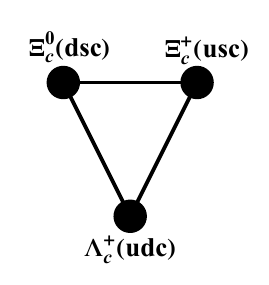}
	\caption{ The $SU_f(3)$ flavor antitriplet of the ground-state singly charmed baryons. }    
    \label{fig:sextet}
\end{figure}


\subsection{Charmed baryon states and mass spectra} 
\label{Masses}

\begin{figure*}
\centering \includegraphics[scale=1] {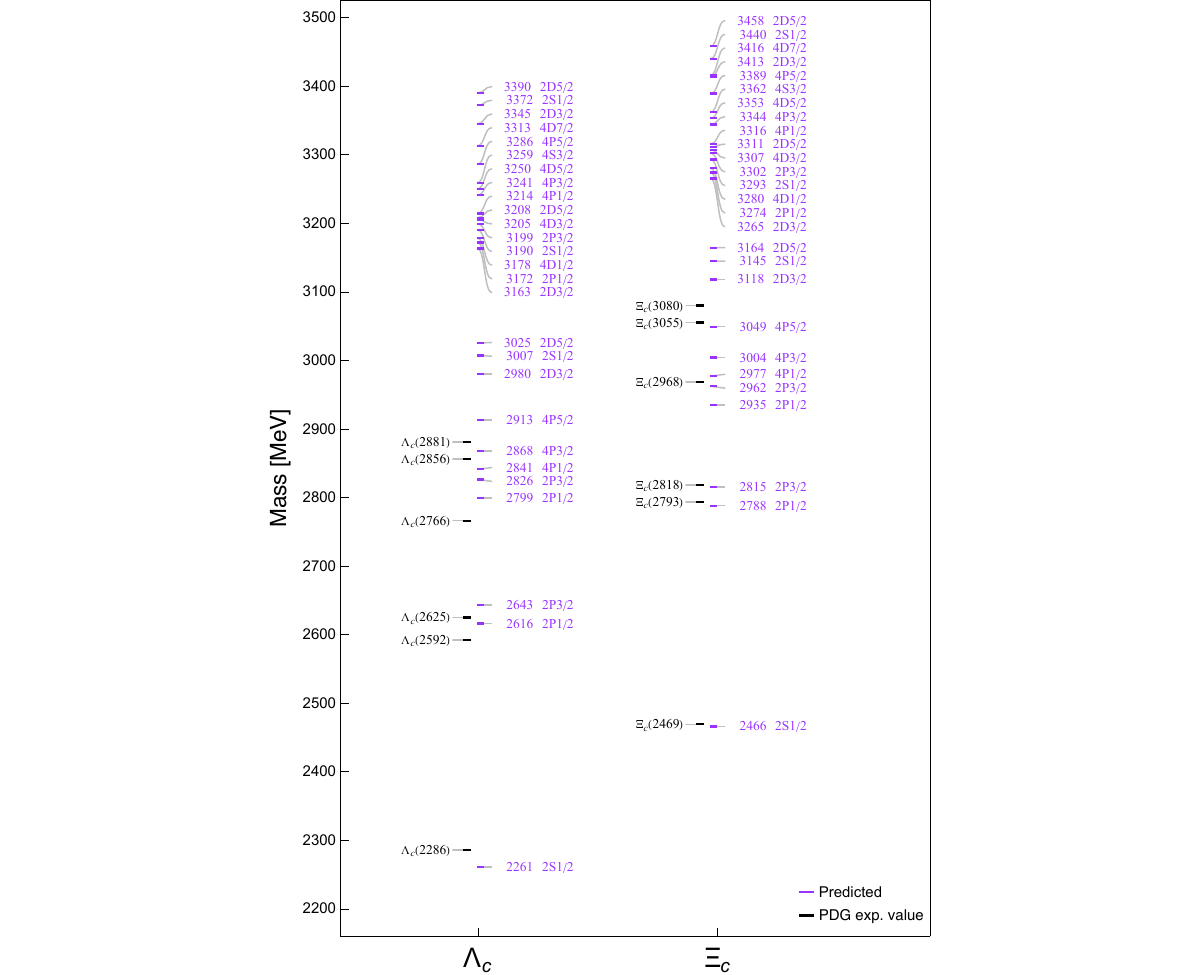}
\caption{ Comparison between the predicted masses of singly charmed baryon belonging to the $\bf {\bar 3}_{\rm F}$-plet, as from Ref.~\cite{Garcia-Tecocoatzi:2022zrf}, 
with the experimental data from the PDG \cite{ParticleDataGroup:2024cfk}. The predicted masses are displayed in purple, while the experimental masses are reported in black \cite{ParticleDataGroup:2024cfk}. }
\label{fig:all-states}
\end{figure*}

First, we establish the masses and quantum states of the charmed baryons, these masses are used to compute the electromagnetic decay widths. Note that the $\Lambda_c$ and $\Xi_c$ states are part of the flavor $\bf {\bar 3}_{\rm F}$-plet (see Fig.~\ref{fig:sextet}). We adopt the masses and assignments of singly charmed baryons obtained in Ref.~\cite{Garcia-Tecocoatzi:2022zrf}. This choice is motivated by the excellent agreement between the calculated $\Xi_c$ masses and the available experimental data. For the $\Lambda_c$ sector, the agreement is also good and reproduces the experimental trend. In Fig.~\ref{fig:all-states}, we present the mass predictions of Ref.~\cite{Garcia-Tecocoatzi:2022zrf} compared to the experimental values reported by the PDG \cite{ParticleDataGroup:2024cfk}. These theoretical masses correspond to the eigenvalues of the Hamiltonian introduced in Refs.~\cite{Santopinto:2018ljf,Giachino:2020dsj}, which models a charmed baryon as a three-quark system, given by 

\begin{eqnarray}
	H &=& H_{\rm h.o.}+P_{\rm S}\; {\bf S}^2_{\rm tot}
 + P_{\rm SL} \; {\bf S}_{\rm tot} \cdot {\bf L}_{\rm tot}+P_{\rm I}  \;  \bm{{\rm I}}^2+P_{\rm F}\; {\bf \hat{C}}_2,
 \nonumber
 \\
	\label{MassFormula}
\end{eqnarray}
where $H_{\rm h.o.}$ is the sum of the constituent masses and the harmonic oscillator Hamiltonian. Moreover, the operators ${\bf S}_{\rm tot}$, ${\bf L}_{\rm tot}$, $\bm{{\rm I}}$, and ${\bf \hat{C}}_2$ correspond to the spin, orbital angular momentum, isospin, and $SU_f(3)$ flavor degrees of freedom, respectively, with the strength of their contributions encoded in the model parameters $P_{\rm S}$, $P_{\rm SL}$, $P_{\rm I}$, and $P_{\rm F}$. In Eq.~\ref{MassFormula}, $H_{\rm h.o.}$ describes the spatial degrees of freedom of the three-quark system using the Jacobi coordinates $\boldsymbol{\rho}$ and $\boldsymbol{\lambda}$, which describe the relative motion between the light quarks, and the relative motion of the light-quark pair and the charm quark $c$, respectively, together with their conjugate momenta $\mathbf{p}_{\rho}$ and $\mathbf{p}_{\lambda}$~\cite{Santopinto:2018ljf,Garcia-Tecocoatzi:2022zrf,Garcia-Tecocoatzi:2023btk,Giachino:2020dsj}:
\begin{eqnarray}
 H_{\rm h.o.} =\sum_{i=1}^3m_i + \frac{\mathbf{p}_{\rho}^2}{2 m_{\rho}} 
+ \frac{\mathbf{p}_{\lambda}^2}{2 m_{\lambda}} 
+\frac{1}{2} m_{\rho} \omega^2_{\rho} \boldsymbol{\rho}^2   
+\frac{1}{2}  m_{\lambda} \omega^2_{\lambda} \boldsymbol{\lambda}^2 ,
	\nonumber \\
\label{eq:Hho}
\end{eqnarray}
where $m_{i}$ with $i=1,2$ are the light quark masses, $m_3$ is the charm quark mass;
$m_\rho=(m_1+m_2)/2$, and $m_\lambda=3m_\rho m_3/(2m_\rho+m_3)$. Eq.~\ref{eq:Hho} effectively describes the dynamics of two decoupled harmonic oscillators, $\rho$- and  $\lambda$. These oscillators are characterized with the quantum numbers $ n_{\rho(\lambda)}= 2 k_{\rho(\lambda)}+l_{\rho(\lambda)}$ where  $k_{\rho(\lambda)}=0,1,...$ is the number of nodes and $l_{\rho(\lambda)}=0,1,...$ is the orbital angular momentum. We define $N=n_\rho+n_\lambda$ as the n$^{th}$-energy band. The $\rho$- and  $\lambda$-oscillator frequencies are $\omega_{\rho(\lambda)}=\sqrt{\frac{3K_c}{m_{\rho(\lambda)}}}$, where $K_c$ is the harmonic oscillator constant that is a further free parameter of the model. The quark masses and the model parameters are fitted to experimental data ~\cite{Garcia-Tecocoatzi:2022zrf}.

The $\Lambda_c$ and $\Xi_c$ states, generally referred to as $A$, are formally described by the state:
 \begin{eqnarray}
|\theta_c,\phi_{A_I}, k_A, J_A,{M_{J_A}}\rangle &
=& 
 | \theta_c \rangle \otimes |\phi_{A_I}, k_A ,J_A,{M_{J_A}}\rangle , \label{eq:statescomp} 
\end{eqnarray}
\noindent
where $| \theta_c \rangle =\frac{1}{\sqrt{6}}(|rgb \rangle  - | rbg \rangle + | gbr \rangle - | grb \rangle + | brg \rangle - | bgr \rangle)$ is the $SU_c(3)$ color singlet. $|\phi_{A_I}, k_A ,J_A,{M_{J_A}}\rangle$ contains the information related to the flavor, spin, and spatial degrees of freedom, and is given by 

\begin{eqnarray}
|\phi_{A_I}, k_A ,J_A,{M_{J_A}}\rangle &
=& 
|\phi_{A_I} \rangle \otimes \sum_{M_{L},M_{S}} \langle L, M_L;S, M_S|J_A,{M_{J_A}}\rangle \nonumber \\
& \times &  \sum_{m_{l_\lambda},m_{l_\rho}} \langle l_\lambda, m_{l_\lambda}; l_\rho, m_{l_\rho}|L,M_L\rangle \nonumber \\
& \times &  \sum_{m_{S_{12}},m_{S_3}} \langle S_{12},m_{S_{12}};S_3,m_{S_3}|S,M_S\rangle \nonumber \\
& \times & \sum_{m_{S_1},m_{S_2}}  \langle S_1, m_{S_1};S_2,m_{S_2} |S_{12},m_{S_{12}} \rangle \nonumber\\
& \times & |S_1,m_{S_1}\rangle \otimes | S_2,m_{S_2}\rangle \otimes | S_3,m_{S_3}\rangle\nonumber  \\
& \otimes & |k_\rho, l_\rho, m_{l_\rho}, k_\lambda, l_\lambda, m_{l_\lambda} \rangle . \label{eq:states}  
\end{eqnarray}
\noindent
 Where $|\phi_{A_I} \rangle$ stands for the flavor wave-function  defined as $|\phi_{A_I} \rangle \equiv  |A,I,M_I\rangle$ for the charmed baryon $A$ with isospin $I$ and isospin projection $M_I$. For the anti-triplet states,  $\Xi_c$, we have two charge states $\vert\Xi^{0}_c,1/2,-1/2 \rangle=\frac{1}{\sqrt{2}}(\vert dsc\rangle-|sdc\rangle)$ and $|\Xi^{+}_c,1/2,1/2 \rangle=\frac{1}{\sqrt{2}}(|usc\rangle-|suc\rangle)$; $\Lambda_c$, there is one possible configuration $|\Lambda^+_c,0,0\rangle=\frac{1}{\sqrt{2}}(|udc\rangle-|duc\rangle)$. \mbox{$k_A = k_\rho + k_\lambda$} is the total number of nodes, $J_A$ represents the total angular momentum, and $M_{J_A}$ is the total angular momentum projection. $|S_i, m_{S_i}\rangle$ denote the spin wave function for each quark ($i=1,2,3$), and $|k_\rho, l_\rho, m_{l_\rho}, k_\lambda, l_\lambda, m_{l_\lambda} \rangle$ is the harmonic-oscillator spatial baryon wave function, that are expressed in terms of $\omega_{\rho}$ and $\omega_{\lambda}$ through the relations $\alpha^2_{\rho(\lambda)}=\omega_{\rho(\lambda)}m_{\rho(\lambda)}$.

\subsection{Classification of antitriplet states}
The $\Lambda_c$ and $\Xi_c$ states are modeled as three-quark systems which contain two light quarks (indistinguishable) and one heavy charm quark. These states are represented as $qqc$, where $q$ are the $s,u,d$ quarks and $c$ is the charm quark. We consider the $\Lambda_c$ and $\Xi_c$ states to belong to the flavor anti-triplet, where the light quarks are in an antisymmetric flavor configuration. Moreover, their color wave function is also antisymmetric. Thus the combined spin-orbital wave function of the light-quark pair must be antisymmetric to obtain a total antisymmetric wave function. The allowed baryon states are given by the Hamiltonian in Eq.~\ref{MassFormula} which defines the total orbital and spin configurations. In this Hamiltonian, ${\bf J_{\rm tot}=} {\bf L}_{\rm tot} + {\bf S}_{\rm tot} $, where ${\bf L_{\rm tot} =l}_{\rho}+{\bf l}_{\lambda}$ and ${\bf S}_{\rm tot} = {\bf S}_{\rm lt}+\frac{1}{2}$, ${\bf S}_{\rm lt}$ is the coupled spin of the two light quarks. We study the energy bands $N=0,1,2$.

For $N=0$, $\bf{L_{\rm _{tot}}}={\bf l}_{\lambda}={\bf l}_{\rho}={\bf 0}$ and the orbital wave function of the two light quarks is symmetric, hence the spin part must be antisymmetric.  This corresponds to $\bf{S}_{\rm lt}=0$, yielding one ground state. 

For $N=1$, with the $\bf{l_\rho=0}$ and $\bf{l_\lambda=1}$ configuration, the light quark spatial wave function is symmetric and the spin part must be antisymmetric. This implies that $\bf{S}_{\rm tot}=1/2$ and $\bf L_{\rm _{tot}}=l_\lambda=1$, giving two $P_\lambda$ states. For the ${
\bf l_\rho=1}$ and $ {\bf l}_{\lambda}={\bf 0}$ configuration, the light quarks present an antisymmetric spatial wave function and a symmetric spin wave function with two possible values $\bf{S}_{\rm tot}=1/2,3/2$. Accordingly, for $\bf L_{\rm tot}={\bf l}_{\rho}=1$ there are five $P_\rho$ states.

For $N=2$, in the case of $\bf{l_\rho=0}$ and $\bf{l_\lambda=2}$, the total spatial wave function is symmetric; thus the light quark pair spin function is antisymmetric with $\bf S_{\rm lt}=0$. This yields two {$D_\lambda$}-wave excitations. Likewise, $N=2$ for two possible radial excited configurations, that is, $k_\rho=0$ and $k_\lambda=1$, and $k_\rho=1$ and $k_\lambda=0$. These configurations generate symmetric spatial wave functions and hence require light quark pair spin antisymmetric configurations $\bf S_{\rm lt}=0$, producing one $\lambda$-radial excitation and one $\rho$-radial excitation. For the case $\bf{l_\rho=1}$ and $\bf{l_\lambda=1}$ the spatial wave functions are antisymmetric and are coupled to ${\bf L_{\rm tot}=0, 1, 2}$. Hence, the spin configuration $\bf{S}_{\rm tot}=1/2,3/2$, coming from the light quark spin configuration, is $\bf S_{\rm lt}=1$. This produces thirteen mixed excited states: six $D$-wave states, five $P$-wave states, and two $S$-wave states. Finally, the symmetric configuration $\bf{l_\rho=2}$ and $\bf{l_\lambda=0}$, combined with the light quark  antisymmetric spin configuration, $\bf S_{\rm lt}=0$, gives two {$D_\rho$} excited states. 

It is worth emphasizing that, although the ground state $\Xi_c$ baryons have a light quark pair spin configuration $\bf S_{\rm lt}=0$, this assignment does not hold for all excited states. For the excited baryon $\Xi_c$, the light-quark pair can also have total spin $\bf S_{\rm lt}=1$, provided that the spatial wave function is antisymmetric to maintain the overall antisymmetry of the total light-quark wave function. The previous considerations and the state constructions are summarized in Table~\ref{tab:classification}.

\begin{table*}[htbp]
\centering

\caption{ 
Classification of the symmetry properties of the two-light-quark spatial and spin wave functions of the antitriplet states. The states are organized by energy bands $N=n_{\rho}+n_{\lambda}$. The first column contains the spatial wave state $\vert k_\lambda,k_\rho, l_\lambda,l_\rho\rangle$ where $l_{\lambda,\rho}$ are the orbital angular momenta and $k_{\lambda,\rho}$ are the number of nodes of the $\lambda$ and $\rho$ oscillators. The second column,  "Orbital symmetry", indicates the symmetry of the two-light-quark spatial wave function under exchange of the light quarks, with $S$ ($A$) denoting symmetric (antisymmetric) configurations. $\bf S_{\rm lt}$ represents the total spin of the two-light-quark subsystem, while ``Spin Symmetry'' specifies the corresponding symmetry of the two-light-quark spin wave function. The fifth column contains the total baryon spin $\bf S_{\rm tot}$. The sixth column contains the total angular momentum $\bf L_{\rm tot}$. The last column lists all possible total spin-parity quantum numbers $\bf J^P$.
}

\label{tab:classification}
\begin{tabular}{c c c c c c c}
\hline
\hline
$\vert k_\lambda,k_\rho, l_\lambda,l_\rho\rangle$ & Orbital & $\bf S_{\rm lt}$ & Spin & $\bf S_{\rm tot}$ & $\bf L_{\rm tot}$ & $\bf J^P$  \\
&Symmetry&& Symmetry&&&\\
\hline
\midrule

$N=0$\\
$\vert0,0, 0,0\rangle$ & $S$ & 0 & $A$ & $\frac12$ & 0 & $\frac12^+$ \\
$N=1$\\
$\vert 0,0, 1,0\rangle$ & $S$ & 0 & $A$ & $\frac12$ & 1 &
$\frac12^-,\ \frac32^-$ \\
\multirow{2}{*}{$\vert 0,0,0,1\rangle$}
& \multirow{2}{*}{$A$}
& \multirow{2}{*}{1}
& \multirow{2}{*}{$S$}
& $\frac12$ & 1 &
$\frac12^-,\ \frac32^-$ \\
& & & & $\frac32$ & 1 &
$\frac12^-,\ \frac32^-,\ \frac52^-$ \\
$N=2$\\
$\vert 0,0, 0,2\rangle$ & $S$ & 0 & $A$ & $\frac12$ & 2 &
$\frac32^+,\ \frac52^+$ \\
$\vert 0,1,0,0\rangle$ & $S$ & 0 & $A$ & $\frac12$ & 0 &
$\frac12^+$ \\
$\vert 1,0, 0,0\rangle $ & $S$ & 0 & $A$ & $\frac12$ & 0 &
$\frac12^+$ \\
\multirow{3}{*}{$\vert 0,0,1,1\rangle$}
& \multirow{3}{*}{$A$}
& \multirow{3}{*}{1}
& \multirow{3}{*}{$S$}
& $\frac12$ & 0 &
$\frac12^+$ \\
& & & & $\frac12$ & 1 &
$\frac12^-,\ \frac32^-$ \\
& & & & $\frac12$ & 2 &
$\frac32^+,\ \frac52^+$ \\
\multirow{3}{*}{$\vert 0,0, 1,1\rangle$}
& \multirow{3}{*}{$A$}
& \multirow{3}{*}{1}
& \multirow{3}{*}{$S$}
& $\frac32$ & 0 &
$\frac32^+$ \\
& & & & $\frac32$ & 1 &
$\frac12^-,\ \frac32^-,\ \frac52^-$ \\
& & & & $\frac32$ & 2 &
$\frac12^+,\ \frac32^+,\ \frac52^+,\ \frac72^+$ \\
$\vert0,0,2,0\rangle$ & $S$ & 0 & $A$ & $\frac12$ & 2 &
$\frac32^+,\ \frac52^+$ \\
\hline
\end{tabular}
\end{table*}


\subsection{Electromagnetic decay widths} 
\label{EMDecThe}
At this point we compute the electromagnetic decay widths of $\Lambda_c$ and $\Xi_c$ states in accordance to the formalism developed in Refs.~\cite{Garcia-Tecocoatzi:2023btk,Garcia-Tecocoatzi:2025fxp,Rivero-Acosta:2025drn,RiveroAcosta:2025nfh}. The formalism treats the emission of left-handed photons in radiative transitions of the form $A \rightarrow A'\gamma$, where  $A$ and $A'$ denote the initial and final baryon states, respectively.  The interaction Hamiltonian describing the electromagnetic
decays is
\begin{eqnarray}
\mathcal{H}_{\rm em} &=& 2\sqrt{\frac{\pi}{k}}\sum^3_{j=1}\mu_j \left[ {\rm k} \, \mathbf{s}_{j,-} e^{-i \mathbf{k} \cdot \mathbf{r}_j} \right. \nonumber \\
&-& \left. \frac{1}{2}\left( \mathbf{p}_{j,-} e^{-i \mathbf{k} \cdot \mathbf{r}_j} + e^{-i \mathbf{k} \cdot \mathbf{r}_j} \mathbf{p}_{j,-} \right) \right],
\label{eq:Hem}
\end{eqnarray}
where $\mu_j $, $\mathbf{s}_{j,-} = s_{j,x} - i s_{j,y}$, and $\mathbf{p}_{j,-} = p_{j,x} - i p_{j,y}$ are the magnetic moment, spin-ladder, and momentum-ladder operators of the $j$-th quark, respectively. The electromagnetic transition amplitude,  
$A_{M_{J_A}}$, of the radiative decay $A \rightarrow A'\gamma$, is computed as, 
\begin{equation}
A_{M_{J_A}} = \langle \phi_{A'_I}, k_{A'}, J_{A'}, M_{J_{A'}}-1 | \mathcal{H}_{\rm em} | \phi_{A_I}, k_{A}, J_{A}, M_{J_{A}} \rangle,
\end{equation}
 where $A'$ is the final charmed baryon. Utilizing this transition amplitude, we obtain the partial decay width
\begin{equation}\label{gammaEM}
\Gamma_{\rm em} = \frac{\Phi}{(2\pi)^2} \frac{2}{2J_A+1} \sum_{M_{J_A}>0} |A_{M_{J_A}}|^2,
\end{equation}
where $\Phi = 4\pi (E_{A'}/m_A) {\rm k}^2$ is the phase-space factor, with ${\rm k} = (m_A^2 - m_{A'}^2)/(2m_A)$ and $E_{A'} = \sqrt{m_{A'}^2 + {\rm k}^2}$. These baryon states are eigenstates of the Hamiltonian given in Eq.~\ref{MassFormula}, and are denoted by $\left| l_{\lambda}, l_{\rho}, k_{\lambda}, k_{\rho} \right\rangle$, which represents a compact form of the state $| \phi_{A_I}, k_{A}, J_{A}, M_{J_{A}} \rangle$.
This notation is adopted to emphasize the spatial degrees of freedom, since the novelty of this work resides in the calculation of electromagnetic decays of spatially excited $D_\rho$-wave, $\rho$–$\lambda$ mixed, and radially excited $\rho$-mode of the $\Lambda_c$ and $\Xi_c$ states.

 Electromagnetic decays are allowed between different flavor multiplets. That is, for decays of anti-triplet excited states, the final flavor wave functions $\vert\phi_{A'_I}\rangle$ may belong to either the anti-triplet or the sextet multiplet. $\vert\phi_{A'_I}\rangle$, for the anti-triplet configurations $\Xi_c$ and $\Lambda_c$, are defined as in Sec.~\ref{Masses}. Moreover, $\vert\phi_{A'_I}\rangle$ for the sextet configurations can represent a $\Xi'_c$ or $\Sigma_c$ charmed baryon. The flavor wave functions of the final states $\Xi'_c$  have two charge states, $|\Xi^{\prime0}_c,1/2,-1/2 \rangle=\frac{1}{\sqrt{2}}(|dsc\rangle+|sdc\rangle)$ and \mbox{$|\Xi^{\prime+}_c,1/2,1/2 \rangle=\frac{1}{\sqrt{2}}(|usc\rangle+|suc\rangle)$}, while for the $\Sigma_c$ states, the flavor wave functions of the final states have three possible charge configurations, $|\Sigma^{++}_c,1,1  \rangle=|uuc\rangle$,
$|\Sigma^0_c,1,-1 \rangle=|ddc\rangle$, and 
$|\Sigma^+_c,1, 0\rangle=\frac{1}{\sqrt{2}}(|udc\rangle+|duc\rangle)$.


\begin{table*}[h!tp]
\caption{ Predicted electromagnetic decay widths (in keV) for $\Lambda_c(nnc)^{+}$ states with isospin 0 belonging to the flavor antitriplet. The first column denotes the baryon name along with its predicted mass in both $S$-wave and $P$-wave configurations, taken from Ref.~\cite{Garcia-Tecocoatzi:2022zrf}, corresponding to the $N = 0$ and $N = 1$ energy bands, respectively, where $N=n_\rho+n_\lambda$. 
 The second column indicates the spin-parity $\mathbf{J}^{\rm P}$, while the third column shows the internal configuration of the state in the three-quark model, written as $\left| l_{\lambda}, l_{\rho}, k_{\lambda}, k_{\rho} \right\rangle$, where $l_{\lambda,\rho}$ are the orbital angular momenta and $k_{\lambda,\rho}$ are the number of nodes of the $\lambda$ and $\rho$ oscillators. The fourth column provides the spectroscopic notation $^{2S+1}L_{x,J}$ associated with each initial state,  where $L$ is the total orbital angular momentum, $S$ is the total spin, $x = \lambda$ or $\rho$ denotes the mode in which the orbital excitation occurs, and the subscript $J$ is the total angular momentum. From the fifth column onward, the predicted electromagnetic decay widths are presented for each decay channel, computed using Eq.~\ref{gammaEM}. The decay products are listed at the top of each column, together with the spectroscopic notation $^{2S+1}L_{x,J}$ of the final states. Zero values correspond to electromagnetic decays that are either forbidden by phase space or have widths too small to be displayed at this scale. Our results are compared with those from Refs.~\cite{Wang:2017kfr}, \cite{Ortiz-Pacheco:2023kjn}, and \cite{Luo:2025pzb}.  The ``$\cdots$'' symbol indicates that no prediction is available for that particular state in previous studies. }

\begin{center}
\scriptsize{
\begingroup
\setlength{\tabcolsep}{1.75pt} 
\renewcommand{\arraystretch}{1.35} 
\begin{tabular}{c c c c  c  c   c  c  c} \hline \hline
  &    &    &    & $\Lambda_{c}^{+} \gamma$  & $\Sigma_{c}^{+} \gamma$  & $\Sigma_{c}^{*+} \gamma$ \\
$\Lambda_c(nnc)$  & $\mathbf{J^P}$  & $\vert l_{\lambda}, l_{\rho}, k_{\lambda}, k_{\rho} \rangle$  & $^{2S+1}L_{x,J}$  & $^2S_{1/2}$  & $^2S_{1/2}$  & $^4S_{3/2}$  \\
\hline
 $N=0$  &  &  &  &  &  \\\ 
$\Lambda_c(2261)$  & $ \mathbf{\frac{1}{2}^+}$ & $\vert \,0\,,\,0\,,\,0\,,\,0 \,\rangle $ &$^{2}S_{1/2}$&0  &0  &0 \\
 $N=1$  &  &  &  &  &  \\\ 
$\Lambda_c(2616)$  & $ \mathbf{\frac{1}{2}^-}$ & $\vert \,1\,,\,0\,,\,0\,,\,0 \,\rangle $ &$^{2}P_{\lambda,1/2}$&$ 3.5 _{- 2.9 }^{+ 3.2 }$    &  $ 1.5 _{- 1.0 }^{+ 1.5 }$    &  $ 0.1 _{- 0.1 }^{+ 0.1 }$   \\
& &  &   & 0.26 & 0.45 & 0.05 &  \cite{Wang:2017kfr} \\
 & &  &   & 0.1 & 1.0 & 0 &  \cite{Ortiz-Pacheco:2023kjn} \\
 & &  &   & ... & $2800^{+2400}_{-1700}$ & $640^{+860}_{-480}$ &  \cite{Luo:2025jpn} \\
$\Lambda_c(2643)$  & $ \mathbf{\frac{3}{2}^-}$ & $\vert \,1\,,\,0\,,\,0\,,\,0 \,\rangle $ &$^{2}P_{\lambda,3/2}$&$ 1.2 _{- 0.8 }^{+ 1.8 }$    &  $ 3.1 _{- 1.7 }^{+ 2.7 }$    &  $ 0.2 _{- 0.1 }^{+ 0.3 }$   \\
& &  &   & 0.3 & 1.17 & 0.26 &  \cite{Wang:2017kfr} \\
 & &  &   & 0.7 & 2.5 & 0.2 &  \cite{Ortiz-Pacheco:2023kjn} \\
  & &  &   & ... & $1900^{+1400}_{-950}$ & $71^{+80}_{-47}$ &  \cite{Luo:2025jpn} \\
$\Lambda_c(2800)$  & $ \mathbf{\frac{1}{2}^-}$ & $\vert \,0\,,\,1\,,\,0\,,\,0 \,\rangle $ &$^{2}P_{\rho,1/2}$&$ 19 _{- 7 }^{+ 7 }$    &  $ 584 _{- 49 }^{+ 51 }$    &  $ 2.6 _{- 1.5 }^{+ 2.0 }$   \\
& &  &   & 1.59 &  41.6 &  0.02 &  \cite{Wang:2017kfr} \\
 & &  &   & 9.6 & 97.3 & 1.5 &  \cite{Ortiz-Pacheco:2023kjn} \\
$\Lambda_c(2842)$  & $ \mathbf{\frac{1}{2}^-}$ & $\vert \,0\,,\,1\,,\,0\,,\,0 \,\rangle $ &$^{4}P_{\rho,1/2}$&$ 12 _{- 4 }^{+ 4 }$    &  $ 10 _{- 5 }^{+ 6 }$    &  $ 90 _{- 26 }^{+ 21 }$   \\
& &  &   & 0.8 & 0.08 & 6.81 &  \cite{Wang:2017kfr} \\
 & &  &   & 5.8  &  5.3  &  6.0 &  \cite{Ortiz-Pacheco:2023kjn} \\
$\Lambda_c(2827)$  & $ \mathbf{\frac{3}{2}^-}$ & $\vert \,0\,,\,1\,,\,0\,,\,0 \,\rangle $ &$^{2}P_{\rho,3/2}$&$ 22 _{- 7 }^{+ 7 }$    &  $ 1299 _{- 232 }^{+ 213 }$    &  $ 3.8 _{- 2.0 }^{+ 2.6 }$   \\
& &  &   & 2.35 &  48.0 &  0.09 &  \cite{Wang:2017kfr} \\
 & &  &   & 11.8  &  447.0  &  2.5 &  \cite{Ortiz-Pacheco:2023kjn} \\
$\Lambda_c(2869)$  & $ \mathbf{\frac{3}{2}^-}$ & $\vert \,0\,,\,1\,,\,0\,,\,0 \,\rangle $ &$^{4}P_{\rho,3/2}$&$ 37 _{- 12 }^{+ 11 }$    &  $ 35 _{- 16 }^{+ 20 }$    &  $ 434 _{- 37 }^{+ 37 }$  \\
& &  &   & 3.29 & 0.55 &  17.4 &  \cite{Wang:2017kfr} \\
 & &  &   & 19.4  &  21.1  &  79.1  &  \cite{Ortiz-Pacheco:2023kjn} \\
$\Lambda_c(2914)$  & $ \mathbf{\frac{5}{2}^-}$ & $\vert \,0\,,\,1\,,\,0\,,\,0 \,\rangle $ &$^{4}P_{\rho,5/2}$&$ 28 _{- 8 }^{+ 8 }$    &  $ 32 _{- 14 }^{+ 16 }$    &  $ 1341 _{- 249 }^{+ 229 }$   \\
& &  &   & ... &  ... &  ... &  \cite{Wang:2017kfr} \\
 & &  &   & 16.1  &  22.2  & 362.8 &  \cite{Ortiz-Pacheco:2023kjn} \\
  & &  &   & ... & ... & $130^{+1000}_{-130}$ &  \cite{Luo:2025jpn} \\
\hline \hline
\end{tabular}

\endgroup
}
\end{center}
\label{lambdasEM}
\end{table*}

\subsection{Uncertainties}
Uncertainties are reported for each electromagnetic decay width and arise from the masses of the charmed baryons involved in the transition $A \rightarrow A'\gamma$, taken from PDG measurements~\cite{ParticleDataGroup:2024cfk}, together with a model uncertainty associated with the mass description in Eq.~(\ref{MassFormula}). Error propagation is performed through a Monte Carlo procedure in which baryon masses are sampled from Gaussian distributions centered on either the PDG values or the theoretical predictions of Ref.~\cite{Garcia-Tecocoatzi:2022zrf}, with widths given by the combined experimental and model uncertainties. For each sampled set, the decay width is computed and the procedure is repeated $10^3$ times. The mean value defines the central result, while the 68\% confidence interval is obtained from quantiles. Fits and uncertainty propagation are performed using \texttt{MINUIT}~\cite{James:1975dr} and \texttt{NUMPY}~\cite{Harris:2020xlr}.

\section{Results and Discussion}
\label{Results}


\begin{turnpage}
\begin{table*}[htp]
\caption{ Predicted electromagnetic decay widths (in keV) for transitions from second shell $\Lambda_c^{+}(snc)$ states with isospin 0 belonging to the flavor antitriplet. The first column lists the baryon name along with its predicted mass in the $N = 2$ energy band, where $N = n_\rho + n_\lambda$, as calculated in Ref.~\cite{Garcia-Tecocoatzi:2022zrf}. 
The second column displays the spin-parity $\mathbf{J}^{\rm P}$, while the third column shows the internal configuration $\left| l_{\lambda}, l_{\rho}, k_{\lambda}, k_{\rho} \right\rangle$ in the three-quark model, where $l_{\lambda,\rho}$ are the orbital angular momenta and $k_{\lambda,\rho}$ denote the number of radial nodes of the $\lambda$ and $\rho$ oscillators. The fourth column presents the spectroscopic notation $^{2S+1}L_{x,J}$ associated with each initial state, where $L$ is the total orbital angular momentum, $S$ is the total spin, the subscript $x$ specifies the orbital excitation and can take the values $x = \lambda$, $\lambda\lambda$, $\rho$, $\rho\rho$, or $\lambda\rho$, and the subscript $J$ is the total angular momentum. From the fifth column onward, the electromagnetic decay widths are shown, computed using Eq.~\ref{gammaEM}. Each column corresponds to a specific decay channel; the final-state particles are indicated at the top of each column, and their spectroscopic notation $^{2S+1}L_{x,J}$ is provided in the second row. Zero values correspond to decay widths that are either forbidden by phase space or too small to be shown at this scale. Our results are compared with those in Refs.~\cite{Yao:2018jmc,Peng:2024pyl}. The symbol ``$\cdots$'' indicates that no prediction is available for the corresponding state in Refs.~\cite{Yao:2018jmc} and~\cite{Peng:2024pyl}. }
\scriptsize{
\begingroup
\setlength{\tabcolsep}{1.75pt} 
\renewcommand{\arraystretch}{1.35} 
\begin{tabular}{c c c c  p{1.0cm}  p{1.0cm}  p{1.0cm}  p{1.0cm}  p{1.0cm}  p{1.0cm}  p{1.0cm}  p{1.0cm}  p{1.0cm}  p{1.0cm}  p{1.0cm}  p{1.0cm}  p{1.0cm}  p{1.0cm}  p{1.0cm}  p{1.0cm}  p{1.0cm}  p{1.0cm}  p{1.0cm}  p{1.0cm}} \hline \hline
  &    &    &    & $\Lambda_{c}^{+} \gamma$  & $\Sigma_{c}^{+} \gamma$  & $\Sigma_{c}^{*+} \gamma$  & $\Lambda_{c}^{+} \gamma$  & $\Lambda_{c}^{+} \gamma$  & $\Lambda_{c}^{+} \gamma$  & $\Lambda_{c}^{+} \gamma$  & $\Lambda_{c}^{+} \gamma$  & $\Lambda_{c}^{+} \gamma$  & $\Lambda_{c}^{+} \gamma$  & $\Sigma_{c}^{+} \gamma$  & $\Sigma_{c}^{+} \gamma$  & $\Sigma_{c}^{+} \gamma$  & $\Sigma_{c}^{+} \gamma$  & $\Sigma_{c}^{+} \gamma$  & $\Sigma_{c}^{+} \gamma$  & $\Sigma_{c}^{+} \gamma$ \\
$\Lambda_c(nnc)$  & $\mathbf{J^P}$  & $\vert l_{\lambda}, l_{\rho}, k_{\lambda}, k_{\rho} \rangle$  & $^{2S+1}L_{x,J}$  & $^2S_{1/2}$  & $^2S_{1/2}$  & $^4S_{3/2}$  & $^2P_{\lambda,1/2}$  & $^2P_{\lambda,3/2}$  & $^2P_{\rho,1/2}$  & $^4P_{\rho,1/2}$  & $^2P_{\rho,3/2}$  & $^4P_{\rho,3/2}$  & $^4P_{\rho,5/2}$  & $^2P_{\lambda,1/2}$  & $^4P_{\lambda,1/2}$  & $^2P_{\lambda,3/2}$  & $^4P_{\lambda,3/2}$  & $^4P_{\lambda,5/2}$  & $^2P_{\rho,1/2}$  & $^2P_{\rho,3/2}$  \\ \hline
 $N=2$  &  &  &  &  &  \\
$\Lambda_c(2981)$  & $ \mathbf{\frac{3}{2}^+}$ & $\vert \,2\,,\,0\,,\,0\,,\,0 \,\rangle $ &$^{2}D_{\lambda\lambda,3/2}$&$ 28 _{- 2 }^{+ 2 }$    &  $ 15 _{- 7 }^{+ 9 }$    &  $ 3.8 _{- 2.0 }^{+ 2.5 }$    &  $ 2.0 _{- 1.9 }^{+ 2.8 }$    &  $ 1.2 _{- 0.9 }^{+ 0.9 }$    &  0  &0  &0  &0  &0  &$ 1.0 _{- 0.7 }^{+ 1.2 }$    &  $ 0.9 _{- 0.8 }^{+ 1.9 }$    &  $ 0.6 _{- 0.5 }^{+ 0.8 }$    &  $ 0.1 _{- 0.1 }^{+ 0.2 }$    &  0  &0  &0 \\
  &  &  & & $...$  & $...$  & $...$ & $0.01$  & $0.07$  & $...$  & $...$  & $...$  & $...$  &  $...$  & $0.23$  & $<0.01$  & $0.01$  & $0.08$  & $<0.01$  & $...$  & $...$  & \cite{Yao:2018jmc} \\
  &  &  & & $41.3$  & $1.4$  & $1.3$ & $0.3$  & $0.1$  & $...$  & $...$  & $...$  & $...$  &  $...$  & $...$  & $...$  & $...$  & $...$  & $0$  & $...$  & $...$  & \cite{Peng:2024pyl} \\
$\Lambda_c(3026)$  & $ \mathbf{\frac{5}{2}^+}$ & $\vert \,2\,,\,0\,,\,0\,,\,0 \,\rangle $ &$^{2}D_{\lambda\lambda,5/2}$&$ 35 _{- 2 }^{+ 3 }$    &  $ 22 _{- 11 }^{+ 13 }$    &  $ 6 _{- 3 }^{+ 4 }$    &  $ 0.2 _{- 0.1 }^{+ 0.1 }$    &  $ 2.2 _{- 1.0 }^{+ 1.7 }$    &  0  &0  &0  &0  &0  &$ 4.7 _{- 3.0 }^{+ 4.5 }$    &  $ 0.2 _{- 0.2 }^{+ 0.4 }$    &  $ 2.6 _{- 1.7 }^{+ 2.8 }$    &  $ 0.7 _{- 0.5 }^{+ 1.1 }$    &  $ 0.2 _{- 0.1 }^{+ 0.3 }$    &  0  &0 \\
  &  &  & & $...$  & $...$  & $...$ & $0.13$  & $0.26$  & $...$  & $...$  & $...$  & $...$  &  $...$  & $0.8$  & $<0.01$  & $0.05$  & $0.13$  & $0.19$  & $...$  & $...$  & \cite{Yao:2018jmc} \\
 &  &  & & $42.8$  & $1.9$  & $2.0$ & $0.3$  & $1.1$  & $...$  & $...$  & $...$  & $...$  &  $...$  & $...$  & $...$  & $...$  & $...$  & $0.1$  & $...$  & $...$  & \cite{Peng:2024pyl} \\
$\Lambda_c(3008)$  & $ \mathbf{\frac{1}{2}^+}$ & $\vert \,0\,,\,0\,,\,1\,,\,0 \,\rangle $ &$^{2}S_{1/2}$&$ 0.4 _{- 0.1 }^{+ 0.2 }$    &  $ 48 _{- 23 }^{+ 28 }$    &  $ 13 _{- 7 }^{+ 9 }$    &  $ 0.4 _{- 0.4 }^{+ 1.2 }$    &  $ 10 _{- 3 }^{+ 3 }$    &  0  &0  &0  &0  &0  &$ 15 _{- 9 }^{+ 14 }$    &  0  &$ 23 _{- 16 }^{+ 27 }$    &  $ 0.1 _{- 0.1 }^{+ 0.1 }$    &  $ 0.4 _{- 0.4 }^{+ 1.3 }$    &  0  &0 \\
 &  &  & & $0$  & $0$  & $3.2$ & $0.3$  & $0.2$  & $...$  & $...$  & $...$  & $...$  &  $...$  & $...$  & $...$  & $...$  & $...$  & $...$  & $...$  & $...$  & \cite{Peng:2024pyl} \\
$\Lambda_c(3375)$  & $ \mathbf{\frac{1}{2}^+}$ & $\vert \,0\,,\,0\,,\,0\,,\,1 \,\rangle $ &$^{2}S_{1/2}$&0  &$ 240 _{- 67 }^{+ 57 }$    &  $ 100 _{- 33 }^{+ 30 }$    &  $ 2.7 _{- 0.7 }^{+ 0.6 }$    &  $ 1.3 _{- 0.4 }^{+ 0.3 }$    &  $ 58 _{- 18 }^{+ 15 }$    &  $ 0.4 _{- 0.1 }^{+ 0.1 }$    &  $ 157 _{- 56 }^{+ 49 }$    &  $ 1.5 _{- 0.6 }^{+ 0.6 }$    &  $ 41 _{- 19 }^{+ 20 }$    &  $ 46 _{- 31 }^{+ 49 }$    &  $ 0.2 _{- 0.2 }^{+ 0.3 }$    &  $ 99 _{- 71 }^{+ 117 }$    &  $ 0.7 _{- 0.5 }^{+ 0.8 }$    &  $ 12 _{- 10 }^{+ 18 }$    &  $ 814 _{- 146 }^{+ 122 }$    &  $ 396 _{- 79 }^{+ 65 }$   \\
$\Lambda_c(3164)$  & $ \mathbf{\frac{3}{2}^+}$ & $\vert \,1\,,\,1\,,\,0\,,\,0 \,\rangle $ &$^{2}D_{\lambda\rho,3/2}$&$ 22 _{- 7 }^{+ 6 }$    &  $ 123 _{- 10 }^{+ 9 }$    &  $ 11 _{- 5 }^{+ 6 }$    &  $ 3.0 _{- 0.8 }^{+ 0.7 }$    &  $ 2.6 _{- 0.5 }^{+ 0.4 }$    &  $ 5 _{- 2 }^{+ 4 }$    &  $ 1.4 _{- 0.9 }^{+ 1.2 }$    &  $ 2.5 _{- 1.3 }^{+ 1.7 }$    &  $ 0.3 _{- 0.2 }^{+ 0.3 }$    &  0  &$ 494 _{- 39 }^{+ 34 }$    &  $ 6 _{- 3 }^{+ 5 }$    &  $ 65 _{- 7 }^{+ 7 }$    &  $ 1.4 _{- 0.8 }^{+ 1.1 }$    &  $ 0.1 _{- 0.1 }^{+ 0.1 }$    &  $ 0.5 _{- 0.4 }^{+ 1.0 }$    &  $ 0.4 _{- 0.3 }^{+ 0.8 }$   \\
$\Lambda_c(3210)$  & $ \mathbf{\frac{5}{2}^+}$ & $\vert \,1\,,\,1\,,\,0\,,\,0 \,\rangle $ &$^{2}D_{\lambda\rho,5/2}$&$ 24 _{- 7 }^{+ 6 }$    &  $ 821 _{- 195 }^{+ 170 }$    &  $ 15 _{- 6 }^{+ 7 }$    &  $ 8 _{- 2 }^{+ 2 }$    &  $ 4.8 _{- 1.1 }^{+ 0.9 }$    &  $ 11 _{- 6 }^{+ 7 }$    &  $ 0.2 _{- 0.1 }^{+ 0.2 }$    &  $ 0.9 _{- 0.9 }^{+ 1.6 }$    &  $ 0.8 _{- 0.4 }^{+ 0.7 }$    &  $ 0.4 _{- 0.3 }^{+ 0.4 }$    &  $ 74 _{- 37 }^{+ 46 }$    &  $ 1.0 _{- 0.6 }^{+ 0.9 }$    &  $ 885 _{- 82 }^{+ 69 }$    &  $ 3.4 _{- 1.8 }^{+ 2.5 }$    &  $ 1.8 _{- 1.1 }^{+ 1.5 }$    &  $ 2.5 _{- 1.9 }^{+ 3.3 }$    &  $ 1.4 _{- 1.1 }^{+ 2.0 }$   \\
$\Lambda_c(3179)$  & $ \mathbf{\frac{1}{2}^+}$ & $\vert \,1\,,\,1\,,\,0\,,\,0 \,\rangle $ &$^{4}D_{\lambda\rho,1/2}$&$ 11 _{- 4 }^{+ 3 }$    &  $ 18 _{- 8 }^{+ 9 }$    &  $ 46 _{- 21 }^{+ 20 }$    &  $ 6 _{- 2 }^{+ 2 }$    &  $ 0.1 _{- 0.1 }^{+ 0.3 }$    &  $ 1.5 _{- 0.8 }^{+ 1.0 }$    &  $ 3.5 _{- 3.2 }^{+ 6.0 }$    &  0  &$ 1.4 _{- 1.0 }^{+ 1.4 }$    &  $ 0.3 _{- 0.2 }^{+ 0.3 }$    &  $ 6 _{- 3 }^{+ 4 }$    &  $ 394 _{- 99 }^{+ 76 }$    &  0  &$ 70 _{- 22 }^{+ 24 }$    &  $ 3.2 _{- 2.4 }^{+ 3.8 }$    &  $ 1.5 _{- 1.3 }^{+ 2.9 }$    &  0 \\
$\Lambda_c(3206)$  & $ \mathbf{\frac{3}{2}^+}$ & $\vert \,1\,,\,1\,,\,0\,,\,0 \,\rangle $ &$^{4}D_{\lambda\rho,3/2}$&$ 24 _{- 8 }^{+ 6 }$    &  $ 40 _{- 17 }^{+ 18 }$    &  $ 109 _{- 21 }^{+ 20 }$    &  $ 16 _{- 4 }^{+ 4 }$    &  $ 1.5 _{- 0.8 }^{+ 1.1 }$    &  $ 4.2 _{- 2.0 }^{+ 2.5 }$    &  $ 5 _{- 3 }^{+ 5 }$    &  $ 0.2 _{- 0.1 }^{+ 0.1 }$    &  $ 0.8 _{- 0.7 }^{+ 1.2 }$    &  $ 0.4 _{- 0.3 }^{+ 0.4 }$    &  $ 17 _{- 9 }^{+ 10 }$    &  $ 316 _{- 30 }^{+ 26 }$    &  $ 0.9 _{- 0.5 }^{+ 0.7 }$    &  $ 199 _{- 28 }^{+ 23 }$    &  $ 18 _{- 7 }^{+ 9 }$    &  $ 6 _{- 4 }^{+ 8 }$    &  $ 0.2 _{- 0.2 }^{+ 0.4 }$   \\
$\Lambda_c(3252)$  & $ \mathbf{\frac{5}{2}^+}$ & $\vert \,1\,,\,1\,,\,0\,,\,0 \,\rangle $ &$^{4}D_{\lambda\rho,5/2}$&$ 36 _{- 10 }^{+ 8 }$    &  $ 66 _{- 26 }^{+ 26 }$    &  $ 216 _{- 41 }^{+ 37 }$    &  $ 7 _{- 2 }^{+ 2 }$    &  $ 14 _{- 4 }^{+ 3 }$    &  $ 2.2 _{- 0.8 }^{+ 0.9 }$    &  $ 6 _{- 3 }^{+ 4 }$    &  $ 4.2 _{- 1.7 }^{+ 2.1 }$    &  $ 2.4 _{- 1.4 }^{+ 2.1 }$    &  $ 0.6 _{- 0.4 }^{+ 0.6 }$    &  $ 9 _{- 4 }^{+ 4 }$    &  $ 45 _{- 23 }^{+ 28 }$    &  $ 17 _{- 7 }^{+ 9 }$    &  $ 572 _{- 53 }^{+ 42 }$    &  $ 109 _{- 10 }^{+ 10 }$    &  $ 4.5 _{- 3.3 }^{+ 5.6 }$    &  $ 7 _{- 5 }^{+ 10 }$   \\
$\Lambda_c(3315)$  & $ \mathbf{\frac{7}{2}^+}$ & $\vert \,1\,,\,1\,,\,0\,,\,0 \,\rangle $ &$^{4}D_{\lambda\rho,7/2}$&$ 25 _{- 6 }^{+ 4 }$    &  $ 52 _{- 19 }^{+ 18 }$    &  $ 882 _{- 202 }^{+ 164 }$    &  $ 2.2 _{- 1.3 }^{+ 1.9 }$    &  $ 17 _{- 3 }^{+ 2 }$    &  $ 0.1 _{- 0.1 }^{+ 0.1 }$    &  $ 1.1 _{- 0.7 }^{+ 1.1 }$    &  $ 7 _{- 3 }^{+ 3 }$    &  $ 10 _{- 5 }^{+ 8 }$    &  $ 1.1 _{- 1.1 }^{+ 2.1 }$    &  $ 1.0 _{- 0.7 }^{+ 1.3 }$    &  $ 16 _{- 10 }^{+ 16 }$    &  $ 26 _{- 10 }^{+ 11 }$    &  $ 83 _{- 42 }^{+ 56 }$    &  $ 966 _{- 104 }^{+ 83 }$    &  $ 0.1 _{- 0.1 }^{+ 0.2 }$    &  $ 18 _{- 12 }^{+ 19 }$   \\
$\Lambda_c(3174)$  & $ \mathbf{\frac{1}{2}^-}$ & $\vert \,1\,,\,1\,,\,0\,,\,0 \,\rangle $ &$^{2}P_{\lambda\rho,1/2}$&0  &$ 24 _{- 7 }^{+ 5 }$    &  0  &0  &0  &$ 1.7 _{- 1.6 }^{+ 2.0 }$    &  $ 1.8 _{- 1.1 }^{+ 1.5 }$    &  $ 0.6 _{- 0.5 }^{+ 0.6 }$    &  $ 1.0 _{- 0.6 }^{+ 0.9 }$    &  $ 0.1 _{- 0.1 }^{+ 0.2 }$    &  $ 616 _{- 67 }^{+ 54 }$    &  $ 7 _{- 4 }^{+ 6 }$    &  $ 63 _{- 4 }^{+ 4 }$    &  $ 4.2 _{- 2.5 }^{+ 3.4 }$    &  $ 0.5 _{- 0.4 }^{+ 0.6 }$    &  0  &0 \\
$\Lambda_c(3201)$  & $ \mathbf{\frac{3}{2}^-}$ & $\vert \,1\,,\,1\,,\,0\,,\,0 \,\rangle $ &$^{2}P_{\lambda\rho,3/2}$&0  &$ 24 _{- 6 }^{+ 5 }$    &  0  &$ 13 _{- 4 }^{+ 4 }$    &  $ 6 _{- 2 }^{+ 2 }$    &  $ 10 _{- 6 }^{+ 7 }$    &  $ 0.8 _{- 0.5 }^{+ 0.7 }$    &  $ 1.3 _{- 0.7 }^{+ 0.9 }$    &  $ 0.8 _{- 0.5 }^{+ 0.7 }$    &  $ 0.2 _{- 0.1 }^{+ 0.2 }$    &  $ 328 _{- 73 }^{+ 77 }$    &  $ 3.4 _{- 1.9 }^{+ 2.7 }$    &  $ 424 _{- 28 }^{+ 29 }$    &  $ 3.5 _{- 2.0 }^{+ 2.9 }$    &  $ 0.8 _{- 0.5 }^{+ 0.8 }$    &  $ 2.8 _{- 2.1 }^{+ 3.8 }$    &  $ 0.7 _{- 0.6 }^{+ 1.2 }$   \\
$\Lambda_c(3215)$  & $ \mathbf{\frac{1}{2}^-}$ & $\vert \,1\,,\,1\,,\,0\,,\,0 \,\rangle $ &$^{4}P_{\lambda\rho,1/2}$&0  &0  &$ 6 _{- 2 }^{+ 1 }$    &  $ 20 _{- 7 }^{+ 7 }$    &  $ 9 _{- 3 }^{+ 3 }$    &  $ 4.3 _{- 2.0 }^{+ 2.5 }$    &  $ 0.4 _{- 0.4 }^{+ 0.5 }$    &  $ 1.6 _{- 0.8 }^{+ 1.1 }$    &  $ 1.5 _{- 1.1 }^{+ 2.2 }$    &  $ 0.4 _{- 0.3 }^{+ 0.5 }$    &  $ 18 _{- 10 }^{+ 12 }$    &  $ 154 _{- 18 }^{+ 14 }$    &  $ 7 _{- 4 }^{+ 5 }$    &  $ 72 _{- 24 }^{+ 19 }$    &  $ 1.1 _{- 0.7 }^{+ 1.0 }$    &  $ 6 _{- 5 }^{+ 9 }$    &  $ 1.6 _{- 1.4 }^{+ 3.1 }$   \\
$\Lambda_c(3243)$  & $ \mathbf{\frac{3}{2}^-}$ & $\vert \,1\,,\,1\,,\,0\,,\,0 \,\rangle $ &$^{4}P_{\lambda\rho,3/2}$&0  &0  &$ 17 _{- 5 }^{+ 4 }$    &  $ 12 _{- 4 }^{+ 4 }$    &  $ 5 _{- 2 }^{+ 2 }$    &  $ 2.9 _{- 1.2 }^{+ 1.5 }$    &  $ 7 _{- 5 }^{+ 6 }$    &  $ 1.1 _{- 0.5 }^{+ 0.7 }$    &  $ 0.9 _{- 0.6 }^{+ 0.8 }$    &  $ 0.2 _{- 0.2 }^{+ 0.3 }$    &  $ 12 _{- 6 }^{+ 8 }$    &  $ 586 _{- 111 }^{+ 98 }$    &  $ 4.9 _{- 2.4 }^{+ 2.9 }$    &  $ 61 _{- 4 }^{+ 4 }$    &  $ 29 _{- 12 }^{+ 9 }$    &  $ 4.9 _{- 3.6 }^{+ 6.3 }$    &  $ 1.5 _{- 1.1 }^{+ 2.3 }$   \\
$\Lambda_c(3288)$  & $ \mathbf{\frac{5}{2}^-}$ & $\vert \,1\,,\,1\,,\,0\,,\,0 \,\rangle $ &$^{4}P_{\lambda\rho,5/2}$&0  &0  &$ 23 _{- 6 }^{+ 4 }$    &  $ 5 _{- 2 }^{+ 1 }$    &  $ 27 _{- 8 }^{+ 8 }$    &  $ 1.5 _{- 0.6 }^{+ 0.7 }$    &  $ 1.5 _{- 0.9 }^{+ 1.2 }$    &  $ 7 _{- 3 }^{+ 4 }$    &  $ 8 _{- 6 }^{+ 7 }$    &  $ 0.9 _{- 0.5 }^{+ 0.8 }$    &  $ 6 _{- 3 }^{+ 3 }$    &  $ 4.9 _{- 2.1 }^{+ 2.7 }$    &  $ 29 _{- 14 }^{+ 16 }$    &  $ 508 _{- 104 }^{+ 99 }$    &  $ 289 _{- 22 }^{+ 22 }$    &  $ 3.7 _{- 2.5 }^{+ 3.9 }$    &  $ 14 _{- 10 }^{+ 18 }$   \\
$\Lambda_c(3261)$  & $ \mathbf{\frac{3}{2}^+}$ & $\vert \,1\,,\,1\,,\,0\,,\,0 \,\rangle $ &$^{4}S_{\lambda\rho,3/2}$&0  &0  &0  &$ 1.7 _{- 0.8 }^{+ 0.6 }$    &  $ 3.7 _{- 1.3 }^{+ 0.8 }$    &  $ 1.6 _{- 0.5 }^{+ 0.5 }$    &  $ 7 _{- 5 }^{+ 5 }$    &  $ 2.8 _{- 1.0 }^{+ 1.0 }$    &  $ 3.4 _{- 2.5 }^{+ 2.8 }$    &  $ 0.1 _{- 0.1 }^{+ 0.1 }$    &  $ 5 _{- 2 }^{+ 1 }$    &  $ 357 _{- 78 }^{+ 70 }$    &  $ 10 _{- 3 }^{+ 3 }$    &  $ 321 _{- 65 }^{+ 58 }$    &  $ 52 _{- 5 }^{+ 5 }$    &  $ 4.3 _{- 2.9 }^{+ 4.3 }$    &  $ 6 _{- 4 }^{+ 7 }$   \\
$\Lambda_c(3192)$  & $ \mathbf{\frac{1}{2}^+}$ & $\vert \,1\,,\,1\,,\,0\,,\,0 \,\rangle $ &$^{2}S_{\lambda\rho,1/2}$&0  &0  &0  &$ 2.1 _{- 0.3 }^{+ 0.2 }$    &  $ 4.2 _{- 0.4 }^{+ 0.3 }$    &  $ 13 _{- 8 }^{+ 9 }$    &  $ 0.4 _{- 0.2 }^{+ 0.2 }$    &  $ 0.5 _{- 0.3 }^{+ 0.5 }$    &  $ 0.1 _{- 0.0 }^{+ 0.0 }$    &  $ 0.3 _{- 0.2 }^{+ 0.3 }$    &  $ 619 _{- 130 }^{+ 125 }$    &  $ 1.4 _{- 0.6 }^{+ 0.8 }$    &  $ 93 _{- 8 }^{+ 8 }$    &  $ 0.2 _{- 0.1 }^{+ 0.1 }$    &  $ 1.1 _{- 0.7 }^{+ 0.9 }$    &  $ 1.3 _{- 1.0 }^{+ 1.9 }$    &  $ 1.4 _{- 1.2 }^{+ 2.5 }$   \\
$\Lambda_c(3348)$  & $ \mathbf{\frac{3}{2}^+}$ & $\vert \,0\,,\,2\,,\,0\,,\,0 \,\rangle $ &$^{2}D_{\rho\rho,3/2}$&$ 17 _{- 4 }^{+ 4 }$    &  $ 90 _{- 27 }^{+ 23 }$    &  $ 37 _{- 14 }^{+ 12 }$    &  $ 13 _{- 4 }^{+ 4 }$    &  $ 9 _{- 3 }^{+ 3 }$    &  $ 8 _{- 3 }^{+ 3 }$    &  $ 22 _{- 10 }^{+ 12 }$    &  $ 8 _{- 3 }^{+ 3 }$    &  $ 6 _{- 3 }^{+ 4 }$    &  $ 0.5 _{- 0.3 }^{+ 0.3 }$    &  $ 2.3 _{- 1.7 }^{+ 2.6 }$    &  $ 0.7 _{- 0.5 }^{+ 1.0 }$    &  $ 3.3 _{- 2.4 }^{+ 4.0 }$    &  $ 0.1 _{- 0.1 }^{+ 0.1 }$    &  $ 0.4 _{- 0.3 }^{+ 0.7 }$    &  $ 1429 _{- 158 }^{+ 120 }$    &  $ 224 _{- 38 }^{+ 31 }$   \\
$\Lambda_c(3393)$  & $ \mathbf{\frac{5}{2}^+}$ & $\vert \,0\,,\,2\,,\,0\,,\,0 \,\rangle $ &$^{2}D_{\rho\rho,5/2}$&$ 15 _{- 3 }^{+ 4 }$    &  $ 100 _{- 28 }^{+ 22 }$    &  $ 42 _{- 14 }^{+ 12 }$    &  $ 5 _{- 2 }^{+ 1 }$    &  $ 18 _{- 6 }^{+ 5 }$    &  $ 19 _{- 6 }^{+ 7 }$    &  $ 3.0 _{- 1.5 }^{+ 1.9 }$    &  $ 14 _{- 5 }^{+ 5 }$    &  $ 12 _{- 5 }^{+ 6 }$    &  $ 8 _{- 4 }^{+ 4 }$    &  $ 3.9 _{- 2.7 }^{+ 4.2 }$    &  $ 1.2 _{- 0.9 }^{+ 1.5 }$    &  $ 6 _{- 4 }^{+ 6 }$    &  $ 0.2 _{- 0.1 }^{+ 0.2 }$    &  $ 0.9 _{- 0.7 }^{+ 1.2 }$    &  $ 7 _{- 4 }^{+ 6 }$    &  $ 1639 _{- 164 }^{+ 134 }$   \\
\hline \hline
\end{tabular}

\endgroup
}
\label{lambdas_EM_2shell}
\end{table*}
\end{turnpage}


\begin{turnpage}
\begin{table*}[htp]
\caption{ Same as table \ref{lambdas_EM_2shell}, but for $\Xi_{c}^{+}$ states with isospin $1/2$.}
\begin{center}
\scriptsize{
\begingroup
\setlength{\tabcolsep}{1.75pt} 
\renewcommand{\arraystretch}{1.35} 
\begin{tabular}{c c c c  p{1.0cm}  p{1.0cm}  p{1.0cm}  p{1.0cm}  p{1.0cm}  p{1.0cm}  p{1.0cm}  p{1.0cm}  p{1.0cm}  p{1.0cm}  p{1.0cm}  p{1.0cm}  p{1.0cm}  p{1.0cm}  p{1.0cm}  p{1.0cm}  p{1.0cm}  p{1.0cm}  p{1.0cm}  p{1.0cm}} \hline \hline
  &    &    &    & $\Xi_{c}^{+} \gamma$  & $\Xi'^{+}_{c} \gamma$  & $\Xi^{*+}_{c} \gamma$  & $\Xi^{+}_{c} \gamma$  & $\Xi^{+}_{c} \gamma$  & $\Xi^{+}_{c} \gamma$  & $\Xi^{+}_{c} \gamma$  & $\Xi^{+}_{c} \gamma$  & $\Xi^{+}_{c} \gamma$  & $\Xi^{+}_{c} \gamma$  & $\Xi'^{+}_{c} \gamma$  & $\Xi'^{+}_{c} \gamma$  & $\Xi'^{+}_{c} \gamma$  & $\Xi'^{+}_{c} \gamma$  & $\Xi'^{+}_{c} \gamma$  & $\Xi'^{+}_{c} \gamma$  & $\Xi'^{+}_{c} \gamma$ \\
$\Xi_c(snc)$  & $\mathbf{J^P}$  & $\vert l_{\lambda}, l_{\rho}, k_{\lambda}, k_{\rho} \rangle$  & $^{2S+1}L_{x,J}$  & $^2S_{1/2}$  & $^2S_{1/2}$  & $^4S_{3/2}$  & $^2P_{\lambda,1/2}$  & $^2P_{\lambda,3/2}$  & $^2P_{\rho,1/2}$  & $^P_{\rho,1/2}$  & $^2P_{\rho,3/2}$  & $^4P_{\rho,3/2}$  & $^4P_{\rho,5/2}$  & $^2P_{\lambda,1/2}$  & $^4P_{\lambda,1/2}$  & $^2P_{\lambda,3/2}$  & $^4P_{\lambda,3/2}$  & $^4P_{\lambda,5/2}$  & $^2P_{\rho,1/2}$  & $^2P_{\rho,3/2}$  \\ \hline
 $N=2$  &  &  &  &  &  \\
$\Xi_c(3118)$  & $ \mathbf{\frac{3}{2}^+}$ & $\vert \,2\,,\,0\,,\,0\,,\,0 \,\rangle $ &$^{2}D_{\lambda\lambda,3/2}$&$ 49 _{- 8 }^{+ 9 }$    &  $ 10 _{- 4 }^{+ 5 }$    &  $ 2.5 _{- 1.0 }^{+ 1.4 }$    &  $ 34 _{- 21 }^{+ 22 }$    &  $ 7 _{- 4 }^{+ 4 }$    &  0  &0  &0  &0  &0  &$ 2.0 _{- 0.9 }^{+ 1.4 }$    &  $ 2.5 _{- 1.7 }^{+ 3.1 }$    &  $ 1.5 _{- 0.8 }^{+ 1.2 }$    &  $ 0.3 _{- 0.2 }^{+ 0.4 }$    &  0  &0  &0 \\
  &  &  & & $...$  & $...$  & $...$ & $1.09$  & $0.57$  & $...$  & $...$  & $...$  & $...$  &  $...$  & $0.06$  & $0.97$  & $0.22$  & $0.66$  & $0.09$  & $...$  & $...$  & \cite{Yao:2018jmc} \\
  &  &  & & $28.2$  & $1.3$  & $1.6$ & $1.2$  & $0.4$  & $...$  & $...$  & $...$  & $...$  &  $...$  & $...$  & $...$  & $...$  & $...$  & $0$  & $...$  & $...$  & \cite{Peng:2024pyl} \\
$\Xi_c(3163)$  & $ \mathbf{\frac{5}{2}^+}$ & $\vert \,2\,,\,0\,,\,0\,,\,0 \,\rangle $ &$^{2}D_{\lambda\lambda,5/2}$&$ 61 _{- 7 }^{+ 8 }$    &  $ 14 _{- 6 }^{+ 7 }$    &  $ 4.0 _{- 1.7 }^{+ 2.0 }$    &  $ 0.2 _{- 0.1 }^{+ 0.1 }$    &  $ 31 _{- 22 }^{+ 24 }$    &  0  &0  &0  &0  &0  &$ 8 _{- 3 }^{+ 4 }$    &  $ 0.5 _{- 0.3 }^{+ 0.5 }$    &  $ 4.7 _{- 2.1 }^{+ 2.9 }$    &  $ 1.7 _{- 0.8 }^{+ 1.2 }$    &  $ 0.6 _{- 0.4 }^{+ 0.6 }$    &  0  &0 \\
  &  &  & & $...$  & $...$  & $...$ & $0.36$  & $0.28$  & $...$  & $...$  & $...$  & $...$  &  $...$  & $0.26$  & $0.21$  & $0.42$  & $0.63$  & $1.24$  & $...$  & $...$  & \cite{Yao:2018jmc} \\
  &  &  & & $27.1$  & $1.6$  & $2.0$ & $0.1$  & $0.6$  & $...$  & $...$  & $...$  & $...$  &  $...$  & $...$  & $...$  & $...$  & $...$  & $0.5$  & $...$  & $...$  & \cite{Peng:2024pyl} \\
$\Xi_c(3145)$  & $ \mathbf{\frac{1}{2}^+}$ & $\vert \,0\,,\,0\,,\,1\,,\,0 \,\rangle $ &$^{2}S_{1/2}$&$ 0.4 _{- 0.1 }^{+ 0.2 }$    &  $ 31 _{- 12 }^{+ 15 }$    &  $ 8 _{- 4 }^{+ 4 }$    &  $ 10 _{- 9 }^{+ 11 }$    &  $ 25 _{- 8 }^{+ 9 }$    &  0  &0  &0  &0  &0  &$ 26 _{- 11 }^{+ 15 }$    &  $ 0.1 _{- 0.1 }^{+ 0.1 }$    &  $ 47 _{- 20 }^{+ 28 }$    &  $ 0.2 _{- 0.1 }^{+ 0.2 }$    &  $ 2.0 _{- 1.4 }^{+ 2.3 }$    &  0  &0 \\
  &  &  & & $0$  & $0.9$  & $1.4$ & $0$  & $0.1$  & $...$  & $...$  & $...$  & $...$  &  $...$  & $...$  & $...$  & $...$  & $...$  & $0$  & $...$  & $...$  & \cite{Peng:2024pyl} \\
$\Xi_c(3440)$  & $ \mathbf{\frac{1}{2}^+}$ & $\vert \,0\,,\,0\,,\,0\,,\,1 \,\rangle $ &$^{2}S_{1/2}$&0  &$ 176 _{- 51 }^{+ 49 }$    &  $ 67 _{- 21 }^{+ 21 }$    &  $ 4.1 _{- 1.5 }^{+ 1.6 }$    &  $ 1.9 _{- 0.7 }^{+ 0.8 }$    &  $ 90 _{- 36 }^{+ 38 }$    &  $ 0.5 _{- 0.2 }^{+ 0.3 }$    &  $ 229 _{- 90 }^{+ 105 }$    &  $ 1.9 _{- 0.9 }^{+ 1.1 }$    &  $ 48 _{- 24 }^{+ 33 }$    &  $ 19 _{- 11 }^{+ 16 }$    &  $ 0.1 _{- 0.1 }^{+ 0.1 }$    &  $ 40 _{- 24 }^{+ 33 }$    &  $ 0.2 _{- 0.2 }^{+ 0.2 }$    &  $ 4.3 _{- 2.9 }^{+ 4.6 }$    &  $ 764 _{- 119 }^{+ 114 }$    &  $ 370 _{- 62 }^{+ 57 }$   \\
$\Xi_c(3265)$  & $ \mathbf{\frac{3}{2}^+}$ & $\vert \,1\,,\,1\,,\,0\,,\,0 \,\rangle $ &$^{2}D_{\lambda\rho,3/2}$&$ 28 _{- 11 }^{+ 13 }$    &  $ 130 _{- 11 }^{+ 10 }$    &  $ 6 _{- 3 }^{+ 3 }$    &  $ 4.5 _{- 1.5 }^{+ 1.8 }$    &  $ 4.3 _{- 1.4 }^{+ 1.6 }$    &  $ 5 _{- 4 }^{+ 4 }$    &  $ 1.3 _{- 0.8 }^{+ 1.2 }$    &  $ 1.9 _{- 1.0 }^{+ 1.3 }$    &  $ 0.3 _{- 0.2 }^{+ 0.2 }$    &  0  &$ 486 _{- 35 }^{+ 33 }$    &  $ 5 _{- 2 }^{+ 3 }$    &  $ 60 _{- 6 }^{+ 6 }$    &  $ 1.2 _{- 0.5 }^{+ 0.8 }$    &  $ 0.1 _{- 0.0 }^{+ 0.1 }$    &  $ 1.0 _{- 0.5 }^{+ 0.9 }$    &  $ 0.7 _{- 0.4 }^{+ 0.7 }$   \\
$\Xi_c(3311)$  & $ \mathbf{\frac{5}{2}^+}$ & $\vert \,1\,,\,1\,,\,0\,,\,0 \,\rangle $ &$^{2}D_{\lambda\rho,5/2}$&$ 34 _{- 13 }^{+ 14 }$    &  $ 663 _{- 153 }^{+ 146 }$    &  $ 9 _{- 3 }^{+ 4 }$    &  $ 12 _{- 4 }^{+ 4 }$    &  $ 8 _{- 2 }^{+ 3 }$    &  $ 13 _{- 6 }^{+ 8 }$    &  $ 0.2 _{- 0.1 }^{+ 0.2 }$    &  $ 19 _{- 13 }^{+ 14 }$    &  $ 0.8 _{- 0.4 }^{+ 0.6 }$    &  $ 0.4 _{- 0.3 }^{+ 0.4 }$    &  $ 60 _{- 22 }^{+ 27 }$    &  $ 0.8 _{- 0.4 }^{+ 0.5 }$    &  $ 863 _{- 90 }^{+ 81 }$    &  $ 3.0 _{- 1.2 }^{+ 1.6 }$    &  $ 1.6 _{- 0.7 }^{+ 0.9 }$    &  $ 3.8 _{- 1.9 }^{+ 2.8 }$    &  $ 2.3 _{- 1.3 }^{+ 1.8 }$   \\
$\Xi_c(3281)$  & $ \mathbf{\frac{1}{2}^+}$ & $\vert \,1\,,\,1\,,\,0\,,\,0 \,\rangle $ &$^{4}D_{\lambda\rho,1/2}$&$ 15 _{- 6 }^{+ 8 }$    &  $ 11 _{- 4 }^{+ 6 }$    &  $ 61 _{- 13 }^{+ 9 }$    &  $ 10 _{- 4 }^{+ 5 }$    &  0  &$ 1.6 _{- 0.8 }^{+ 1.4 }$    &  $ 2.0 _{- 2.0 }^{+ 4.3 }$    &  0  &$ 3.4 _{- 2.0 }^{+ 3.2 }$    &  $ 0.3 _{- 0.2 }^{+ 0.4 }$    &  $ 5 _{- 2 }^{+ 3 }$    &  $ 388 _{- 68 }^{+ 58 }$    &  0  &$ 62 _{- 15 }^{+ 19 }$    &  $ 2.5 _{- 1.5 }^{+ 2.6 }$    &  $ 2.5 _{- 1.7 }^{+ 3.3 }$    &  0 \\
$\Xi_c(3308)$  & $ \mathbf{\frac{3}{2}^+}$ & $\vert \,1\,,\,1\,,\,0\,,\,0 \,\rangle $ &$^{4}D_{\lambda\rho,3/2}$&$ 34 _{- 14 }^{+ 16 }$    &  $ 26 _{- 10 }^{+ 12 }$    &  $ 94 _{- 17 }^{+ 16 }$    &  $ 26 _{- 9 }^{+ 11 }$    &  $ 1.5 _{- 0.6 }^{+ 0.9 }$    &  $ 4.7 _{- 2.2 }^{+ 3.1 }$    &  $ 8 _{- 5 }^{+ 7 }$    &  $ 0.2 _{- 0.1 }^{+ 0.2 }$    &  $ 1.6 _{- 1.3 }^{+ 2.0 }$    &  $ 0.9 _{- 0.5 }^{+ 0.8 }$    &  $ 15 _{- 6 }^{+ 8 }$    &  $ 304 _{- 31 }^{+ 29 }$    &  $ 0.8 _{- 0.3 }^{+ 0.4 }$    &  $ 193 _{- 18 }^{+ 19 }$    &  $ 16 _{- 5 }^{+ 6 }$    &  $ 8 _{- 5 }^{+ 8 }$    &  $ 0.4 _{- 0.2 }^{+ 0.4 }$   \\
$\Xi_c(3353)$  & $ \mathbf{\frac{5}{2}^+}$ & $\vert \,1\,,\,1\,,\,0\,,\,0 \,\rangle $ &$^{4}D_{\lambda\rho,5/2}$&$ 55 _{- 21 }^{+ 22 }$    &  $ 44 _{- 16 }^{+ 17 }$    &  $ 186 _{- 35 }^{+ 32 }$    &  $ 11 _{- 4 }^{+ 4 }$    &  $ 23 _{- 8 }^{+ 9 }$    &  $ 2.6 _{- 1.0 }^{+ 1.3 }$    &  $ 7 _{- 4 }^{+ 6 }$    &  $ 4.9 _{- 2.0 }^{+ 2.6 }$    &  $ 11 _{- 7 }^{+ 8 }$    &  $ 1.5 _{- 1.0 }^{+ 1.2 }$    &  $ 8 _{- 3 }^{+ 3 }$    &  $ 36 _{- 15 }^{+ 18 }$    &  $ 15 _{- 5 }^{+ 6 }$    &  $ 556 _{- 57 }^{+ 51 }$    &  $ 104 _{- 9 }^{+ 9 }$    &  $ 6 _{- 3 }^{+ 4 }$    &  $ 10 _{- 5 }^{+ 7 }$   \\
$\Xi_c(3416)$  & $ \mathbf{\frac{7}{2}^+}$ & $\vert \,1\,,\,1\,,\,0\,,\,0 \,\rangle $ &$^{4}D_{\lambda\rho,7/2}$&$ 42 _{- 15 }^{+ 15 }$    &  $ 36 _{- 12 }^{+ 13 }$    &  $ 724 _{- 165 }^{+ 157 }$    &  $ 1.5 _{- 0.9 }^{+ 1.4 }$    &  $ 32 _{- 10 }^{+ 10 }$    &  $ 0.1 _{- 0.1 }^{+ 0.1 }$    &  $ 1.2 _{- 0.7 }^{+ 1.1 }$    &  $ 9 _{- 3 }^{+ 4 }$    &  $ 12 _{- 7 }^{+ 9 }$    &  $ 21 _{- 14 }^{+ 16 }$    &  $ 0.5 _{- 0.3 }^{+ 0.6 }$    &  $ 10 _{- 5 }^{+ 8 }$    &  $ 24 _{- 7 }^{+ 9 }$    &  $ 64 _{- 29 }^{+ 33 }$    &  $ 944 _{- 106 }^{+ 93 }$    &  $ 0.2 _{- 0.1 }^{+ 0.2 }$    &  $ 22 _{- 11 }^{+ 15 }$   \\
$\Xi_c(3275)$  & $ \mathbf{\frac{1}{2}^-}$ & $\vert \,1\,,\,1\,,\,0\,,\,0 \,\rangle $ &$^{2}P_{\lambda\rho,1/2}$&0  &$ 18 _{- 5 }^{+ 5 }$    &  0  &0  &0  &$ 17 _{- 11 }^{+ 12 }$    &  $ 1.7 _{- 0.9 }^{+ 1.5 }$    &  $ 3.0 _{- 1.9 }^{+ 2.1 }$    &  $ 0.9 _{- 0.5 }^{+ 0.8 }$    &  $ 0.1 _{- 0.1 }^{+ 0.1 }$    &  $ 591 _{- 68 }^{+ 60 }$    &  $ 7 _{- 3 }^{+ 4 }$    &  $ 63 _{- 6 }^{+ 5 }$    &  $ 3.7 _{- 1.5 }^{+ 2.1 }$    &  $ 0.5 _{- 0.3 }^{+ 0.4 }$    &  0  &0 \\
$\Xi_c(3302)$  & $ \mathbf{\frac{3}{2}^-}$ & $\vert \,1\,,\,1\,,\,0\,,\,0 \,\rangle $ &$^{2}P_{\lambda\rho,3/2}$&0  &$ 18 _{- 5 }^{+ 5 }$    &  0  &$ 17 _{- 6 }^{+ 7 }$    &  $ 7 _{- 3 }^{+ 3 }$    &  $ 22 _{- 12 }^{+ 14 }$    &  $ 0.8 _{- 0.5 }^{+ 0.6 }$    &  $ 10 _{- 7 }^{+ 7 }$    &  $ 0.8 _{- 0.4 }^{+ 0.6 }$    &  $ 0.2 _{- 0.1 }^{+ 0.2 }$    &  $ 307 _{- 52 }^{+ 55 }$    &  $ 3.0 _{- 1.3 }^{+ 1.7 }$    &  $ 412 _{- 35 }^{+ 32 }$    &  $ 3.1 _{- 1.2 }^{+ 1.6 }$    &  $ 0.7 _{- 0.4 }^{+ 0.5 }$    &  $ 4.4 _{- 2.3 }^{+ 3.5 }$    &  $ 1.4 _{- 0.7 }^{+ 1.1 }$   \\
$\Xi_c(3317)$  & $ \mathbf{\frac{1}{2}^-}$ & $\vert \,1\,,\,1\,,\,0\,,\,0 \,\rangle $ &$^{4}P_{\lambda\rho,1/2}$&0  &0  &$ 4.4 _{- 1.4 }^{+ 1.2 }$    &  $ 28 _{- 11 }^{+ 14 }$    &  $ 12 _{- 5 }^{+ 6 }$    &  $ 4.7 _{- 2.2 }^{+ 2.9 }$    &  $ 4.3 _{- 2.8 }^{+ 3.0 }$    &  $ 1.7 _{- 0.8 }^{+ 1.2 }$    &  $ 0.2 _{- 0.2 }^{+ 0.6 }$    &  $ 0.4 _{- 0.3 }^{+ 0.5 }$    &  $ 16 _{- 6 }^{+ 8 }$    &  $ 148 _{- 17 }^{+ 15 }$    &  $ 6 _{- 3 }^{+ 3 }$    &  $ 72 _{- 16 }^{+ 12 }$    &  $ 1.1 _{- 0.6 }^{+ 0.8 }$    &  $ 8 _{- 5 }^{+ 8 }$    &  $ 2.8 _{- 1.7 }^{+ 2.8 }$   \\
$\Xi_c(3344)$  & $ \mathbf{\frac{3}{2}^-}$ & $\vert \,1\,,\,1\,,\,0\,,\,0 \,\rangle $ &$^{4}P_{\lambda\rho,3/2}$&0  &0  &$ 13 _{- 4 }^{+ 3 }$    &  $ 17 _{- 7 }^{+ 8 }$    &  $ 8 _{- 3 }^{+ 4 }$    &  $ 3.3 _{- 1.3 }^{+ 2.0 }$    &  $ 25 _{- 16 }^{+ 18 }$    &  $ 1.2 _{- 0.5 }^{+ 0.8 }$    &  $ 1.4 _{- 0.9 }^{+ 1.0 }$    &  $ 0.4 _{- 0.3 }^{+ 0.5 }$    &  $ 10 _{- 4 }^{+ 5 }$    &  $ 555 _{- 90 }^{+ 88 }$    &  $ 4.2 _{- 1.6 }^{+ 1.9 }$    &  $ 58 _{- 5 }^{+ 5 }$    &  $ 30 _{- 7 }^{+ 5 }$    &  $ 7 _{- 4 }^{+ 5 }$    &  $ 2.3 _{- 1.3 }^{+ 1.9 }$   \\
$\Xi_c(3389)$  & $ \mathbf{\frac{5}{2}^-}$ & $\vert \,1\,,\,1\,,\,0\,,\,0 \,\rangle $ &$^{4}P_{\lambda\rho,5/2}$&0  &0  &$ 18 _{- 5 }^{+ 4 }$    &  $ 8 _{- 3 }^{+ 3 }$    &  $ 40 _{- 14 }^{+ 16 }$    &  $ 1.8 _{- 0.7 }^{+ 0.9 }$    &  $ 1.9 _{- 1.1 }^{+ 1.7 }$    &  $ 8 _{- 3 }^{+ 5 }$    &  $ 25 _{- 15 }^{+ 18 }$    &  $ 6 _{- 4 }^{+ 5 }$    &  $ 6 _{- 2 }^{+ 2 }$    &  $ 4.8 _{- 1.8 }^{+ 2.2 }$    &  $ 25 _{- 9 }^{+ 10 }$    &  $ 477 _{- 84 }^{+ 87 }$    &  $ 282 _{- 21 }^{+ 23 }$    &  $ 4.4 _{- 2.3 }^{+ 2.9 }$    &  $ 18 _{- 9 }^{+ 13 }$   \\
$\Xi_c(3362)$  & $ \mathbf{\frac{3}{2}^+}$ & $\vert \,1\,,\,1\,,\,0\,,\,0 \,\rangle $ &$^{4}S_{\lambda\rho,3/2}$&0  &0  &0  &$ 6 _{- 1 }^{+ 1 }$    &  $ 12 _{- 2 }^{+ 2 }$    &  $ 2.2 _{- 0.7 }^{+ 0.8 }$    &  $ 20 _{- 12 }^{+ 14 }$    &  $ 3.5 _{- 1.4 }^{+ 1.5 }$    &  $ 13 _{- 8 }^{+ 9 }$    &  $ 0.6 _{- 0.4 }^{+ 0.5 }$    &  $ 5 _{- 1 }^{+ 1 }$    &  $ 331 _{- 61 }^{+ 64 }$    &  $ 9 _{- 3 }^{+ 2 }$    &  $ 295 _{- 48 }^{+ 48 }$    &  $ 47 _{- 5 }^{+ 5 }$    &  $ 5 _{- 2 }^{+ 3 }$    &  $ 8 _{- 4 }^{+ 5 }$   \\
$\Xi_c(3293)$  & $ \mathbf{\frac{1}{2}^+}$ & $\vert \,1\,,\,1\,,\,0\,,\,0 \,\rangle $ &$^{2}S_{\lambda\rho,1/2}$&0  &0  &0  &$ 5 _{- 1 }^{+ 1 }$    &  $ 9 _{- 3 }^{+ 3 }$    &  $ 36 _{- 19 }^{+ 22 }$    &  $ 0.4 _{- 0.2 }^{+ 0.3 }$    &  $ 0.4 _{- 0.4 }^{+ 0.7 }$    &  $ 0.1 _{- 0.0 }^{+ 0.0 }$    &  $ 0.2 _{- 0.2 }^{+ 0.3 }$    &  $ 572 _{- 103 }^{+ 102 }$    &  $ 1.3 _{- 0.5 }^{+ 0.6 }$    &  $ 84 _{- 8 }^{+ 8 }$    &  $ 0.2 _{- 0.1 }^{+ 0.1 }$    &  $ 1.0 _{- 0.5 }^{+ 0.6 }$    &  $ 2.1 _{- 1.1 }^{+ 1.4 }$    &  $ 2.6 _{- 1.4 }^{+ 2.2 }$   \\
$\Xi_c(3413)$  & $ \mathbf{\frac{3}{2}^+}$ & $\vert \,0\,,\,2\,,\,0\,,\,0 \,\rangle $ &$^{2}D_{\rho\rho,3/2}$&$ 63 _{- 7 }^{+ 7 }$    &  $ 63 _{- 19 }^{+ 19 }$    &  $ 24 _{- 8 }^{+ 9 }$    &  $ 17 _{- 7 }^{+ 8 }$    &  $ 11 _{- 5 }^{+ 6 }$    &  $ 11 _{- 5 }^{+ 6 }$    &  $ 26 _{- 12 }^{+ 18 }$    &  $ 11 _{- 5 }^{+ 6 }$    &  $ 7 _{- 4 }^{+ 5 }$    &  $ 0.5 _{- 0.3 }^{+ 0.4 }$    &  $ 0.9 _{- 0.6 }^{+ 0.9 }$    &  $ 0.3 _{- 0.2 }^{+ 0.3 }$    &  $ 1.3 _{- 0.8 }^{+ 1.3 }$    &  0  &$ 0.1 _{- 0.1 }^{+ 0.2 }$    &  $ 1357 _{- 153 }^{+ 128 }$    &  $ 211 _{- 30 }^{+ 28 }$   \\
$\Xi_c(3458)$  & $ \mathbf{\frac{5}{2}^+}$ & $\vert \,0\,,\,2\,,\,0\,,\,0 \,\rangle $ &$^{2}D_{\rho\rho,5/2}$ & $ 60 _{- 6 }^{+ 6 }$    &  $ 75 _{- 20 }^{+ 20 }$    &  $ 29 _{- 9 }^{+ 9 }$    &  $ 8 _{- 3 }^{+ 3 }$    &  $ 25 _{- 10 }^{+ 11 }$    &  $ 26 _{- 11 }^{+ 13 }$    &  $ 3.4 _{- 1.7 }^{+ 2.3 }$    &  $ 20 _{- 8 }^{+ 9 }$    &  $ 14 _{- 7 }^{+ 8 }$    &  $ 9 _{- 4 }^{+ 6 }$    &  $ 1.6 _{- 0.9 }^{+ 1.3 }$    &  $ 0.5 _{- 0.3 }^{+ 0.5 }$    &  $ 2.4 _{- 1.4 }^{+ 2.1 }$    &  $ 0.1 _{- 0.0 }^{+ 0.1 }$    &  $ 0.3 _{- 0.2 }^{+ 0.3 }$    &  $ 6 _{- 3 }^{+ 3 }$    &  $ 1563 _{- 169 }^{+ 143 }$   \\
\hline \hline
\end{tabular}

\endgroup
}
\end{center}
\label{cascades-_EM_2shellp}
\end{table*}
\end{turnpage}

\begin{turnpage}
\begin{table*}[htp]
\caption{ Same as table \ref{lambdas_EM_2shell}, but for $\Xi_{c}^{0}$ states with isospin $1/2$.}
\begin{center}
\scriptsize{
\begingroup
\setlength{\tabcolsep}{1.75pt} 
\renewcommand{\arraystretch}{1.35} 
\begin{tabular}{c c c c|  p{1.0cm}  p{1.0cm}  p{1.0cm}  p{1.0cm}  p{1.0cm}  p{1.0cm}  p{1.0cm}  p{1.0cm}  p{1.0cm}  p{1.0cm}  p{1.0cm}  p{1.0cm}  p{1.0cm}  p{1.0cm}  p{1.0cm}  p{1.0cm}  p{1.0cm}  p{1.0cm}  p{1.0cm}  p{1.0cm}} \hline \hline
  &    &    &    & $\Xi_{c}^{0} \gamma$  & $\Xi'^{0}_{c} \gamma$  & $\Xi^{*0}_{c} \gamma$  & $\Xi^{0}_{c} \gamma$  & $\Xi^{0}_{c} \gamma$  & $\Xi^{0}_{c} \gamma$  & $\Xi^{0}_{c} \gamma$  & $\Xi^{0}_{c} \gamma$  & $\Xi^{0}_{c} \gamma$  & $\Xi^{0}_{c} \gamma$  & $\Xi'^{0}_{c} \gamma$  & $\Xi'^{0}_{c} \gamma$  & $\Xi'^{0}_{c} \gamma$  & $\Xi'^{0}_{c} \gamma$  & $\Xi'^{0}_{c} \gamma$  & $\Xi'^{0}_{c} \gamma$  & $\Xi'^{0}_{c} \gamma$ \\
$\Xi_c(snc)$  & $\mathbf{J^P}$  & $\vert l_{\lambda}, l_{\rho}, k_{\lambda}, k_{\rho} \rangle$  & $^{2S+1}L_{x,J}$  & $^2S_{1/2}$  & $^2S_{1/2}$  & $^4S_{3/2}$  & $^2P_{\lambda,1/2}$  & $^2P_{\lambda,3/2}$  & $^2P_{\rho,1/2}$  & $^4P_{\rho,1/2}$  & $^2P_{\rho,3/2}$  & $^4P_{\rho,3/2}$  & $^4P_{\rho,5/2}$  & $^2P_{\lambda,1/2}$  & $^4P_{\lambda,1/2}$  & $^2P_{\lambda,3/2}$  & $^4P_{\lambda,3/2}$  & $^4P_{\lambda,5/2}$  & $^2P_{\rho,1/2}$  & $^2P_{\rho,3/2}$  \\ \hline
 $N=2$  &  &  &  &  &  \\\ 
$\Xi_c(3118)$  & $ \mathbf{\frac{3}{2}^+}$ & $\vert \,2\,,\,0\,,\,0\,,\,0 \,\rangle $ &$^{2}D_{\lambda\lambda,3/2}$&$ 22 _{- 8 }^{+ 8 }$    &  $ 0.2 _{- 0.1 }^{+ 0.2 }$    &  $ 0.1 _{- 0.0 }^{+ 0.0 }$    &  $ 543 _{- 32 }^{+ 31 }$    &  $ 74 _{- 5 }^{+ 5 }$    &  0  &0  &0  &0  &0  &0  &$ 0.1 _{- 0.0 }^{+ 0.1 }$    &  0  &0  &0  &0  &0 \\
  &  &  & & $...$  & $...$  & $...$ & $79.0$  & $21.1$  & $...$  & $...$  & $...$  & $...$  &  $...$  & $0$  & $0$  & $0$  & $0$  & $0$  & $...$  & $...$  & \cite{Yao:2018jmc} \\
  &  &  & & $17.8$  & $0$  & $0$ & $270.4$  & $46.0$  & $...$  & $...$  & $...$  & $...$  &  $...$  & $...$  & $...$  & $...$  & $...$  & $0$  & $...$  & $...$  & \cite{Peng:2024pyl} \\
$\Xi_c(3163)$  & $ \mathbf{\frac{5}{2}^+}$ & $\vert \,2\,,\,0\,,\,0\,,\,0 \,\rangle $ &$^{2}D_{\lambda\lambda,5/2}$&$ 18 _{- 8 }^{+ 7 }$    &  $ 0.3 _{- 0.2 }^{+ 0.2 }$    &  $ 0.1 _{- 0.0 }^{+ 0.1 }$    &  $ 2.3 _{- 0.7 }^{+ 0.7 }$    &  $ 673 _{- 42 }^{+ 41 }$    &  0  &0  &0  &0  &0  &$ 0.2 _{- 0.1 }^{+ 0.1 }$    &  0  &$ 0.1 _{- 0.1 }^{+ 0.1 }$    &  0  &0  &0  &0 \\
  &  &  & & $...$  & $...$  & $...$ & $7.62$  & $85.1$  & $...$  & $...$  & $...$  & $...$  &  $...$  & $0$  & $0$  & $0$  & $0$  & $0$  & $...$  & $...$  & \cite{Yao:2018jmc} \\
  &  &  & &  $21.4$  & $0$ & $0$   & $0.6$  & $298.8$  & $...$  & $...$  & $...$  & $...$  &  $...$  & $...$  & $...$  & $...$  & $...$  & $0$  & $...$  & $...$  & \cite{Peng:2024pyl} \\
$\Xi_c(3145)$  & $ \mathbf{\frac{1}{2}^+}$ & $\vert \,0\,,\,0\,,\,1\,,\,0 \,\rangle $ &$^{2}S_{1/2}$&$ 0.4 _{- 0.1 }^{+ 0.2 }$    &  $ 0.7 _{- 0.4 }^{+ 0.5 }$    &  $ 0.2 _{- 0.1 }^{+ 0.1 }$    &  $ 327 _{- 29 }^{+ 30 }$    &  $ 94 _{- 11 }^{+ 10 }$    &  0  &0  &0  &0  &0  &$ 0.5 _{- 0.3 }^{+ 0.4 }$    &  0  &$ 1.0 _{- 0.6 }^{+ 0.8 }$    &  0  &0  &0  &0 \\
  &  &  & & $0$  & $0.1$  & $0$ & $45.9$  & $68.1$  & $...$  & $...$  & $...$  & $...$  &  $...$  & $...$  & $...$  & $...$  & $...$  & $0$  & $...$  & $...$  & \cite{Peng:2024pyl} \\
$\Xi_c(3440)$  & $ \mathbf{\frac{1}{2}^+}$ & $\vert \,0\,,\,0\,,\,0\,,\,1 \,\rangle $ &$^{2}S_{1/2}$&0  &$ 3.7 _{- 1.7 }^{+ 1.9 }$    &  $ 1.4 _{- 0.7 }^{+ 0.8 }$    &  $ 6 _{- 2 }^{+ 2 }$    &  $ 2.5 _{- 0.8 }^{+ 0.8 }$    &  $ 124 _{- 38 }^{+ 39 }$    &  $ 0.7 _{- 0.3 }^{+ 0.3 }$    &  $ 313 _{- 104 }^{+ 111 }$    &  $ 2.6 _{- 1.1 }^{+ 1.2 }$    &  $ 65 _{- 30 }^{+ 39 }$    &  $ 0.4 _{- 0.3 }^{+ 0.5 }$    &  0  &$ 0.9 _{- 0.6 }^{+ 1.0 }$    &  0  &$ 0.1 _{- 0.1 }^{+ 0.1 }$    &  $ 16 _{- 6 }^{+ 6 }$    &  $ 8 _{- 3 }^{+ 3 }$   \\
$\Xi_c(3265)$  & $ \mathbf{\frac{3}{2}^+}$ & $\vert \,1\,,\,1\,,\,0\,,\,0 \,\rangle $ &$^{2}D_{\lambda\rho,3/2}$&$ 39 _{- 12 }^{+ 14 }$    &  $ 2.7 _{- 0.8 }^{+ 0.8 }$    &  $ 0.1 _{- 0.1 }^{+ 0.1 }$    &  $ 6 _{- 2 }^{+ 2 }$    &  $ 6 _{- 2 }^{+ 2 }$    &  $ 212 _{- 11 }^{+ 12 }$    &  $ 0.7 _{- 0.5 }^{+ 0.7 }$    &  $ 27 _{- 1 }^{+ 2 }$    &  $ 0.2 _{- 0.1 }^{+ 0.2 }$    &  0  &$ 10 _{- 3 }^{+ 3 }$    &  $ 0.1 _{- 0.1 }^{+ 0.1 }$    &  $ 1.3 _{- 0.4 }^{+ 0.4 }$    &  0  &0  &0  &0 \\
$\Xi_c(3311)$  & $ \mathbf{\frac{5}{2}^+}$ & $\vert \,1\,,\,1\,,\,0\,,\,0 \,\rangle $ &$^{2}D_{\lambda\rho,5/2}$&$ 47 _{- 15 }^{+ 15 }$    &  $ 14 _{- 6 }^{+ 6 }$    &  $ 0.2 _{- 0.1 }^{+ 0.1 }$    &  $ 16 _{- 4 }^{+ 4 }$    &  $ 11 _{- 3 }^{+ 3 }$    &  $ 15 _{- 6 }^{+ 9 }$    &  $ 0.1 _{- 0.1 }^{+ 0.1 }$    &  $ 342 _{- 25 }^{+ 23 }$    &  $ 0.4 _{- 0.3 }^{+ 0.4 }$    &  $ 0.2 _{- 0.1 }^{+ 0.2 }$    &  $ 1.3 _{- 0.7 }^{+ 0.9 }$    &  0  &$ 18 _{- 6 }^{+ 6 }$    &  $ 0.1 _{- 0.0 }^{+ 0.0 }$    &  0  &$ 0.1 _{- 0.0 }^{+ 0.1 }$    &  $ 0.1 _{- 0.0 }^{+ 0.0 }$   \\
$\Xi_c(3281)$  & $ \mathbf{\frac{1}{2}^+}$ & $\vert \,1\,,\,1\,,\,0\,,\,0 \,\rangle $ &$^{4}D_{\lambda\rho,1/2}$&$ 21 _{- 7 }^{+ 8 }$    &  $ 0.2 _{- 0.1 }^{+ 0.2 }$    &  $ 1.3 _{- 0.3 }^{+ 0.4 }$    &  $ 13 _{- 4 }^{+ 5 }$    &  0  &$ 0.8 _{- 0.5 }^{+ 0.7 }$    &  $ 166 _{- 29 }^{+ 17 }$    &  0  &$ 23 _{- 6 }^{+ 8 }$    &  $ 0.4 _{- 0.3 }^{+ 0.5 }$    &  $ 0.1 _{- 0.1 }^{+ 0.1 }$    &  $ 8 _{- 2 }^{+ 2 }$    &  0  &$ 1.3 _{- 0.6 }^{+ 0.7 }$    &  $ 0.1 _{- 0.0 }^{+ 0.1 }$    &  $ 0.1 _{- 0.0 }^{+ 0.1 }$    &  0 \\
$\Xi_c(3308)$  & $ \mathbf{\frac{3}{2}^+}$ & $\vert \,1\,,\,1\,,\,0\,,\,0 \,\rangle $ &$^{4}D_{\lambda\rho,3/2}$&$ 46 _{- 15 }^{+ 17 }$    &  $ 0.6 _{- 0.3 }^{+ 0.4 }$    &  $ 2.0 _{- 0.8 }^{+ 0.8 }$    &  $ 35 _{- 10 }^{+ 12 }$    &  $ 2.1 _{- 0.7 }^{+ 1.0 }$    &  $ 2.3 _{- 1.3 }^{+ 1.8 }$    &  $ 123 _{- 9 }^{+ 8 }$    &  $ 0.1 _{- 0.1 }^{+ 0.1 }$    &  $ 81 _{- 7 }^{+ 5 }$    &  $ 5 _{- 1 }^{+ 2 }$    &  $ 0.3 _{- 0.2 }^{+ 0.2 }$    &  $ 6 _{- 2 }^{+ 2 }$    &  0  &$ 4.0 _{- 1.1 }^{+ 1.2 }$    &  $ 0.3 _{- 0.2 }^{+ 0.2 }$    &  $ 0.2 _{- 0.1 }^{+ 0.2 }$    &  0 \\
$\Xi_c(3353)$  & $ \mathbf{\frac{5}{2}^+}$ & $\vert \,1\,,\,1\,,\,0\,,\,0 \,\rangle $ &$^{4}D_{\lambda\rho,5/2}$&$ 75 _{- 23 }^{+ 24 }$    &  $ 0.9 _{- 0.5 }^{+ 0.6 }$    &  $ 3.9 _{- 1.5 }^{+ 1.6 }$    &  $ 16 _{- 4 }^{+ 4 }$    &  $ 32 _{- 9 }^{+ 9 }$    &  $ 1.2 _{- 0.6 }^{+ 0.8 }$    &  $ 13 _{- 7 }^{+ 9 }$    &  $ 2.2 _{- 1.2 }^{+ 1.5 }$    &  $ 223 _{- 17 }^{+ 14 }$    &  $ 44 _{- 3 }^{+ 2 }$    &  $ 0.2 _{- 0.1 }^{+ 0.1 }$    &  $ 0.8 _{- 0.4 }^{+ 0.6 }$    &  $ 0.3 _{- 0.2 }^{+ 0.2 }$    &  $ 12 _{- 4 }^{+ 4 }$    &  $ 2.2 _{- 0.6 }^{+ 0.7 }$    &  $ 0.1 _{- 0.1 }^{+ 0.1 }$    &  $ 0.2 _{- 0.1 }^{+ 0.2 }$   \\
$\Xi_c(3416)$  & $ \mathbf{\frac{7}{2}^+}$ & $\vert \,1\,,\,1\,,\,0\,,\,0 \,\rangle $ &$^{4}D_{\lambda\rho,7/2}$&$ 58 _{- 17 }^{+ 15 }$    &  $ 0.8 _{- 0.4 }^{+ 0.5 }$    &  $ 15 _{- 6 }^{+ 7 }$    &  $ 2.1 _{- 1.2 }^{+ 1.7 }$    &  $ 45 _{- 11 }^{+ 10 }$    &  $ 0.1 _{- 0.1 }^{+ 0.1 }$    &  $ 1.6 _{- 0.9 }^{+ 1.3 }$    &  $ 3.7 _{- 1.9 }^{+ 2.3 }$    &  $ 22 _{- 10 }^{+ 13 }$    &  $ 386 _{- 40 }^{+ 36 }$    &  0  &$ 0.2 _{- 0.1 }^{+ 0.2 }$    &  $ 0.5 _{- 0.2 }^{+ 0.3 }$    &  $ 1.4 _{- 0.8 }^{+ 1.0 }$    &  $ 20 _{- 6 }^{+ 6 }$    &  0  &$ 0.5 _{- 0.3 }^{+ 0.4 }$   \\
$\Xi_c(3275)$  & $ \mathbf{\frac{1}{2}^-}$ & $\vert \,1\,,\,1\,,\,0\,,\,0 \,\rangle $ &$^{2}P_{\lambda\rho,1/2}$&0  &$ 0.4 _{- 0.2 }^{+ 0.2 }$    &  0  &0  &0  &$ 258 _{- 23 }^{+ 24 }$    &  $ 0.9 _{- 0.6 }^{+ 1.0 }$    &  $ 34 _{- 4 }^{+ 4 }$    &  $ 0.5 _{- 0.3 }^{+ 0.5 }$    &  $ 0.1 _{- 0.0 }^{+ 0.1 }$    &  $ 12 _{- 4 }^{+ 4 }$    &  $ 0.1 _{- 0.1 }^{+ 0.1 }$    &  $ 1.3 _{- 0.4 }^{+ 0.4 }$    &  $ 0.1 _{- 0.0 }^{+ 0.1 }$    &  0  &0  &0 \\
$\Xi_c(3302)$  & $ \mathbf{\frac{3}{2}^-}$ & $\vert \,1\,,\,1\,,\,0\,,\,0 \,\rangle $ &$^{2}P_{\lambda\rho,3/2}$&0  &$ 0.4 _{- 0.2 }^{+ 0.2 }$    &  0  &$ 23 _{- 7 }^{+ 8 }$    &  $ 10 _{- 3 }^{+ 3 }$    &  $ 116 _{- 20 }^{+ 22 }$    &  $ 0.4 _{- 0.3 }^{+ 0.4 }$    &  $ 199 _{- 14 }^{+ 13 }$    &  $ 0.4 _{- 0.3 }^{+ 0.4 }$    &  $ 0.1 _{- 0.1 }^{+ 0.1 }$    &  $ 6 _{- 2 }^{+ 3 }$    &  $ 0.1 _{- 0.0 }^{+ 0.1 }$    &  $ 9 _{- 3 }^{+ 3 }$    &  $ 0.1 _{- 0.0 }^{+ 0.0 }$    &  0  &$ 0.1 _{- 0.1 }^{+ 0.1 }$    &  0 \\
$\Xi_c(3317)$  & $ \mathbf{\frac{1}{2}^-}$ & $\vert \,1\,,\,1\,,\,0\,,\,0 \,\rangle $ &$^{4}P_{\lambda\rho,1/2}$&0  &0  &$ 0.1 _{- 0.0 }^{+ 0.0 }$    &  $ 38 _{- 12 }^{+ 14 }$    &  $ 16 _{- 5 }^{+ 7 }$    &  $ 2.4 _{- 1.4 }^{+ 1.9 }$    &  $ 65 _{- 7 }^{+ 7 }$    &  $ 0.9 _{- 0.5 }^{+ 0.7 }$    &  $ 35 _{- 8 }^{+ 6 }$    &  $ 0.7 _{- 0.5 }^{+ 0.8 }$    &  $ 0.3 _{- 0.2 }^{+ 0.3 }$    &  $ 3.1 _{- 1.0 }^{+ 1.0 }$    &  $ 0.1 _{- 0.1 }^{+ 0.1 }$    &  $ 1.5 _{- 0.4 }^{+ 0.4 }$    &  0  &$ 0.2 _{- 0.1 }^{+ 0.2 }$    &  $ 0.1 _{- 0.0 }^{+ 0.1 }$   \\
$\Xi_c(3344)$  & $ \mathbf{\frac{3}{2}^-}$ & $\vert \,1\,,\,1\,,\,0\,,\,0 \,\rangle $ &$^{4}P_{\lambda\rho,3/2}$&0  &0  &$ 0.3 _{- 0.1 }^{+ 0.1 }$    &  $ 24 _{- 7 }^{+ 8 }$    &  $ 10 _{- 3 }^{+ 4 }$    &  $ 1.6 _{- 0.9 }^{+ 1.1 }$    &  $ 237 _{- 41 }^{+ 46 }$    &  $ 0.6 _{- 0.4 }^{+ 0.5 }$    &  $ 25 _{- 1 }^{+ 1 }$    &  $ 18 _{- 2 }^{+ 1 }$    &  $ 0.2 _{- 0.1 }^{+ 0.2 }$    &  $ 12 _{- 4 }^{+ 4 }$    &  $ 0.1 _{- 0.0 }^{+ 0.1 }$    &  $ 1.2 _{- 0.4 }^{+ 0.4 }$    &  $ 0.6 _{- 0.2 }^{+ 0.2 }$    &  $ 0.1 _{- 0.1 }^{+ 0.2 }$    &  0 \\
$\Xi_c(3389)$  & $ \mathbf{\frac{5}{2}^-}$ & $\vert \,1\,,\,1\,,\,0\,,\,0 \,\rangle $ &$^{4}P_{\lambda\rho,5/2}$&0  &0  &$ 0.4 _{- 0.2 }^{+ 0.2 }$    &  $ 11 _{- 3 }^{+ 3 }$    &  $ 54 _{- 17 }^{+ 17 }$    &  $ 0.9 _{- 0.5 }^{+ 0.6 }$    &  $ 3.9 _{- 1.9 }^{+ 2.6 }$    &  $ 3.9 _{- 2.1 }^{+ 2.8 }$    &  $ 204 _{- 37 }^{+ 41 }$    &  $ 135 _{- 9 }^{+ 7 }$    &  $ 0.1 _{- 0.1 }^{+ 0.1 }$    &  $ 0.1 _{- 0.0 }^{+ 0.1 }$    &  $ 0.5 _{- 0.3 }^{+ 0.3 }$    &  $ 10 _{- 4 }^{+ 4 }$    &  $ 6 _{- 2 }^{+ 2 }$    &  $ 0.1 _{- 0.1 }^{+ 0.1 }$    &  $ 0.4 _{- 0.2 }^{+ 0.4 }$   \\
$\Xi_c(3362)$  & $ \mathbf{\frac{3}{2}^+}$ & $\vert \,1\,,\,1\,,\,0\,,\,0 \,\rangle $ &$^{4}S_{\lambda\rho,3/2}$&0  &0  &0  &$ 8 _{- 1 }^{+ 1 }$    &  $ 16 _{- 2 }^{+ 2 }$    &  $ 0.8 _{- 0.4 }^{+ 0.4 }$    &  $ 137 _{- 30 }^{+ 32 }$    &  $ 1.4 _{- 0.7 }^{+ 0.8 }$    &  $ 122 _{- 20 }^{+ 19 }$    &  $ 21 _{- 1 }^{+ 1 }$    &  $ 0.1 _{- 0.0 }^{+ 0.1 }$    &  $ 7 _{- 3 }^{+ 3 }$    &  $ 0.2 _{- 0.1 }^{+ 0.1 }$    &  $ 6 _{- 2 }^{+ 2 }$    &  $ 1.0 _{- 0.3 }^{+ 0.3 }$    &  $ 0.1 _{- 0.1 }^{+ 0.1 }$    &  $ 0.2 _{- 0.1 }^{+ 0.2 }$   \\
$\Xi_c(3293)$  & $ \mathbf{\frac{1}{2}^+}$ & $\vert \,1\,,\,1\,,\,0\,,\,0 \,\rangle $ &$^{2}S_{\lambda\rho,1/2}$&0  &0  &0  &$ 7 _{- 1 }^{+ 1 }$    &  $ 13 _{- 3 }^{+ 2 }$    &  $ 219 _{- 34 }^{+ 38 }$    &  $ 0.2 _{- 0.1 }^{+ 0.1 }$    &  $ 44 _{- 2 }^{+ 3 }$    &  0  &$ 0.1 _{- 0.1 }^{+ 0.1 }$    &  $ 12 _{- 4 }^{+ 5 }$    &  0  &$ 1.8 _{- 0.5 }^{+ 0.6 }$    &  0  &0  &0  &$ 0.1 _{- 0.0 }^{+ 0.1 }$   \\
$\Xi_c(3413)$  & $ \mathbf{\frac{3}{2}^+}$ & $\vert \,0\,,\,2\,,\,0\,,\,0 \,\rangle $ &$^{2}D_{\rho\rho,3/2}$&$ 87 _{- 4 }^{+ 4 }$    &  $ 1.4 _{- 0.6 }^{+ 0.7 }$    &  $ 0.5 _{- 0.3 }^{+ 0.3 }$    &  $ 23 _{- 8 }^{+ 8 }$    &  $ 15 _{- 6 }^{+ 6 }$    &  $ 15 _{- 5 }^{+ 6 }$    &  $ 36 _{- 16 }^{+ 21 }$    &  $ 15 _{- 6 }^{+ 7 }$    &  $ 9 _{- 4 }^{+ 6 }$    &  $ 0.7 _{- 0.4 }^{+ 0.4 }$    &  0  &0  &0  &0  &0  &$ 29 _{- 9 }^{+ 9 }$    &  $ 4.5 _{- 1.6 }^{+ 1.6 }$   \\
$\Xi_c(3458)$  & $ \mathbf{\frac{5}{2}^+}$ & $\vert \,0\,,\,2\,,\,0\,,\,0 \,\rangle $ &$^{2}D_{\rho\rho,5/2}$&$ 84 _{- 5 }^{+ 5 }$    &  $ 1.6 _{- 0.7 }^{+ 0.8 }$    &  $ 0.6 _{- 0.3 }^{+ 0.3 }$    &  $ 10 _{- 3 }^{+ 3 }$    &  $ 34 _{- 11 }^{+ 12 }$    &  $ 36 _{- 12 }^{+ 14 }$    &  $ 4.6 _{- 2.1 }^{+ 2.5 }$    &  $ 28 _{- 9 }^{+ 10 }$    &  $ 19 _{- 8 }^{+ 10 }$    &  $ 12 _{- 6 }^{+ 7 }$    &  0  &0  &$ 0.1 _{- 0.0 }^{+ 0.1 }$    &  0  &0  &$ 0.1 _{- 0.1 }^{+ 0.1 }$    &  $ 33 _{- 10 }^{+ 10 }$   \\
\hline \hline
\end{tabular}

\endgroup
}
\end{center}
\label{cascades-_EM_2shell}
\end{table*}
\end{turnpage}


\subsection{Results}
This section presents our numerical results for the electromagnetic decay widths of the singly charmed baryons $\Lambda_c$ and $\Xi_c$, shown in Tables~\ref{lambdasEM}-\ref{cascades-_EM_2shell}. The baryon names and their predicted masses are taken from Ref.~\cite{Garcia-Tecocoatzi:2022zrf}. Additionally, we provide the total angular momentum and parity $\mathbf{J^{P}}$, the internal configuration of the baryon state given by $\left| l_{\lambda}, l_{\rho}, k_{\lambda}, k_{\rho} \right\rangle$, and the corresponding spectroscopic notation $^{2S+1}L_{x,J}$ for the initial states, where $L$ denotes the total orbital angular momentum, $S$ the total spin, the subscript $x$ the type of orbital excitation, and the subscript $J$ the total angular momentum. Here, $x$ can take the values $\lambda$, $\lambda\lambda$, $\rho$, $\rho\rho$, and $\lambda\rho$. The value $\lambda$ corresponds to a single $\lambda$-mode excitation, while $\rho$ corresponds to a single $\rho$-mode excitation. The $\lambda\lambda$ and $\rho\rho$ configurations represent double excitations in their respective modes. Finally, $\lambda\rho$ denotes a mixed excitation of  both the $\lambda$- and $\rho$-modes.

In Tables~\ref{lambdasEM}-\ref{cascades-_EM_2shell}, the predictions obtained in this work are displayed starting from the fifth column. Each column represents a particular decay channel, with the corresponding final-state baryon identified in the table header. The second row specifies the spectroscopic notation $^{2S+1}L_{x,J}$ associated with each final state as defined before. Moreover, our predictions are compared with previous theoretical results, including those obtained using LC-QCD~\cite{Luo:2025pzb} and constituent quark models (CQM)~\cite{Wang:2017kfr,Ortiz-Pacheco:2023kjn}. Results from Ref.~\cite{Cheng:1992xi,Banuls:1999br,Jiang:2015xqa,Wang:2018cre,Zhu:1998ih,Wang:2010xfj,Wang:2009cd,Aliev:2009jt,Aliev:2011bm,Aliev:2014bma,Aliev:2016xvq,Bernotas:2013eia,Shah:2016nxi,Gandhi:2019xfw,Gandhi:2019bju,Yang:2019tst,Kim:2021xpp,Chow:1995nw,Gamermann:2010ga,Zhu:2000py,Bijker:2020tns,Ivanov:1998wj,Ivanov:1999bk,Tawfiq:1999cf} are not included, since those studies considered only a limited number of decay channels.

In this paper, we present results for the $\Lambda_{c}$ charmed baryons with initial states belonging to the energy bands $N=0,1,2$ in the decay process. In contrast, for the $\Xi_{c}$ baryons, we present electromagnetic decay widths only for initial states in the $N=2$ energy band, since the lower-energy bands were previously discussed in Ref.~\cite{Garcia-Tecocoatzi:2025fxp}. For $N=0$, we compute $S$-wave to $S$-wave transitions, which are presented in Table~\ref{lambdasEM} for the $\Lambda_c$ baryons. For the energy band $N=1$, we evaluate $P$-wave to $S$-wave transitions, also shown in Table~\ref{lambdasEM}. Finally, for $N=2$, we consider the decays of the $S$-, \mbox{$P$-,} and $D$-wave states to the $S$-, and $P$-wave final states, which are presented in Tables~\ref{lambdas_EM_2shell}–\ref{cascades-_EM_2shell} for the $\Lambda_c$ and $\Xi_c$ baryons, respectively. Tables~\ref{lambdas_EM_2shell}--\ref{cascades-_EM_2shell} are organized according to the electric charge of the baryons.  Our predictions are compared with some of the results in Refs.~\cite{Yao:2018jmc, Peng:2024pyl}, however, a one-to-one comparison is not possible due to the fact that those works use different basis quantum states.

también decaen a sextet states definidos en Appendix A


\subsection{Discussion}

Electromagnetic decay widths are generally smaller than the dominant strong decay widths. Nevertheless, they can serve as powerful tools since they allow the estimation of observable branching ratios in collider experiments, particularly when resonances exhibit similar masses and strong decay widths.

In particular, we focus on the electromagnetic decay widths of the $\Xi_{c}(3055)^{+,0}$ and $\Xi_{c}(3080)^{+,0}$ states. In Ref.~\cite{Garcia-Tecocoatzi:2022zrf}, these resonances were interpreted as members of the flavor sextet. These assignments were proposed in a context where angular-distribution measurements were unavailable and several nearby resonances had been reported in the literature, making their identification especially challenging.
Fortunately, recent experimental measurements by the LHCb collaboration~\cite{LHCb:2024eyx} have established the quantum numbers of the charmed $\Xi_{c}(3055)^{+,0}$ states. Motivated by these results, we revisit their assignment and interpret them instead as members of the flavor antitriplet.
In the same spirit, we also discuss the $\Xi_{c}(3080)^{+,0}$ states, whose quantum numbers have not yet been experimentally determined, and likewise consider them as antitriplet states.
These states are displayed in the mass spectra shown in Fig.~\ref{fig:all-states}. In the following discussion, we provide information that may help establish their quantum number assignments based on the electromagnetic decay widths calculated in this work.

\subsubsection{Assignment of the $\Xi_c(3055)^{+,0}$ baryon}

Following the recent determination of the spin-parity of the $\Xi_c(3055)^{+,0}$ baryons by the LHCb Collaboration~\cite{LHCb:2024eyx}, we discuss the interpretation of this state based on the theoretical results of Ref.~\cite{Garcia-Tecocoatzi:2022zrf}. The $\Xi_c(3055)^{+,0}$ is identified as a $D_\lambda$-wave excitation with quantum numbers $\mathbf{J^{P}}=3/2^{+}$ and spin $S=1/2$, belonging to the flavor antitriplet, in agreement with the measurement of Ref.~\cite{LHCb:2024eyx}. 

In Tables~\ref{cascades-_EM_2shellp} and~\ref{cascades-_EM_2shell} the electromagnetic widths are calculated with theory masses of Ref.~\cite{Garcia-Tecocoatzi:2022zrf}. Although the theoretical mass for this state is slightly overestimated, we show that the framework employed in this paper still gives useful results by using the experimental mass to calculate the electromagnetic decay widths. That is, the experimental mass reported by the PDG is $3055.9$ MeV, while the theory mass in Ref.~\cite{Garcia-Tecocoatzi:2022zrf} $3118.9$ MeV.
We make a comparison of the electromagnetic decay widths calculated for both masses for the states $\Xi_c(3055)^{+,0}$

\begin{itemize}
\item $\Xi_c(3055)^{+}$

\begin{enumerate}
\item $\Gamma_{em}[\Xi_c(3118)^{+} \to \Xi_c^{+} \, \gamma ] = 49 _{- 8 }^{+ 9 } $ keV,

$\Gamma_{em}[\Xi_c(3055)^{+} \to \Xi_c^{+} \, \gamma ] = 43 _{- 8 }^{+ 8 } $ keV,

\item $\Gamma_{em}[\Xi_c(3118)^{+} \to \Xi_c'^{+}  \, \gamma ]  = 10 _{- 4 }^{+ 5 } $ keV,

$\Gamma_{em}[\Xi_c(3055)^{+} \to \Xi_c'^{+}  \, \gamma ]  = 5 _{- 2 }^{+ 2 } $ keV,

\item $\Gamma_{em}[\Xi_c(3118)^{+} \to \Xi_c^{*+}  \, \gamma  ] = 2.5 _{- 1.0 }^{+ 1.4 } $ keV,

$\Gamma_{em}[\Xi_c(3055)^{+} \to \Xi_c^{*+}  \, \gamma  ] = 1.1 _{- 0.4 }^{+ 0.4 } $ keV,

\item $\Gamma_{em}[\Xi_c(3118)^{+} \to \Xi_c^{+} \, ^2P_{\lambda,1/2} \, \gamma ] = 34 _{- 21 }^{+ 22 } $ keV,

$\Gamma_{em}[\Xi_c(3055)^{+} \to \Xi_c^{+} \, ^2P_{\lambda,1/2} \, \gamma ] = 34 _{- 21 }^{+ 21 } $ keV,

\item $\Gamma_{em}[\Xi_c(3118)^{+} \to \Xi_c^{+} \, ^4P_{\lambda,3/2} \, \gamma ] = 7 _{- 4 }^{+ 4 } $ keV,

$\Gamma_{em}[\Xi_c(3055)^{+} \to \Xi_c^{+} \, ^4P_{\lambda,3/2} \, \gamma ] = 6 _{- 3 }^{+ 3 } $ keV,

\item $\Gamma_{em}[\Xi_c(3118)^{+} \to \Xi_c^{'+} \, ^2P_{\lambda,1/2} \, \gamma ] = 2.0 _{- 0.9 }^{+ 1.4 } $ keV,

$\Gamma_{em}[\Xi_c(3055)^{+} \to \Xi_c^{'+} \, ^2P_{\lambda,1/2} \, \gamma ] = 0.4 _{- 0.1 }^{+ 0.1 } $ keV,

\item $\Gamma_{em}[\Xi_c(3118)^{+} \to \Xi_c^{'+} \, ^4P_{\lambda,1/2} \, \gamma ] = 2.5 _{- 1.7 }^{+ 3.1 } $ keV,

$\Gamma_{em}[\Xi_c(3055)^{+} \to \Xi_c^{'+} \, ^4P_{\lambda,1/2} \, \gamma ] = 0.5 _{- 0.2 }^{+ 0.2 } $ keV,

\item $\Gamma_{em}[\Xi_c(3118)^{+} \to \Xi_c^{'+} \, ^2P_{\lambda,3/2} \, \gamma ] = 1.5 _{- 0.8 }^{+ 1.2 } $ keV,

$\Gamma_{em}[\Xi_c(3055)^{+} \to \Xi_c^{'+} \, ^2P_{\lambda,3/2} \, \gamma ] = 0.12 _{- 0.03 }^{+ 0.03 } $ keV,

\item $\Gamma_{em}[\Xi_c(3118)^{+} \to \Xi_c^{'+} \, ^4P_{\lambda,3/2} \, \gamma ] = 0.3 _{- 0.2 }^{+ 0.4 } $ keV,

$\Gamma_{em}[\Xi_c(3055)^{+} \to \Xi_c^{'+} \, ^4P_{\lambda,3/2} \, \gamma ] = 0.02 _{- 0.01 }^{+ 0.01 } $ keV.

\end{enumerate}

\item $\Xi_c(3055)^{0}$
\begin{enumerate}
\item $\Gamma_{em}[\Xi_c(3118)^{0} \to \Xi_c^{0}  \, \gamma ] = 22 _{- 8 }^{+ 8 } $ keV,

$\Gamma_{em}[\Xi_c(3055)^{0} \to \Xi_c^{0}  \, \gamma ] = 20 _{- 8 }^{+ 8 } $ keV,

\item $\Gamma_{em}[\Xi_c(3118)^{0} \to \Xi_c'^{0}  \, \gamma ]  = 0.2 _{- 0.1 }^{+ 0.2 } $ keV,

$\Gamma_{em}[\Xi_c(3055)^{0} \to \Xi_c'^{0}  \, \gamma ]  = 0.1 _{- 0.1 }^{+ 0.1 } $ keV,

\item $\Gamma_{em}[\Xi_c(3118)^{0} \to \Xi_c^{*0}  \, \gamma ] = 0.1 _{- 0.0 }^{+ 0.0 } $ keV ,

$\Gamma_{em}[\Xi_c(3055)^{0} \to \Xi_c^{*0}  \, \gamma ] = 0.02 _{- 0.0 }^{+ 0.0 } $ keV ,

\item $\Gamma_{em}[\Xi_c(3118)^{0} \to \Xi_c^{0} \, ^2P_{\lambda,1/2} \, \gamma ] = 543 _{- 32 }^{+ 31 } $ keV,

$\Gamma_{em}[\Xi_c(3055)^{0} \to \Xi_c^{0} \, ^2P_{\lambda,1/2} \, \gamma ] = 495_{- 30 }^{+ 30 } $ keV,

\item $\Gamma_{em}[\Xi_c(3118)^{0} \to \Xi_c^{0} \, ^2P_{\lambda,3/2} \, \gamma ] = 74 _{- 5 }^{+ 5 } $ keV,

$\Gamma_{em}[\Xi_c(3055)^{0} \to \Xi_c^{0} \, ^2P_{\lambda,3/2} \, \gamma ] = 65 _{- 5 }^{+ 5 } $ keV,

\item $\Gamma_{em}[\Xi_c(3118)^{0} \to \Xi_c^{'0} \, ^4P_{\lambda,1/2} \, \gamma ] = 0.1 _{- 0.0 }^{+ 0.1 } $ keV,

$\Gamma_{em}[\Xi_c(3055)^{0} \to \Xi_c^{'0} \, ^4P_{\lambda,1/2} \, \gamma ] = 0.01 _{- 0.0 }^{+ 0.0 } $ keV.
\end{enumerate}
\end{itemize}

With these decay widths we can get several branching ratios. If we analyze, for example the following cases
\begin{align}
\frac{\Gamma_{em} [\Xi_c (3055)^{+}   \to \Xi_c^{+} \gamma]}
{\Gamma_{em} [\Xi_c (3055) ^{+}  \to \Xi_c^{'+} \gamma]}
&= 8.6^{+1.2}_{-1.2} \, ,
\label{eq:ratio1} \\[2mm]
\frac{\Gamma_{em} [\Xi_c (3055)^{+}   \to \Xi_c^{+} \gamma]}
{\Gamma_{em} [\Xi_c (3055)^{+}   \to \Xi_c^{*+} \gamma]}
&= 39^{+5}_{-5} \, ,
\label{eq:ratio2} \\[2mm]
\frac{\Gamma_{em} [\Xi_c (3055)^{+}   \to \Xi_c^{'+} \gamma]}
{\Gamma_{em} [\Xi_c (3055)^{+}   \to \Xi_c^{*+} \gamma]}
&= 4.5^{+1}_{-1} \, .
\label{eq:ratio3}
\end{align}

\begin{align}
\frac{\Gamma_{em} [\Xi_c (3055)^{0}   \to \Xi_c^{0} \gamma]}
{\Gamma_{em} [\Xi_c (3055)^{0}   \to \Xi_c^{'0} \gamma]}
&= 200^{+25}_{-25} \, ,
\label{eq:ratio4} \\[2mm]
\frac{\Gamma_{em} [\Xi_c (3055)^{0}   \to \Xi_c^{0} \gamma]}
{\Gamma_{em} [\Xi_c (3055) ^{0}  \to \Xi_c^{*0} \gamma]}
&= 1000^{+50}_{-50} \, ,
\label{eq:ratio5} \\[2mm]
\frac{\Gamma_{em} [\Xi_c (3055)^{0}   \to \Xi_c^{'0} \gamma]}
{\Gamma_{em} [\Xi_c (3055)^{0}   \to \Xi_c^{*0} \gamma]}
&= 5^{+0.5}_{-0.5} \, .
\label{eq:ratio6}
\end{align}

In particular, the electromagnetic decay branching ratios are highly sensitive to the underlying quantum numbers, making them valuable observables for discriminating between competing assignments. Since the phase-space dependence cancels out in the ratios, the resulting observables are governed by the underlying spectroscopic amplitudes. Consequently, the branching ratios given in Eqs. (\ref{eq:ratio1}-\ref{eq:ratio6}) can provide an independent probe of the internal structure and quantum-number assignments of the $\Xi_c(3055)^{+,0}$ states, complementary to the angular-distribution analyses performed by the LHCb collaboration in Ref.~\cite{LHCb:2024eyx}. 

\subsubsection{Assignment of the $\Xi_c(3080)^{+,0}$ baryon }

The identification of the $\Xi_c(3055)^{+,0}$ baryons by the LHCb Collaboration as $1D$ states with $\mathbf{J^{P}}=3/2^{+}$~\cite{LHCb:2024eyx} provides valuable insight into the interpretation of the nearby $\Xi_c(3080)^{+,0}$ states. In particular, the theoretical study of Ref.~\cite{Garcia-Tecocoatzi:2022zrf} predicts a mass splitting of approximately 30 MeV between the $1D$ $\mathbf{J^{P}}=3/2^{+}$ and $\mathbf{J^{P}}=5/2^{+}$ states, which naturally supports assigning the $\Xi_c(3080)^{+,0}$ baryons to the corresponding $1D$ $\mathbf{J^{P}}=5/2^{+}$ doublet. According to the PDG~\cite{ParticleDataGroup:2024cfk}, the measured mass is $3077.2$ MeV for the positively charged state, and $3079.9$ MeV for the neutral state. Nevertheless, their quantum numbers have not yet been experimentally established. The theoretical mass predicted in Ref.~\cite{Garcia-Tecocoatzi:2022zrf} differs from the experimental value by about $2.5\%$. Motivated by these considerations, in the following we compare the numerical predictions for the electromagnetic decay widths obtained using both the theoretical and experimental masses, assuming that the $\Xi_c(3080)^{+,0}$ states belong to the $1D$ $\mathbf{J^{P}}=5/2^{+}$ configuration.

\begin{itemize}
\item $\Xi_c(3080)^{+}$ as $\mathbf{J^{P}}=5/2^{+}$

\begin{enumerate}
\item $\Gamma_{em}[\Xi_c(3163)^{+} \to \Xi_c^{+} \, \gamma ] = 61 _{- 7 }^{+ 8 } $ keV,

$\Gamma_{em}[\Xi_c(3080)^{+} \to \Xi_c^{+} \, \gamma ] = 51 _{- 8 }^{+ 8 } $ keV,

\item $\Gamma_{em}[\Xi_c(3163)^{+} \to \Xi_c'^{+}  \, \gamma ]  = 14 _{- 6 }^{+ 7 } $ keV,

$\Gamma_{em}[\Xi_c(3080)^{+} \to \Xi_c'^{+}  \, \gamma ]  = 6 _{- 2 }^{+ 2 } $ keV,

\item $\Gamma_{em}[\Xi_c(3163)^{+} \to \Xi_c^{*+}  \, \gamma  ] = 4.0 _{- 1.7 }^{+ 2.0 } $ keV,

$\Gamma_{em}[\Xi_c(3080)^{+} \to \Xi_c^{*+}  \, \gamma  ] = 1.4 _{- 1.0 }^{+ 0.8 } $ keV,

\item $\Gamma_{em}[\Xi_c(3163)^{+} \to \Xi_c^{+} \, ^2P_{\lambda,1/2} \, \gamma ] = 0.2 _{- 0.1 }^{+ 0.1 } $ keV,

$\Gamma_{em}[\Xi_c(3080)^{+} \to \Xi_c^{+} \, ^2P_{\lambda,1/2} \, \gamma ] =  0.05_{- 0.02 }^{+ 0.02 } $ keV,

\item $\Gamma_{em}[\Xi_c(3163)^{+} \to \Xi_c^{+} \, ^2P_{\lambda,3/2} \, \gamma ] = 31 _{- 22 }^{+ 24 } $ keV,

$\Gamma_{em}[\Xi_c(3080)^{+} \to \Xi_c^{+} \, ^2P_{\lambda,3/2} \, \gamma ] = 36 _{- 8 }^{+ 8 } $ keV,

\item $\Gamma_{em}[\Xi_c(3163)^{+} \to \Xi_c^{'+} \, ^2P_{\lambda,1/2} \, \gamma ] = 8 _{- 3 }^{+ 4 } $ keV,

$\Gamma_{em}[\Xi_c(3080)^{+} \to \Xi_c^{'+} \, ^2P_{\lambda,1/2} \, \gamma ] = 1.5 _{- 0.9 }^{+ 1.0 } $ keV,

\item $\Gamma_{em}[\Xi_c(3163)^{+} \to \Xi_c^{'+} \, ^4P_{\lambda,1/2} \, \gamma ] = 0.5 _{- 0.3 }^{+ 0.5 } $ keV,

$\Gamma_{em}[\Xi_c(3080)^{+} \to \Xi_c^{'+} \, ^4P_{\lambda,1/2} \, \gamma ] = 0.1 _{- 0.1 }^{+ 0.1 } $ keV,

\item $\Gamma_{em}[\Xi_c(3163)^{+} \to \Xi_c^{'+} \, ^2P_{\lambda,3/2} \, \gamma ] = 4.7 _{- 2.1 }^{+ 2.9 } $ keV,

$\Gamma_{em}[\Xi_c(3080)^{+} \to \Xi_c^{'+} \, ^2P_{\lambda,3/2} \, \gamma ] = 0.3 _{- 0.1 }^{+ 0.1 } $ keV,

\item $\Gamma_{em}[\Xi_c(3163)^{+} \to \Xi_c^{'+} \, ^4P_{\lambda,3/2} \, \gamma ] = 1.7 _{- 0.8 }^{+ 1.2 } $ keV,

$\Gamma_{em}[\Xi_c(3080)^{+} \to \Xi_c^{'+} \, ^4P_{\lambda,3/2} \, \gamma ] = 0.3 _{- 0.1 }^{+ 0.1 } $ keV,

\item $\Gamma_{em}[\Xi_c(3163)^{+} \to \Xi_c^{'+} \, ^4P_{\lambda,5/2} \, \gamma ] = 0.6 _{- 0.4 }^{+ 0.6 } $ keV,

$\Gamma_{em}[\Xi_c(3080)^{+} \to \Xi_c^{'+} \, ^4P_{\lambda,5/2} \, \gamma ] = 0.1 _{- 0.1 }^{+ 0.1 } $ keV.

\end{enumerate}

\item $\Xi_c(3080)^{0}$ as $\mathbf{J^{P}}=5/2^{+}$

\begin{enumerate}
\item $\Gamma_{em}[\Xi_c(3163)^{0} \to \Xi_c^{0}  \, \gamma ] = 18 _{- 8 }^{+ 7 } $ keV,

$\Gamma_{em}[\Xi_c(3080)^{0} \to \Xi_c^{0}  \, \gamma ] = 17 _{- 8 }^{+ 8 } $ keV,

\item $\Gamma_{em}[\Xi_c(3163)^{0} \to \Xi_c'^{0}  \, \gamma ]  = 0.3 _{- 0.2 }^{+ 0.2 } $ keV,

$\Gamma_{em}[\Xi_c(3080)^{0} \to \Xi_c'^{0}  \, \gamma ]  = 0.13 _{- 0.1 }^{+ 0.1 } $ keV,

\item $\Gamma_{em}[\Xi_c(3163)^{0} \to \Xi_c^{*0}  \, \gamma ] = 0.1 _{- 0.7 }^{+ 0.7 } $ keV ,

$\Gamma_{em}[\Xi_c(3080)^{0} \to \Xi_c^{*0}  \, \gamma ] = 0.05 _{- 0.0 }^{+ 0.0 } $ keV ,

\item $\Gamma_{em}[\Xi_c(3163)^{0} \to \Xi_c^{0} \, ^2P_{\lambda,1/2} \, \gamma ] = 2.3 _{- 0.7 }^{+ 0.7 } $ keV,

$\Gamma_{em}[\Xi_c(3080)^{0} \to \Xi_c^{0} \, ^2P_{\lambda,1/2} \, \gamma ] = 0.7 _{- 0.2 }^{+ 0.2 } $ keV,

\item $\Gamma_{em}[\Xi_c(3163)^{0} \to \Xi_c^{0} \, ^2P_{\lambda,3/2} \, \gamma ] = 673 _{- 42 }^{+ 41 } $ keV,

$\Gamma_{em}[\Xi_c(3080)^{0} \to \Xi_c^{0} \, ^2P_{\lambda,3/2} \, \gamma ] = 586 _{- 35 }^{+ 35 } $ keV, 

\item $\Gamma_{em}[\Xi_c(3163)^{0} \to \Xi_c^{'0} \, ^2P_{\lambda,1/2} \, \gamma ] = 0.2 _{- 0.1 }^{+ 0.1 } $ keV,

$\Gamma_{em}[\Xi_c(3080)^{0} \to \Xi_c^{'0} \, ^2P_{\lambda,1/2} \, \gamma ] = 0.03 _{- 0.01 }^{+ 0.01 } $ keV.
\end{enumerate}
\end{itemize}
From our previous results, we observe that the theoretical predictions for the electromagnetic decay widths are essentially independent of the mass once the uncertainties in the calculation, estimated through the bootstrap Monte Carlo method, are taken into account. Considering the electromagnetic decays calculated with the experimental mass, we compute some relevant branching ratios assuming the assignment \mbox{$\mathbf{J^{P}}=5/2^{+}$} for the $\Xi_c(3080)^{+,0}$:

\begin{align}
\frac{\Gamma_{em} [\Xi_c (3080)^{+}   \to \Xi_c^{+} \gamma]}
{\Gamma_{em} [\Xi_c (3080)^{+}   \to \Xi_c^{'+} \gamma]}
&= 8.5^{+1.5}_{-1.5} \, ,
\label{eq:ratio3080_1} \\[2mm]
\frac{\Gamma_{em} [\Xi_c (3080)^{+}   \to \Xi_c^{+} \gamma]}
{\Gamma_{em} [\Xi_c (3080)^{+}   \to \Xi_c^{*+} \gamma]}
&= 36^{+4}_{-4} \, ,
\label{eq:ratio3080_2}
\end{align}

\begin{align}
\frac{\Gamma_{em} [\Xi_c (3080)^{0}   \to \Xi_c^{0} \gamma]}
{\Gamma_{em} [\Xi_c (3080)^{0}   \to \Xi_c^{'0} \gamma]}
&= 130^{+15}_{-15} \, ,
\label{eq:ratio3080_3} \\[2mm]
\frac{\Gamma_{em} [\Xi_c (3080)^{0}   \to \Xi_c^{0} \gamma]}
{\Gamma_{em} [\Xi_c (3080)^{0}   \to \Xi_c^{*0} \gamma]}
&= 340^{+40}_{-40} \, .
\label{eq:ratio3080_4}
\end{align}

An alternative interpretation of the $\Xi_c(3080)^{+,0}$ state is the assignment $\mathbf{J^{P}}=1/2^{+}$. In Ref.~\cite{Chen:2017aqm}, Chen \textit{et al.} argued that the $\mathbf{J^{P}}=5/2^{+}$ assignment is not strongly supported by their analysis.  Therefore, we also present the  electromagnetic decay widths for the $\Xi_c(3080)^{+,0}$ states assuming a $\mathbf{J^{P}}=1/2^{+}$ assignment: 
\begin{itemize}
\item $\Xi_c(3080)^{+}$ as $\mathbf{J^{P}}=1/2^{+}$

\begin{enumerate}
\item 

$\Gamma_{em}[\Xi_c(3080)^{+} \to \Xi_c^{+} \, \gamma ] = 0.2 _{- 0.1 }^{+ 0.1 } $ keV,

\item 

$\Gamma_{em}[\Xi_c(3080)^{+} \to \Xi_c'^{+}  \, \gamma ]  = 15 _{- 5 }^{+ 5 } $ keV,

\item 

$\Gamma_{em}[\Xi_c(3080)^{+} \to \Xi_c^{*+}  \, \gamma  ] = 3.5 _{- 1 }^{+ 1 } $ keV,

\item 

$\Gamma_{em}[\Xi_c(3080)^{+} \to \Xi_c^{+} \, ^2P_{\lambda,1/2} \, \gamma ] =  10.7_{- 3 }^{+ 3 } $ keV,

\item 

$\Gamma_{em}[\Xi_c(3080)^{+} \to \Xi_c^{+} \, ^2P_{\lambda,3/2} \, \gamma ] = 36 _{- 4 }^{+ 4 } $ keV,

\item 

$\Gamma_{em}[\Xi_c(3080)^{+} \to \Xi_c^{'+} \, ^2P_{\lambda,1/2} \, \gamma ] = 6.6 _{- 2 }^{+ 2 } $ keV,

\item 

$\Gamma_{em}[\Xi_c(3080)^{+} \to \Xi_c^{'+} \, ^4P_{\lambda,1/2} \, \gamma ] = 0.02 _{- 0.01 }^{+ 0.01 } $ keV,

\item 

$\Gamma_{em}[\Xi_c(3080)^{+} \to \Xi_c^{'+} \, ^2P_{\lambda,3/2} \, \gamma ] = 6 _{- 1 }^{+ 1 } $ keV,

\item 

$\Gamma_{em}[\Xi_c(3080)^{+} \to \Xi_c^{'+} \, ^4P_{\lambda,3/2} \, \gamma ] = 0.02 _{- 0.01 }^{+ 0.01 } $ keV,

\item 

$\Gamma_{em}[\Xi_c(3080)^{+} \to \Xi_c^{'+} \, ^4P_{\lambda,5/2} \, \gamma ] = 0.1 _{- 0.1 }^{+ 0.1 } $ keV.

\end{enumerate}

\item $\Xi_c(3080)^{0}$ as $\mathbf{J^{P}}=1/2^{+}$

\begin{enumerate}
\item 

$\Gamma_{em}[\Xi_c(3080)^{0} \to \Xi_c^{0}  \, \gamma ] = 0.2 _{- 0.1 }^{+ 0.1 } $ keV,

\item 

$\Gamma_{em}[\Xi_c(3080)^{0} \to \Xi_c'^{0}  \, \gamma ]  = 0.3 _{- 0.1 }^{+ 0.1 } $ keV,

\item 

$\Gamma_{em}[\Xi_c(3080)^{0} \to \Xi_c^{*0}  \, \gamma ] = 0.07 _{- 0.02 }^{+ 0.02 } $ keV ,

\item 

$\Gamma_{em}[\Xi_c(3080)^{0} \to \Xi_c^{0} \, ^2P_{\lambda,1/2} \, \gamma ] = 266 _{- 30 }^{+ 30 } $ keV,

\item 

$\Gamma_{em}[\Xi_c(3080)^{0} \to \Xi_c^{0} \, ^2P_{\lambda,3/2} \, \gamma ] = 91 _{- 28 }^{+ 28 } $ keV, 

\item 

$\Gamma_{em}[\Xi_c(3080)^{0} \to \Xi_c^{'0} \, ^2P_{\lambda,1/2} \, \gamma ] = 0.13 _{- 0.04 }^{+ 0.04 } $ keV.
\end{enumerate}
\end{itemize}

To help discriminate between these possible configurations, it is useful to calculate the corresponding branching-ratio predictions for the $\Xi_c(3080)^{+,0}$ states under the $\mathbf{J^{P}}=1/2^{+}$ assignment and compare them with those obtained for the $\mathbf{J^{P}}=5/2^{+}$ scenario. Such comparisons may provide valuable guidance for future experimental studies aimed at determining the quantum numbers of  $\Xi_c(3080)^{+,0}$ states.

\begin{align}
\frac{\Gamma_{em} [\Xi_c (3080)^{+}   \to \Xi_c^{+} \gamma]}
{\Gamma_{em} [\Xi_c (3080)^{+}   \to \Xi_c^{'+} \gamma]}
&= 0.013^{+0.003}_{-0.003} \, ,
\label{eq:ratio3080_alt1} \\[2mm]
\frac{\Gamma_{em} [\Xi_c (3080)^{+}   \to \Xi_c^{+} \gamma]}
{\Gamma_{em} [\Xi_c (3080)^{+}   \to \Xi_c^{*+} \gamma]}
&= 0.06^{+0.01}_{-0.01} \, .
\label{eq:ratio3080_alt2}
\end{align}

\begin{align}
\frac{\Gamma_{em} [\Xi_c (3080)^{0}   \to \Xi_c^{0} \gamma]}
{\Gamma_{em} [\Xi_c (3080)^{0}   \to \Xi_c^{'0} \gamma]}
&= 0.7^{+0.2}_{-0.2} \, ,
\label{eq:ratio3080_alt3} \\[2mm]
\frac{\Gamma_{em} [\Xi_c (3080)^{0}   \to \Xi_c^{0} \gamma]}
{\Gamma_{em} [\Xi_c (3080)^{0}   \to \Xi_c^{*0} \gamma]}
&= 3^{+0.5}_{-0.5} \, .
\label{eq:ratio3080_alt4}
\end{align}

We observe that the electromagnetic branching ratios for the decays into $\Xi_c^{+,0}\gamma$, $\Xi_c^{\prime +,0}\gamma$, and $\Xi_c^{*+,0}\gamma$ differ substantially between the assignment $\mathbf{J^{P}}=5/2^{+}$ (see Eqs.~\eqref{eq:ratio3080_1}--\eqref{eq:ratio3080_2}) and the assignment $\mathbf{J^{P}}=1/2^{+}$ (see Eqs.~\eqref{eq:ratio3080_alt1}--\eqref{eq:ratio3080_alt4}). Hence, these branching ratios constitute particularly useful observables for discriminating between possible quantum-number assignments of the $\Xi_c(3080)^{+,0}$ states.

\subsubsection{Mixing effects}
Configuration mixing is not currently introduced in our framework through the mass Hamiltonian, in light of the predicted spectrum accurately reproducing the available experimental masses. However, it should be noted that electromagnetic decay transitions between states sharing the same total angular momentum $J$, but differing in their total light–quark spin content, can potentially be affected by mixing effects - via mixing angles - as their transitions are expected to vary significantly, making the associated observables mixing sensitive. Nonetheless,  to reliably determine these mixing angles, precise experimental measurements are needed. Some experimental data are available for the $\Xi_c$ states, e.g., those discussed in Ref.~\cite{Garcia-Tecocoatzi:2025fxp}. Nevertheless, the data and analyses of electromagnetic decay widths referred to in Ref.~\cite{Garcia-Tecocoatzi:2025fxp} solely correspond to the $P$-wave sector; up to now, there is no data available for the $D$-wave sector, which is the main focus of the present work for the $\Xi_c$ states. In addition, the currently available experimental measurements are not sufficiently accurate to establish the presence of mixing effects, since predictions in Ref.~\cite{Garcia-Tecocoatzi:2025fxp} remain within the experimental bounds. Consequently, more accurate experimental data are required to establish the presence or absence of mixing effects and, correspondingly, it is not currently possible to reliably determine mixing angles. In this sense, electromagnetic decay widths hold promise as valuable probes of mixing in the near future, when more precise experimental data become available, motivating the formalism herein to be naturally extended to account for such effects.
\section{Conclusions}
\label{Conclusions}
We calculate the electromagnetic decay widths for the $\Lambda_c$ and $\Xi_c$ charmed baryons belonging to the flavor \mbox{${\mathbf{\bar{3}}}_{\rm F}$-plet.} We consider transitions from ground and $P$-wave states to ground states for the $\Lambda_c$ baryons, since transitions involving the $\Xi_c$ baryons at this level were already addressed in our previous work~\cite{Garcia-Tecocoatzi:2025fxp}. We also analyze transitions from all second-shell $\Lambda_c$ and $\Xi_c$ states to both ground and $P$-wave final states. We calculate for the first time the electromagnetic decays of $D_\rho$-wave states, $\rho \lambda$ mixed states, and radially excited $\rho$-mode states for singly charmed baryons belonging to the flavor antitriplet. The EMDs of $P$-wave states have been investigated in Refs.~\cite{Chow:1995nw,Gamermann:2010ga,Zhu:2000py,Luo:2025pzb,Bijker:2020tns,Ortiz-Pacheco:2023kjn,Wang:2017kfr,Peng:2024pyl,Ivanov:1998wj,Ivanov:1999bk,Tawfiq:1999cf}. The EMDs of second shell singly charmed baryons were studied in Refs.~\cite{Yao:2018jmc,Peng:2024pyl}; however, in both cases only a subset of second shell states was considered.



We discuss the importance of the electromagnetic decay widths of charmed baryons for determining their quantum configurations when different states exhibit similar masses and strong decay widths. We interpret the $\Xi_c(3055)^{+,-}$ baryons as $D_\lambda$-wave excitations with $\mathbf{J^P}=3/2^{+}$ and $S=1/2$, belonging to the flavor antitriplet, being consistent with the recent LHCb determination~\cite{LHCb:2024eyx}. Our results indicate that the electromagnetic branching ratios of the $\Xi_c(3080)^{+,0}$ states are highly sensitive to their quantum-number assignments, making them powerful observables for distinguishing between the $\mathbf{J^{P}}=5/2^{+}$ and $\mathbf{J^{P}}=1/2^{+}$ scenarios. In this sense, they can provide additional information for future angular-distribution-based experimental analyses aimed at determining the quantum numbers of the $\Xi_c(3080)^{+,0}$ states.

Finally, for the present work, we have accounted for the propagation of parameter uncertainties using a Monte Carlo bootstrap method. The inclusion of uncertainties, often missed in similar and related research studies, is essential for a statistical significant comparison between theory and experiment to take place, driving future discussion.


\section*{Acknowledgements}
We acknowledge Alejandra D\'avila-Rivera for useful discussions. C.A.V.-A. is supported by the Secretar\'ia de Ciencia, Humanidades, Tecnolog\'ia e Innovaci\'on (Secihti) Investigadoras e Investigadores por M\'exico project 749 and SNII 58928. A.G.-R. is supported by Universidad Aut\'onoma de Zacatecas through the project CBF-2026-310.

\clearpage


\begin{thebibliography}{109}
\providecommand{\natexlab}[1]{#1}
\providecommand{\url}[1]{\texttt{#1}}
\expandafter\ifx\csname urlstyle\endcsname\relax
  \providecommand{\doi}[1]{doi: #1}\else
  \providecommand{\doi}{doi: \begingroup \urlstyle{rm}\Url}\fi

\bibitem[Ablikim et~al.(2021)]{BESIII:2020kap}
Medina Ablikim et~al.
\newblock {Determination of the $\Lambda_c^+$ spin via the reaction
  $e^+e^-\to\Lambda_c^+\bar\Lambda_c^-$}.
\newblock \emph{Phys. Rev. D}, 103\penalty0 (9):\penalty0 L091101, 2021.
\newblock \doi{10.1103/PhysRevD.103.L091101}.

\bibitem[Moon et~al.(2021)]{Belle:2020tom}
T.~J. Moon et~al.
\newblock {First determination of the spin and parity of the charmed-strange
  baryon $\Xi_{c}(2970)^+$}.
\newblock \emph{Phys. Rev. D}, 103\penalty0 (11):\penalty0 L111101, 2021.
\newblock \doi{10.1103/PhysRevD.103.L111101}.

\bibitem[Aaij et~al.(2025{\natexlab{a}})]{LHCb:2024eyx}
Roel Aaij et~al.
\newblock {First Determination of the Spin-Parity of $\Xi_c(3055)^{+,0}$
  Baryons}.
\newblock \emph{Phys. Rev. Lett.}, 134\penalty0 (8):\penalty0 081901,
  2025{\natexlab{a}}.
\newblock \doi{10.1103/PhysRevLett.134.081901}.

\bibitem[Aaij et~al.(2021)]{LHCb:2021ptx}
Roel Aaij et~al.
\newblock {Observation of excited $\Omega_c^0$ baryons in $\Omega_b^- \to
  \Xi_c^+ K^-\pi^-$decays}.
\newblock \emph{Phys. Rev. D}, 104\penalty0 (9):\penalty0 L091102, 2021.
\newblock \doi{10.1103/PhysRevD.104.L091102}.

\bibitem[Aubert et~al.(2008)]{BaBar:2008get}
Bernard Aubert et~al.
\newblock {Measurements of $\mathcal{B}(\bar{B}^0 \to \Lambda_c^+ \bar{p})$ and
  $\mathcal{B}(\bar{B}^- \to \Lambda_c^+ \bar{p} \pi^-)$ and studies of
  $\Lambda_c^+ \pi^-$ resonances}.
\newblock \emph{Phys. Rev. D}, 78:\penalty0 112003, 2008.
\newblock \doi{10.1103/PhysRevD.78.112003}.

\bibitem[D{\'a}vila-Rivera et~al.(2026)D{\'a}vila-Rivera,
  Garc{\'\i}a-Tecocoatzi, Ramirez-Morales, Rivero-Acosta, Santopinto, and
  Vaquera-Araujo]{Davila-Rivera:2025exk}
A.~D{\'a}vila-Rivera, H.~Garc{\'\i}a-Tecocoatzi, A.~Ramirez-Morales, Ailier
  Rivero-Acosta, E.~Santopinto, and Carlos~Alberto Vaquera-Araujo.
\newblock {Radiative decays of the $\Sigma_c$, $\Xi'_c$, and $\Omega_c$ charmed
  baryons}.
\newblock \emph{Phys. Rev. D}, 113\penalty0 (9):\penalty0 093001, 2026.
\newblock \doi{10.1103/srjs-mrs3}.

\bibitem[Klempt and Richard(2010)]{klempt2010baryon}
Eberhard Klempt and Jean-Marc Richard.
\newblock Baryon spectroscopy.
\newblock \emph{Reviews of Modern Physics}, 82\penalty0 (2):\penalty0
  1095--1153, 2010.

\bibitem[Cheng(2022)]{cheng2022charmed}
Hai-Yang Cheng.
\newblock Charmed baryon physics circa 2021.
\newblock \emph{Chinese Journal of Physics}, 78:\penalty0 324--362, 2022.

\bibitem[Chen et~al.(2017{\natexlab{a}})Chen, Chen, Liu, Liu, and
  Zhu]{chen2017review}
Hua-Xing Chen, Wei Chen, Xiang Liu, Yan-Rui Liu, and Shi-Lin Zhu.
\newblock A review of the open charm and open bottom systems.
\newblock \emph{Reports on Progress in Physics}, 80\penalty0 (7):\penalty0
  076201, 2017{\natexlab{a}}.

\bibitem[Cazzoli et~al.(1975)Cazzoli, Cnops, Connolly, Louttit, Murtagh,
  Palmer, Samios, Tso, and Williams]{Cazzoli:1975et}
E.~G. Cazzoli, A.~M. Cnops, P.~L. Connolly, R.~I. Louttit, M.~J. Murtagh, R.~B.
  Palmer, N.~P. Samios, T.~T. Tso, and H.~H. Williams.
\newblock {Evidence for $\Delta S = - \Delta Q$ Currents or Charmed-Baryon
  Production by Neutrinos}.
\newblock \emph{Phys. Rev. Lett.}, 34:\penalty0 1125--1128, 1975.
\newblock \doi{10.1103/PhysRevLett.34.1125}.

\bibitem[Knapp et~al.(1976)]{Knapp:1976qw}
B.~Knapp et~al.
\newblock {Observation of a Narrow Antibaryon State at 2.26 GeV/c$^2$}.
\newblock \emph{Phys. Rev. Lett.}, 37:\penalty0 882, 1976.
\newblock \doi{10.1103/PhysRevLett.37.882}.

\bibitem[Calicchio et~al.(1980)]{BEBCTSTNeutrino:1980ktj}
M.~Calicchio et~al.
\newblock {First Observation of the Production and Decay of the
  $\Sigma_{c}^{+}$}.
\newblock \emph{Phys. Lett. B}, 93:\penalty0 521--524, 1980.
\newblock \doi{10.1016/0370-2693(80)90379-2}.

\bibitem[Biagi et~al.(1983)]{Biagi:1983en}
S.~F. Biagi et~al.
\newblock {Observation of a Narrow State at 2.46 GeV/$c^2$: A Candidate for the
  Charmed Strange Baryon $A^+$}.
\newblock \emph{Phys. Lett. B}, 122:\penalty0 455, 1983.
\newblock \doi{10.1016/0370-2693(83)91601-5}.

\bibitem[Biagi et~al.(1985)]{Biagi:1984mu}
S.~F. Biagi et~al.
\newblock {Properties of the Charmed Strange Baryon $A^+$ and Evidence for the
  Charmed Doubly Strange Baryon $T^0$ at 2.74 GeV/$c^2$}.
\newblock \emph{Z. Phys. C}, 28:\penalty0 175, 1985.
\newblock \doi{10.1007/BF01575721}.

\bibitem[Albrecht et~al.(1993)]{ARGUS:1993vtm}
H.~Albrecht et~al.
\newblock {Observation of a new charmed baryon}.
\newblock \emph{Phys. Lett. B}, 317:\penalty0 227--232, 1993.
\newblock \doi{10.1016/0370-2693(93)91598-H}.

\bibitem[Edwards et~al.(1995)]{CLEO:1994oxm}
K.~W. Edwards et~al.
\newblock {Observation of excited charmed baryon states decaying to
  $\Lambda_c^+ \pi^+ \pi^-$}.
\newblock \emph{Phys. Rev. Lett.}, 74:\penalty0 3331--3335, 1995.
\newblock \doi{10.1103/PhysRevLett.74.3331}.

\bibitem[Albrecht et~al.(1997)]{ARGUS:1997snv}
H.~Albrecht et~al.
\newblock {Evidence for $\Lambda_c (2593)^{+}$ production}.
\newblock \emph{Phys. Lett. B}, 402:\penalty0 207--212, 1997.
\newblock \doi{10.1016/S0370-2693(97)00503-0}.

\bibitem[Alexander et~al.(1999)]{CLEO:1999msf}
J.~P. Alexander et~al.
\newblock {Evidence of new states decaying into $\Xi_c^{*} \pi$}.
\newblock \emph{Phys. Rev. Lett.}, 83:\penalty0 3390--3393, 1999.
\newblock \doi{10.1103/PhysRevLett.83.3390}.

\bibitem[Artuso et~al.(2001)]{CLEO:2000mbh}
M.~Artuso et~al.
\newblock {Observation of new states decaying into $\Lambda_c^{+} \pi^-
  \pi^+$}.
\newblock \emph{Phys. Rev. Lett.}, 86:\penalty0 4479--4482, 2001.
\newblock \doi{10.1103/PhysRevLett.86.4479}.

\bibitem[Csorna et~al.(2001)]{CLEO:2000ibb}
S.~E. Csorna et~al.
\newblock {Evidence of new states decaying into $\Xi'_c \pi$}.
\newblock \emph{Phys. Rev. Lett.}, 86:\penalty0 4243--4246, 2001.
\newblock \doi{10.1103/PhysRevLett.86.4243}.

\bibitem[Chistov et~al.(2006)]{Belle:2006edu}
R.~Chistov et~al.
\newblock {Observation of new states decaying into $\Lambda_c^{+} K^{-}
  \pi^{+}$ and $\Lambda_c^{+} K^{0}_{S} \pi^{-}$ }.
\newblock \emph{Phys. Rev. Lett.}, 97:\penalty0 162001, 2006.
\newblock \doi{10.1103/PhysRevLett.97.162001}.

\bibitem[Lesiak et~al.(2008)]{Belle:2008yxs}
T.~Lesiak et~al.
\newblock {Measurement of masses of the $\Xi_c (2645)$ and $\Xi_c(2815)$
  baryons and observation of $\Xi_c(2980) \to \Xi_c(2645) \pi$}.
\newblock \emph{Phys. Lett. B}, 665:\penalty0 9--15, 2008.
\newblock \doi{10.1016/j.physletb.2008.05.055}.

\bibitem[Aaltonen et~al.(2011)]{CDF:2011zbc}
T.~Aaltonen et~al.
\newblock {Measurements of the properties of $\Lambda_c(2595)$,
  $\Lambda_c(2625)$, $\Sigma_c(2455)$, and $\Sigma_c(2520)$ baryons}.
\newblock \emph{Phys. Rev. D}, 84:\penalty0 012003, 2011.
\newblock \doi{10.1103/PhysRevD.84.012003}.

\bibitem[{R.~Aaij \textit{et al.} [LHCb Collaboration]}(2017)]{LHCb:2017uwr}
{R.~Aaij \textit{et al.} [LHCb Collaboration]}.
\newblock {Observation of five new narrow $\Omega_c^0$ states decaying to
  $\Xi_c^+ K^-$}.
\newblock \emph{Phys. Rev. Lett.}, 118\penalty0 (18):\penalty0 182001, 2017.
\newblock \doi{10.1103/PhysRevLett.118.182001}.

\bibitem[Yelton et~al.(2018)]{Belle:2017ext}
J.~Yelton et~al.
\newblock {Observation of Excited $\Omega_c$ Charmed Baryons in $e^+e^-$
  Collisions}.
\newblock \emph{Phys. Rev. D}, 97\penalty0 (5):\penalty0 051102, 2018.
\newblock \doi{10.1103/PhysRevD.97.051102}.

\bibitem[{R.~Aaij \textit{et al.} [LHCb Collaboration]}(2020)]{LHCb:2020iby}
{R.~Aaij \textit{et al.} [LHCb Collaboration]}.
\newblock {Observation of New $\Xi_c^0$ Baryons Decaying to $\Lambda_c^+ K^-$}.
\newblock \emph{Phys. Rev. Lett.}, 124\penalty0 (22):\penalty0 222001, 2020.
\newblock \doi{10.1103/PhysRevLett.124.222001}.

\bibitem[{R.~Aaij \textit{et al.} [LHCb Collaboration]}(2023)]{LHCb:2023sxp}
{R.~Aaij \textit{et al.} [LHCb Collaboration]}.
\newblock {Observation of New $\Omega_c^{0}$ States Decaying to the
  $\Xi_c^{+}K^{-}$ Final State}.
\newblock \emph{Phys. Rev. Lett.}, 131\penalty0 (13):\penalty0 131902, 2023.
\newblock \doi{10.1103/PhysRevLett.131.131902}.

\bibitem[Aaij et~al.(2025{\natexlab{b}})]{LHCb:2025mge}
Roel Aaij et~al.
\newblock {Observation of a New Charmed Baryon Decaying to $\Xi_c^{+} \pi^{-}
  \pi^{+}$}.
\newblock \emph{Phys. Rev. Lett.}, 135\penalty0 (16):\penalty0 161901,
  2025{\natexlab{b}}.
\newblock \doi{10.1103/gghl-m6fm}.

\bibitem[Groote et~al.(1997)Groote, Korner, and Yakovlev]{Groote:1996em}
S.~Groote, J.~G. Korner, and Oleg~I. Yakovlev.
\newblock {QCD sum rules for heavy baryons at next-to-leading order in
  $\alpha_s$ }.
\newblock \emph{Phys. Rev. D}, 55:\penalty0 3016--3026, 1997.
\newblock \doi{10.1103/PhysRevD.55.3016}.

\bibitem[Gerasyuta and Ivanov(1999)]{Gerasyuta:1999pc}
S.~M. Gerasyuta and D.~V. Ivanov.
\newblock {Charmed baryons in bootstrap quark model}.
\newblock \emph{Nuovo Cim. A}, 112:\penalty0 261--276, 1999.
\newblock \doi{10.1007/BF03035848}.

\bibitem[Wang and Huang(2003)]{Wang:2003zp}
Dao-Wei Wang and Ming-Qiu Huang.
\newblock {Excited heavy baryon masses to order $\Lambda_{QCD} / m_Q$ from QCD
  sum rules}.
\newblock \emph{Phys. Rev. D}, 68:\penalty0 034019, 2003.
\newblock \doi{10.1103/PhysRevD.68.034019}.

\bibitem[Brown et~al.(2014)Brown, Detmold, Meinel, and Orginos]{Brown:2014ena}
Zachary~S. Brown, William Detmold, Stefan Meinel, and Kostas Orginos.
\newblock {Charmed bottom baryon spectroscopy from lattice QCD}.
\newblock \emph{Phys. Rev. D}, 90\penalty0 (9):\penalty0 094507, 2014.
\newblock \doi{10.1103/PhysRevD.90.094507}.

\bibitem[Padmanath et~al.(2013)Padmanath, Edwards, Mathur, and
  Peardon]{Padmanath:2013bla}
M.~Padmanath, Robert~G. Edwards, Nilmani Mathur, and Michael Peardon.
\newblock {Excited-state spectroscopy of singly, doubly and triply-charmed
  baryons from lattice QCD}.
\newblock In \emph{{6th International Workshop on Charm Physics}}, 11 2013.

\bibitem[Bahtiyar et~al.(2020)Bahtiyar, Can, Erkol, Gubler, Oka, and
  Takahashi]{Bahtiyar:2020uuj}
Huseyin Bahtiyar, Kadir~Utku Can, Guray Erkol, Philipp Gubler, Makoto Oka, and
  Toru~T. Takahashi.
\newblock {Charmed baryon spectrum from lattice QCD near the physical point}.
\newblock \emph{Phys. Rev. D}, 102\penalty0 (5):\penalty0 054513, 2020.
\newblock \doi{10.1103/PhysRevD.102.054513}.

\bibitem[Yang and Chen(2021)]{Yang:2021lce}
Hui-Min Yang and Hua-Xing Chen.
\newblock {$P$-wave charmed baryons of the $SU(3)$ flavor $6_F$}.
\newblock \emph{Phys. Rev. D}, 104\penalty0 (3):\penalty0 034037, 2021.
\newblock \doi{10.1103/PhysRevD.104.034037}.

\bibitem[Pan and Pan(2024)]{Pan:2023hwt}
Ji-Hai Pan and Jisi Pan.
\newblock {Investigation of the mass spectra of singly heavy baryons
  $\Sigma_Q$, $\Xi'_Q$, and $\Omega_Q$ $(Q=c,b)$ in the Regge trajectory
  model}.
\newblock \emph{Phys. Rev. D}, 109\penalty0 (7):\penalty0 076010, 2024.
\newblock \doi{10.1103/PhysRevD.109.076010}.

\bibitem[Yu et~al.(2023{\natexlab{a}})Yu, Li, Wang, Lu, and Yan]{Yu:2022ymb}
Guo-Liang Yu, Zhen-Yu Li, Zhi-Gang Wang, Jie Lu, and Meng Yan.
\newblock {Systematic analysis of single heavy baryons $\Lambda_Q$, $\Sigma_Q$
  and $\Omega_Q$}.
\newblock \emph{Nucl. Phys. B}, 990:\penalty0 116183, 2023{\natexlab{a}}.
\newblock \doi{10.1016/j.nuclphysb.2023.116183}.

\bibitem[Wang et~al.(2023)Wang, Lu, and Liu]{Wang:2023wii}
Zhi-Gang Wang, Fei Lu, and Yang Liu.
\newblock {Analysis of the D-wave $\Sigma $-type charmed baryon states with the
  QCD sum rules}.
\newblock \emph{Eur. Phys. J. C}, 83\penalty0 (8):\penalty0 689, 2023.
\newblock \doi{10.1140/epjc/s10052-023-11852-w}.

\bibitem[Duraes and Nielsen(2007)]{Duraes:2007te}
Francisco~O. Duraes and Marina Nielsen.
\newblock {QCD sum rules study of $\Xi_c$ and $\Xi_b$ baryons}.
\newblock \emph{Phys. Lett. B}, 658:\penalty0 40--44, 2007.
\newblock \doi{10.1016/j.physletb.2007.10.054}.

\bibitem[Pirjol and Yan(1997)]{Pirjol:1997nh}
Dan Pirjol and Tung-Mow Yan.
\newblock {Predictions for s-wave and p-wave heavy baryons from sum rules and
  constituent quark model. 1. Strong interactions}.
\newblock \emph{Phys. Rev. D}, 56:\penalty0 5483--5510, 1997.
\newblock \doi{10.1103/PhysRevD.56.5483}.

\bibitem[Chiladze and Falk(1997)]{Chiladze:1997ev}
George Chiladze and Adam~F. Falk.
\newblock {Phenomenology of new baryons with charm and strangeness}.
\newblock \emph{Phys. Rev. D}, 56:\penalty0 R6738--R6741, 1997.
\newblock \doi{10.1103/PhysRevD.56.R6738}.

\bibitem[Tawfiq et~al.(1998)Tawfiq, O'Donnell, and Korner]{Tawfiq:1998nk}
Salam Tawfiq, Patrick~J. O'Donnell, and J.~G. Korner.
\newblock {Charmed baryon strong coupling constants in a light front quark
  model}.
\newblock \emph{Phys. Rev. D}, 58:\penalty0 054010, 1998.
\newblock \doi{10.1103/PhysRevD.58.054010}.

\bibitem[Ivanov et~al.(1998)Ivanov, Korner, Lyubovitskij, and
  Rusetsky]{Ivanov:1998qe}
Mikhail~A. Ivanov, J.~G. Korner, Valery~E. Lyubovitskij, and A.~G. Rusetsky.
\newblock {One pion charm baryon transitions in a relativistic three quark
  model}.
\newblock \emph{Phys. Lett. B}, 442:\penalty0 435--442, 1998.
\newblock \doi{10.1016/S0370-2693(98)01245-3}.

\bibitem[Albertus et~al.(2005)Albertus, Hernandez, Nieves, and
  Verde-Velasco]{Albertus:2005zy}
C.~Albertus, E.~Hernandez, J.~Nieves, and J.~M. Verde-Velasco.
\newblock {Study of the strong $\Sigma_c \to \Lambda_c \pi$, $\Sigma_c^{*} \to
  \Lambda_c \pi$ and $\Xi_c^{*} \to \Xi_c \pi$ decays in a nonrelativistic
  quark model}.
\newblock \emph{Phys. Rev. D}, 72:\penalty0 094022, 2005.
\newblock \doi{10.1103/PhysRevD.72.094022}.

\bibitem[Cheng and Chua(2007)]{Cheng:2006dk}
Hai-Yang Cheng and Chun-Khiang Chua.
\newblock {Strong Decays of Charmed Baryons in Heavy Hadron Chiral Perturbation
  Theory}.
\newblock \emph{Phys. Rev. D}, 75:\penalty0 014006, 2007.
\newblock \doi{10.1103/PhysRevD.75.014006}.

\bibitem[Guo et~al.(2008)Guo, Wei, and Wu]{Guo:2007qu}
Xin-Heng Guo, Ke-Wei Wei, and Xing-Hua Wu.
\newblock {Strong decays of heavy baryons in Bethe-Salpeter formalism}.
\newblock \emph{Phys. Rev. D}, 77:\penalty0 036003, 2008.
\newblock \doi{10.1103/PhysRevD.77.036003}.

\bibitem[Ye et~al.(2017)Ye, Zhao, and Zhang]{Ye:2017yvl}
Dan-Dan Ye, Ze~Zhao, and Ailin Zhang.
\newblock {Study of $P$-wave excitations of observed charmed strange baryons}.
\newblock \emph{Phys. Rev. D}, 96\penalty0 (11):\penalty0 114009, 2017.
\newblock \doi{10.1103/PhysRevD.96.114009}.

\bibitem[Zhao et~al.(2017)Zhao, Ye, and Zhang]{Zhao:2017fov}
Ze~Zhao, Dan-Dan Ye, and Ailin Zhang.
\newblock {Hadronic decay properties of newly observed $\Omega_c$ baryons}.
\newblock \emph{Phys. Rev. D}, 95\penalty0 (11):\penalty0 114024, 2017.
\newblock \doi{10.1103/PhysRevD.95.114024}.

\bibitem[Yang and Chen(2024)]{Yang:2023fsc}
Hui-Min Yang and Hua-Xing Chen.
\newblock {2P-wave charmed baryons from QCD sum rules}.
\newblock \emph{Phys. Rev. D}, 109\penalty0 (3):\penalty0 036032, 2024.
\newblock \doi{10.1103/PhysRevD.109.036032}.

\bibitem[Yu et~al.(2023{\natexlab{b}})Yu, Meng, Li, Wang, and Jie]{Yu:2023bxn}
Guo-Liang Yu, Yan Meng, Zhen-Yu Li, Zhi-Gang Wang, and Lu~Jie.
\newblock {Strong decay properties of single heavy baryons $\Lambda_Q$,
  $\Sigma_Q$ and $\Omega_Q$}.
\newblock \emph{Int. J. Mod. Phys. A}, 38\penalty0 (15n16):\penalty0 2350082,
  2023{\natexlab{b}}.
\newblock \doi{10.1142/S0217751X23500823}.

\bibitem[Wang et~al.(2025)Wang, Luo, Chen, Cui, Tan, and Zhou]{Wang:2024rai}
Yi-Jie Wang, Xuan Luo, Hua-Xing Chen, Er-Liang Cui, Wei-Han Tan, and Zhi-Yong
  Zhou.
\newblock {Strong decay properties of P-wave single bottom baryons of the SU(3)
  flavor antitriplet ${\bar{3}}_{\rm F}$}.
\newblock \emph{Phys. Rev. D}, 111\penalty0 (7):\penalty0 076003, 2025.
\newblock \doi{10.1103/PhysRevD.111.076003}.

\bibitem[Blechman et~al.(2003)Blechman, Falk, Pirjol, and
  Yelton]{Blechman:2003mq}
Andrew~E. Blechman, Adam~F. Falk, Dan Pirjol, and John~M. Yelton.
\newblock {Threshold effects in excited charmed baryon decays}.
\newblock \emph{Phys. Rev. D}, 67:\penalty0 074033, 2003.
\newblock \doi{10.1103/PhysRevD.67.074033}.

\bibitem[Ortega et~al.(2013)Ortega, Entem, and Fernandez]{Ortega:2012cx}
P.~G. Ortega, D.~R. Entem, and F.~Fernandez.
\newblock {Quark model description of the $\Lambda_c(2940)^+$ as a molecular
  $D^*N$ state and the possible existence of the$ \Lambda_b(6248)$}.
\newblock \emph{Phys. Lett. B}, 718:\penalty0 1381--1384, 2013.
\newblock \doi{10.1016/j.physletb.2012.12.025}.

\bibitem[Cheng and Chua(2015)]{Cheng:2015naa}
Hai-Yang Cheng and Chun-Khiang Chua.
\newblock {Strong Decays of Charmed Baryons in Heavy Hadron Chiral Perturbation
  Theory: An Update}.
\newblock \emph{Phys. Rev. D}, 92\penalty0 (7):\penalty0 074014, 2015.
\newblock \doi{10.1103/PhysRevD.92.074014}.

\bibitem[Aliev et~al.(2018)Aliev, Azizi, and Sundu]{Aliev:2018ube}
T.~M. Aliev, K.~Azizi, and H.~Sundu.
\newblock {On the nature of $\Xi_{c}(2930)$}.
\newblock \emph{Eur. Phys. J. A}, 54\penalty0 (9):\penalty0 159, 2018.
\newblock \doi{10.1140/epja/i2018-12593-3}.

\bibitem[L{\"u}(2020)]{Lu:2020ivo}
Qi-Fang L{\"u}.
\newblock {Canonical interpretations of the newly observed $\Xi _c(2923)^0$,
  $\Xi _c(2939)^0$, and $\Xi _c(2965)^0$ resonances}.
\newblock \emph{Eur. Phys. J. C}, 80\penalty0 (10):\penalty0 921, 2020.
\newblock \doi{10.1140/epjc/s10052-020-08488-5}.

\bibitem[Chen et~al.(2021)Chen, Luo, and Liu]{Chen:2021eyk}
Bing Chen, Si-Qiang Luo, and Xiang Liu.
\newblock {Universal behavior of mass gaps existing in the single heavy baryon
  family}.
\newblock \emph{Eur. Phys. J. C}, 81\penalty0 (5):\penalty0 474, 2021.
\newblock \doi{10.1140/epjc/s10052-021-09234-1}.

\bibitem[Zhang et~al.(2023)Zhang, Liu, Luo, Wang, Wang, and Xu]{Zhang:2022pxc}
Zi-Le Zhang, Zhan-Wei Liu, Si-Qiang Luo, Fu-Lai Wang, Bo~Wang, and Hao Xu.
\newblock {$\Lambda_c(2910)$ and $\Lambda_c(2940)$ as conventional baryons
  dressed with the $D^{*}N$ channel}.
\newblock \emph{Phys. Rev. D}, 107\penalty0 (3):\penalty0 034036, 2023.
\newblock \doi{10.1103/PhysRevD.107.034036}.

\bibitem[Jessop et~al.(1999)]{CLEO:1998wvk}
C.~P. Jessop et~al.
\newblock {Observation of two narrow states decaying into $\Xi^{+}_c \gamma$
  and $\Xi^{0}_c \gamma$}.
\newblock \emph{Phys. Rev. Lett.}, 82:\penalty0 492--496, 1999.
\newblock \doi{10.1103/PhysRevLett.82.492}.

\bibitem[Aubert et~al.(2006)]{BaBar:2006pve}
Bernard Aubert et~al.
\newblock {Observation of an excited charm baryon $\Omega_c^*$ decaying to
  $\Omega_c^0 \gamma$}.
\newblock \emph{Phys. Rev. Lett.}, 97:\penalty0 232001, 2006.
\newblock \doi{10.1103/PhysRevLett.97.232001}.

\bibitem[Solovieva et~al.(2009)]{Solovieva:2008fw}
E.~Solovieva et~al.
\newblock {Study of $\Omega_c^{0}$ and $\Omega_c^{*0}$ Baryons at Belle}.
\newblock \emph{Phys. Lett. B}, 672:\penalty0 1--5, 2009.
\newblock \doi{10.1016/j.physletb.2008.12.062}.

\bibitem[Yelton et~al.(2020)]{Belle:2020ozq}
J.~Yelton et~al.
\newblock {Study of electromagnetic decays of orbitally excited $\Xi_c$
  baryons}.
\newblock \emph{Phys. Rev. D}, 102\penalty0 (7):\penalty0 071103, 2020.
\newblock \doi{10.1103/PhysRevD.102.071103}.

\bibitem[Cheng et~al.(1993)Cheng, Cheung, Lin, Lin, Yan, and Yu]{Cheng:1992xi}
Hai-Yang Cheng, Chi-Yee Cheung, Guey-Lin Lin, Y.~C. Lin, Tung-Mow Yan, and
  Hoi-Lai Yu.
\newblock Chiral lagrangians for radiative decays of heavy hadrons.
\newblock \emph{Phys. Rev. D}, 47:\penalty0 1030--1042, Feb 1993.
\newblock \doi{10.1103/PhysRevD.47.1030}.
\newblock URL \url{https://link.aps.org/doi/10.1103/PhysRevD.47.1030}.

\bibitem[{Ba\~nuls, M. C. and Pich, A. and Scimemi, I.}(2000)]{Banuls:1999br}
{Ba\~nuls, M. C. and Pich, A. and Scimemi, I.}
\newblock Electromagnetic decays of heavy baryons.
\newblock \emph{Phys. Rev. D}, 61:\penalty0 094009, Apr 2000.
\newblock \doi{10.1103/PhysRevD.61.094009}.
\newblock URL \url{https://link.aps.org/doi/10.1103/PhysRevD.61.094009}.

\bibitem[Jiang et~al.(2015)Jiang, Chen, and Zhu]{Jiang:2015xqa}
Nan Jiang, Xiao-Lin Chen, and Shi-Lin Zhu.
\newblock Electromagnetic decays of the charmed and bottom baryons in chiral
  perturbation theory.
\newblock \emph{Phys. Rev. D}, 92:\penalty0 054017, Sep 2015.
\newblock \doi{10.1103/PhysRevD.92.054017}.
\newblock URL \url{https://link.aps.org/doi/10.1103/PhysRevD.92.054017}.

\bibitem[Wang et~al.(2019)Wang, Meng, and Zhu]{Wang:2018cre}
Guang-Juan Wang, Lu~Meng, and Shi-Lin Zhu.
\newblock {Radiative decays of the singly heavy baryons in chiral perturbation
  theory}.
\newblock \emph{Phys. Rev. D}, 99\penalty0 (3):\penalty0 034021, 2019.
\newblock \doi{10.1103/PhysRevD.99.034021}.

\bibitem[Zhu and Dai(1999)]{Zhu:1998ih}
Shi-Lin Zhu and Yuan-Ben Dai.
\newblock {Radiative decays of heavy hadrons from light-cone QCD sum rules in
  the leading order of HQET}.
\newblock \emph{Phys. Rev. D}, 59:\penalty0 114015, May 1999.
\newblock \doi{10.1103/PhysRevD.59.114015}.
\newblock URL \url{https://link.aps.org/doi/10.1103/PhysRevD.59.114015}.

\bibitem[Wang(2010{\natexlab{a}})]{Wang:2010xfj}
Zhi-Gang Wang.
\newblock {Analysis of the vertexes $\Xi_{Q}^{*} \Xi'_{Q}V$, $\Sigma_{Q}^{*}
  \Sigma_{Q} V$ and radiative decays $\Xi_Q^{*} \to \Xi'_Q \gamma$,
  $\Sigma_Q^{*} \to \Sigma_Q \gamma$ }.
\newblock \emph{Eur. Phys. J. A}, 44:\penalty0 105--117, 2010{\natexlab{a}}.
\newblock \doi{10.1140/epja/i2010-10952-8}.

\bibitem[Wang(2010{\natexlab{b}})]{Wang:2009cd}
Zhi-Gang Wang.
\newblock {Analysis of the vertices ${\Omega_Q}^{*} \Omega_Q \phi$ and
  radiative decays ${\Omega_Q}^{*} \to \Omega_Q \gamma$}.
\newblock \emph{Phys. Rev. D}, 81:\penalty0 036002, Feb 2010{\natexlab{b}}.
\newblock \doi{10.1103/PhysRevD.81.036002}.
\newblock URL \url{https://link.aps.org/doi/10.1103/PhysRevD.81.036002}.

\bibitem[Aliev et~al.(2009)Aliev, Azizi, and Ozpineci]{Aliev:2009jt}
T.~M. Aliev, K.~Azizi, and A.~Ozpineci.
\newblock {Radiative decays of the heavy flavored baryons in light cone QCD sum
  rules}.
\newblock \emph{Phys. Rev. D}, 79:\penalty0 056005, Mar 2009.
\newblock \doi{10.1103/PhysRevD.79.056005}.
\newblock URL \url{https://link.aps.org/doi/10.1103/PhysRevD.79.056005}.

\bibitem[{T.~M.~Aliev, M.~Savc{\i} and V.~S.~Zamiralov}(2012)]{Aliev:2011bm}
{T.~M.~Aliev, M.~Savc{\i} and V.~S.~Zamiralov}.
\newblock Vector meson dominance and radiative decays of heavy spin-3/2 baryons
  to heavy spin-1/2 baryons.
\newblock \emph{Modern Physics Letters A}, 27\penalty0 (11):\penalty0 1250054,
  2012.
\newblock \doi{10.1142/S021773231250054X}.
\newblock URL \url{https://doi.org/10.1142/S021773231250054X}.

\bibitem[Aliev et~al.(2015)Aliev, Azizi, and Sundu]{Aliev:2014bma}
T.~M. Aliev, K.~Azizi, and H.~Sundu.
\newblock {Radiative $\Omega _{Q}^{*}\rightarrow \Omega _{Q}\gamma $ and $\Xi
  _{Q}^{*}\rightarrow \Xi ^{\prime}_Q \gamma $ transitions in light cone QCD}.
\newblock \emph{Eur. Phys. J. C}, 75\penalty0 (1):\penalty0 14, 2015.
\newblock \doi{10.1140/epjc/s10052-014-3229-0}.

\bibitem[Aliev et~al.(2016)Aliev, Barakat, and Savc{\i}]{Aliev:2016xvq}
T.~M. Aliev, T.~Barakat, and M.~Savc{\i}.
\newblock Analysis of the radiative decays
  ${\mathrm{\ensuremath{\Sigma}}}_{Q}\ensuremath{\rightarrow}{\mathrm{\ensuremath{\Lambda}}}_{Q}\ensuremath{\gamma}$
  and
  ${\mathrm{\ensuremath{\Xi}}}_{Q}^{\ensuremath{'}}\ensuremath{\rightarrow}{\mathrm{\ensuremath{\Xi}}}_{Q}\ensuremath{\gamma}$
  in light cone sum rules.
\newblock \emph{Phys. Rev. D}, 93:\penalty0 056007, Mar 2016.
\newblock \doi{10.1103/PhysRevD.93.056007}.
\newblock URL \url{https://link.aps.org/doi/10.1103/PhysRevD.93.056007}.

\bibitem[Bernotas and \ifmmode~\check{S}\else
  \v{S}\fi{}imonis(2013)]{Bernotas:2013eia}
A.~Bernotas and V.~\ifmmode~\check{S}\else \v{S}\fi{}imonis.
\newblock {Radiative M1 transitions of heavy baryons in the bag model}.
\newblock \emph{Phys. Rev. D}, 87:\penalty0 074016, Apr 2013.
\newblock \doi{10.1103/PhysRevD.87.074016}.
\newblock URL \url{https://link.aps.org/doi/10.1103/PhysRevD.87.074016}.

\bibitem[Shah et~al.(2016)Shah, Thakkar, Rai, and Vinodkumar]{Shah:2016nxi}
Zalak Shah, Kaushal Thakkar, Ajay~Kumar Rai, and P.~C. Vinodkumar.
\newblock {Mass spectra and Regge trajectories of $\Lambda_{c}^{+}$,
  $\Sigma_{c}^{0}$, $\Xi_{c}^{0}$ and $\Omega_{c}^{0}$ baryons}.
\newblock \emph{Chin. Phys. C}, 40\penalty0 (12):\penalty0 123102, 2016.
\newblock \doi{10.1088/1674-1137/40/12/123102}.

\bibitem[Gandhi et~al.(2020)Gandhi, Shah, and Rai]{Gandhi:2019xfw}
Keval Gandhi, Zalak Shah, and Ajay~Kumar Rai.
\newblock {Spectrum of nonstrange singly charmed baryons in the constituent
  quark model}.
\newblock \emph{Int. J. Theor. Phys.}, 59\penalty0 (4):\penalty0 1129--1156,
  2020.
\newblock \doi{10.1007/s10773-020-04394-4}.

\bibitem[Gandhi and Rai(2020)]{Gandhi:2019bju}
Keval Gandhi and Ajay~Kumar Rai.
\newblock {Spectrum of strange singly charmed baryons in the constituent quark
  model}.
\newblock \emph{Eur. Phys. J. Plus}, 135\penalty0 (2):\penalty0 213, 2020.
\newblock \doi{10.1140/epjp/s13360-020-00141-0}.

\bibitem[Yang and Kim(2020)]{Yang:2019tst}
Ghil-Seok Yang and Hyun-Chul Kim.
\newblock {Magnetic transitions and radiative decays of singly heavy baryons}.
\newblock \emph{Phys. Lett. B}, 801:\penalty0 135142, 2020.
\newblock \doi{10.1016/j.physletb.2019.135142}.

\bibitem[Kim et~al.(2021)Kim, Kim, Yang, and Oka]{Kim:2021xpp}
June-Young Kim, Hyun-Chul Kim, Ghil-Seok Yang, and Makoto Oka.
\newblock {Electromagnetic transitions of the singly charmed baryons with spin
  3/2}.
\newblock \emph{Phys. Rev. D}, 103\penalty0 (7):\penalty0 074025, 2021.
\newblock \doi{10.1103/PhysRevD.103.074025}.

\bibitem[Chow(1996)]{Chow:1995nw}
Chi-Keung Chow.
\newblock Radiative decays of excited ${\ensuremath{\Lambda}}_{Q}$ baryons in
  the bound state picture.
\newblock \emph{Phys. Rev. D}, 54:\penalty0 3374--3376, Sep 1996.
\newblock \doi{10.1103/PhysRevD.54.3374}.
\newblock URL \url{https://link.aps.org/doi/10.1103/PhysRevD.54.3374}.

\bibitem[Gamermann et~al.(2011)Gamermann, Jim\'enez-Tejero, and
  Ramos]{Gamermann:2010ga}
D.~Gamermann, C.~E. Jim\'enez-Tejero, and A.~Ramos.
\newblock Radiative decays of dynamically generated charmed baryons.
\newblock \emph{Phys. Rev. D}, 83:\penalty0 074018, Apr 2011.
\newblock \doi{10.1103/PhysRevD.83.074018}.
\newblock URL \url{https://link.aps.org/doi/10.1103/PhysRevD.83.074018}.

\bibitem[Zhu(2000)]{Zhu:2000py}
Shi-Lin Zhu.
\newblock {Strong and electromagnetic decays of p-wave heavy baryons
  $\Lambda_{c1}$, $\Lambda^{*}_{c1}$}.
\newblock \emph{Phys. Rev. D}, 61:\penalty0 114019, 2000.
\newblock \doi{10.1103/PhysRevD.61.114019}.

\bibitem[Luo et~al.(2025{\natexlab{a}})Luo, Wang, and Chen]{Luo:2025jpn}
Xuan Luo, Yi-Jie Wang, and Hua-Xing Chen.
\newblock {P-wave single charmed baryons of the SU(3) flavor ${\bar{3}}_{\rm
  F}$}.
\newblock \emph{Phys. Rev. D}, 111\penalty0 (9):\penalty0 094039,
  2025{\natexlab{a}}.
\newblock \doi{10.1103/p48x-mbnm}.

\bibitem[Luo et~al.(2025{\natexlab{b}})Luo, Chen, Cui, Yang, Zhou, and
  Zhou]{Luo:2025pzb}
Xuan Luo, Hua-Xing Chen, Er-Liang Cui, Hui-Min Yang, Dan Zhou, and Zhi-Yong
  Zhou.
\newblock {Radiative decays of $P$-wave charmed baryons in the $SU(3)$ flavor
  $\bf6_F$ representation}.
\newblock \emph{Phys. Rev. D}, 112\penalty0 (9):\penalty0 096028,
  2025{\natexlab{b}}.
\newblock \doi{10.1103/5w6h-xr48}.

\bibitem[Bijker et~al.(2022)Bijker, Garc\'\i{}a-Tecocoatzi, Giachino,
  Ortiz-Pacheco, and Santopinto]{Bijker:2020tns}
Roelof Bijker, Hugo Garc\'\i{}a-Tecocoatzi, Alessandro Giachino, Emmanuel
  Ortiz-Pacheco, and Elena Santopinto.
\newblock {Masses and decay widths of \ensuremath{\Xi}c/b and
  \ensuremath{\Xi}c/b' baryons}.
\newblock \emph{Phys. Rev. D}, 105\penalty0 (7):\penalty0 074029, 2022.
\newblock \doi{10.1103/PhysRevD.105.074029}.

\bibitem[Ortiz-Pacheco and Bijker(2023)]{Ortiz-Pacheco:2023kjn}
Emmanuel Ortiz-Pacheco and Roelof Bijker.
\newblock {Masses and radiative decay widths of $S$- and $P$-wave singly,
  doubly, and triply heavy charm and bottom baryons}.
\newblock \emph{Phys. Rev. D}, 108:\penalty0 054014, Sep 2023.
\newblock \doi{10.1103/PhysRevD.108.054014}.
\newblock URL \url{https://link.aps.org/doi/10.1103/PhysRevD.108.054014}.

\bibitem[Garc{\'\i}a-Tecocoatzi et~al.(2025)Garc{\'\i}a-Tecocoatzi,
  Ramirez-Morales, Rivero-Acosta, Santopinto, and
  Vaquera-Araujo]{Garcia-Tecocoatzi:2025fxp}
H.~Garc{\'\i}a-Tecocoatzi, A.~Ramirez-Morales, A.~Rivero-Acosta, E.~Santopinto,
  and C.~A. Vaquera-Araujo.
\newblock {$\Xi_c(2790)^{+/0}$ and $\Xi_c(2815)^{+/0}$ radiative decays}.
\newblock \emph{Phys. Lett. B}, 868:\penalty0 139666, 2025.
\newblock \doi{10.1016/j.physletb.2025.139666}.

\bibitem[Wang et~al.(2017{\natexlab{a}})Wang, Xiao, Zhong, and
  Zhao]{Wang:2017hej}
Kai-Lei Wang, Li-Ye Xiao, Xian-Hui Zhong, and Qiang Zhao.
\newblock {Understanding the newly observed $\Omega_c$ states through their
  decays}.
\newblock \emph{Phys. Rev. D}, 95\penalty0 (11):\penalty0 116010,
  2017{\natexlab{a}}.
\newblock \doi{10.1103/PhysRevD.95.116010}.

\bibitem[Wang et~al.(2017{\natexlab{b}})Wang, Yao, Zhong, and
  Zhao]{Wang:2017kfr}
Kai-Lei Wang, Ya-Xiong Yao, Xian-Hui Zhong, and Qiang Zhao.
\newblock {Strong and radiative decays of the low-lying $S$- and $P$-wave
  singly heavy baryons}.
\newblock \emph{Phys. Rev. D}, 96\penalty0 (11):\penalty0 116016,
  2017{\natexlab{b}}.
\newblock \doi{10.1103/PhysRevD.96.116016}.

\bibitem[Yao et~al.(2018)Yao, Wang, and Zhong]{Yao:2018jmc}
Ya-Xiong Yao, Kai-Lei Wang, and Xian-Hui Zhong.
\newblock {Strong and radiative decays of the low-lying $D$-wave singly heavy
  baryons}.
\newblock \emph{Phys. Rev. D}, 98\penalty0 (7):\penalty0 076015, 2018.
\newblock \doi{10.1103/PhysRevD.98.076015}.

\bibitem[Peng et~al.(2024)Peng, Luo, and Liu]{Peng:2024pyl}
Yu-Xin Peng, Si-Qiang Luo, and Xiang Liu.
\newblock {Refining radiative decay studies in singly heavy baryons}.
\newblock \emph{Phys. Rev. D}, 110\penalty0 (7):\penalty0 074034, 2024.
\newblock \doi{10.1103/PhysRevD.110.074034}.

\bibitem[Ivanov et~al.(1999{\natexlab{a}})Ivanov, Körner, and
  Lyubovitskij]{Ivanov:1998wj}
M.A. Ivanov, J.G. Körner, and V.E. Lyubovitskij.
\newblock One-photon transitions between heavy baryons in a relativistic
  three-quark model.
\newblock \emph{Physics Letters B}, 448\penalty0 (1):\penalty0 143--151,
  1999{\natexlab{a}}.
\newblock ISSN 0370-2693.
\newblock \doi{https://doi.org/10.1016/S0370-2693(99)00029-5}.
\newblock URL
  \url{https://www.sciencedirect.com/science/article/pii/S0370269399000295}.

\bibitem[Ivanov et~al.(1999{\natexlab{b}})Ivanov, K\"orner, Lyubovitskij, and
  Rusetsky]{Ivanov:1999bk}
M.~A. Ivanov, J.~G. K\"orner, V.~E. Lyubovitskij, and A.~G. Rusetsky.
\newblock Strong and radiative decays of heavy flavored baryons.
\newblock \emph{Phys. Rev. D}, 60:\penalty0 094002, Sep 1999{\natexlab{b}}.
\newblock \doi{10.1103/PhysRevD.60.094002}.
\newblock URL \url{https://link.aps.org/doi/10.1103/PhysRevD.60.094002}.

\bibitem[Tawfiq et~al.(2001)Tawfiq, K\"orner, and O'Donnell]{Tawfiq:1999cf}
Salam Tawfiq, J.~G. K\"orner, and Patrick~J. O'Donnell.
\newblock {Electromagnetic transitions of heavy baryons in the $SU(2N_{f} )
  \otimes O(3)$ symmetry}.
\newblock \emph{Phys. Rev. D}, 63:\penalty0 034005, Jan 2001.
\newblock \doi{10.1103/PhysRevD.63.034005}.
\newblock URL \url{https://link.aps.org/doi/10.1103/PhysRevD.63.034005}.

\bibitem[Ferraris et~al.(1995)Ferraris, Giannini, Pizzo, Santopinto, and
  Tiator]{Ferraris:1995ui}
M.~Ferraris, M.~M. Giannini, M.~Pizzo, E.~Santopinto, and L.~Tiator.
\newblock {A Three body force model for the baryon spectrum}.
\newblock \emph{Phys. Lett. B}, 364:\penalty0 231--238, 1995.
\newblock \doi{10.1016/0370-2693(95)01091-2}.

\bibitem[Santopinto et~al.(1997)Santopinto, Iachello, and
  Giannini]{Santopinto:1997jz}
E.~Santopinto, F.~Iachello, and M.~M. Giannini.
\newblock {Exactly solvable models of baryon spectroscopy}.
\newblock \emph{Nucl. Phys. A}, 623:\penalty0 100C--109C, 1997.
\newblock \doi{10.1016/S0375-9474(97)00427-2}.

\bibitem[Santopinto et~al.(1998)Santopinto, Iachello, and
  Giannini]{Santopinto:1998ma}
E.~Santopinto, F.~Iachello, and M.~M. Giannini.
\newblock {Nucleon form-factors in a simple three-body quark model}.
\newblock \emph{Eur. Phys. J. A}, 1:\penalty0 307--315, 1998.
\newblock \doi{10.1007/s100500050065}.

\bibitem[Giannini and Santopinto(2015)]{Giannini:2015zia}
M.~M. Giannini and E.~Santopinto.
\newblock {The hypercentral Constituent Quark Model and its application to
  baryon properties}.
\newblock \emph{Chin. J. Phys.}, 53:\penalty0 020301, 2015.
\newblock \doi{10.6122/CJP.20150120}.

\bibitem[Close and Copley(1970)]{Close:1970kt}
F.~E. Close and L.~A. Copley.
\newblock {Electromagnetic interactions of weakly bound composite systems}.
\newblock \emph{Nucl. Phys. B}, 19:\penalty0 477--500, 1970.
\newblock \doi{10.1016/0550-3213(70)90362-7}.

\bibitem[Garc\'\i{}a-Tecocoatzi et~al.(2024)Garc\'\i{}a-Tecocoatzi, Giachino,
  Ramirez-Morales, Rivero-Acosta, Santopinto, and
  Vaquera-Araujo]{Garcia-Tecocoatzi:2023btk}
H.~Garc\'\i{}a-Tecocoatzi, A.~Giachino, A.~Ramirez-Morales, Ailier
  Rivero-Acosta, E.~Santopinto, and Carlos~Alberto Vaquera-Araujo.
\newblock {Strong decay widths and mass spectra of the 1D, 2P and 2S singly
  bottom baryons}.
\newblock \emph{Phys. Rev. D}, 110\penalty0 (11):\penalty0 114005, 2024.
\newblock \doi{10.1103/PhysRevD.110.114005}.

\bibitem[Rivero-Acosta et~al.(2025)Rivero-Acosta, Garc\'\i{}a-Tecocoatzi,
  Ramirez-Morales, Santopinto, and Vaquera-Araujo]{Rivero-Acosta:2025drn}
Ailier Rivero-Acosta, H.~Garc\'\i{}a-Tecocoatzi, A.~Ramirez-Morales,
  E.~Santopinto, and Carlos~Alberto Vaquera-Araujo.
\newblock {Radiative decays of the second shell $\Lambda_b$ and $\Xi_b$ bottom
  baryons}.
\newblock \emph{Phys. Rev. D}, 112\penalty0 (7):\penalty0 072014, 5 2025.
\newblock \doi{10.1103/ng36-kn1p}.

\bibitem[Garcia-Tecocoatzi et~al.(2023)Garcia-Tecocoatzi, Giachino, Li,
  Ramirez-Morales, and Santopinto]{Garcia-Tecocoatzi:2022zrf}
H.~Garcia-Tecocoatzi, A.~Giachino, J.~Li, A.~Ramirez-Morales, and
  E.~Santopinto.
\newblock {Strong decay widths and mass spectra of charmed baryons}.
\newblock \emph{Phys. Rev. D}, 107\penalty0 (3):\penalty0 034031, 2023.
\newblock \doi{10.1103/PhysRevD.107.034031}.

\bibitem[Navas et~al.(2024)]{ParticleDataGroup:2024cfk}
S.~Navas et~al.
\newblock {Review of particle physics}.
\newblock \emph{Phys. Rev. D}, 110\penalty0 (3):\penalty0 030001, 2024.
\newblock \doi{10.1103/PhysRevD.110.030001}.

\bibitem[Santopinto et~al.(2019)Santopinto, Giachino, Ferretti,
  Garc\'\i{}a-Tecocoatzi, Bedolla, Bijker, and
  Ortiz-Pacheco]{Santopinto:2018ljf}
E.~Santopinto, A.~Giachino, J.~Ferretti, H.~Garc\'\i{}a-Tecocoatzi, M.~A.
  Bedolla, R.~Bijker, and E.~Ortiz-Pacheco.
\newblock {The $\varOmega _{ c}$-puzzle solved by means of quark model
  predictions}.
\newblock \emph{Eur. Phys. J. C}, 79\penalty0 (12):\penalty0 1012, 2019.
\newblock \doi{10.1140/epjc/s10052-019-7527-4}.

\bibitem[Giachino(2020)]{Giachino:2020dsj}
Alessandro Giachino.
\newblock \emph{{Multiquark States and Exotic Spectroscopy}}.
\newblock PhD thesis, Universita' Di Genova, Genoa U., 2020.

\bibitem[Rivero~Acosta(2025)]{RiveroAcosta:2025nfh}
Ailier Rivero~Acosta.
\newblock \emph{{Study of Heavy Baryons Phenomenology}}.
\newblock PhD thesis, Universit{\`a} degli studi di Genova, Italy and
  Universidad de Guanajuato, Mexico, 2025.

\bibitem[James and Roos(1975)]{James:1975dr}
F.~James and M.~Roos.
\newblock {Minuit: A System for Function Minimization and Analysis of the
  Parameter Errors and Correlations}.
\newblock \emph{Comput. Phys. Commun.}, 10:\penalty0 343--367, 1975.
\newblock \doi{10.1016/0010-4655(75)90039-9}.

\bibitem[Harris et~al.(2020)]{Harris:2020xlr}
Charles~R. Harris et~al.
\newblock {Array programming with NumPy}.
\newblock \emph{Nature}, 585\penalty0 (7825):\penalty0 357--362, 2020.
\newblock \doi{10.1038/s41586-020-2649-2}.

\bibitem[Chen et~al.(2017{\natexlab{b}})Chen, Liu, and Zhang]{Chen:2017aqm}
Bing Chen, Xiang Liu, and Ailin Zhang.
\newblock {Newly observed $\Lambda_c(2860)^+$ at LHCb and its
  \emph{D}-wave partners $\Lambda_c(2880)^+$, $\Xi_c(3055)^+$
  and $\Xi_c(3080)^+$}.
\newblock \emph{Phys. Rev. D}, 95\penalty0 (7):\penalty0 074022,
  2017{\natexlab{b}}.
\newblock \doi{10.1103/PhysRevD.95.074022}.

\end{thebibliography}


\end{document}